%% file: left_dim8_basis_v3.tex
\documentclass[11pt,a4paper]{article}
             
\usepackage{float}
\usepackage{jheppub} 

\usepackage{braket,slashed,bm}
\usepackage{array,multirow}
\usepackage[normalem]{ulem}
\usepackage{cancel}
\usepackage[table]{xcolor}
\usepackage[vcentermath]{youngtab}

\usepackage[T1]{fontenc} 

\usepackage{mathrsfs}

\usepackage{booktabs}
\usepackage{adjustbox}
\usepackage{mathtools}

\usepackage{soul}

\usepackage{arydshln}
\usepackage[utf8]{inputenc}

\def\nn{\nonumber\\ }

\renewcommand{\O}{\mathcal{O}}
\newcommand{\hc}{\mathrm{h.c.}}

\begin{document}

\title{Low-Energy Effective Field Theory below the Electroweak Scale: Dimension-8 Operators}

\author{Christopher W.~Murphy}
\affiliation{homepoint, Ann Arbor, MI 48105, U.S.A.}

\abstract{
We construct a complete basis of dimension-8 operators in the Low-Energy Effective Field Theory below the Electroweak Scale (LEFT).
We find there are 35058 dimension-8 operators in the LEFT for two generations of up-type quarks and three generations of down-type quarks, charged leptons, and left-handed neutrinos.
The existence of this operator basis is a necessary prerequisite for matching to the Standard Model Effective Field Theory at the dimension-8 level.
}
\maketitle


\section{Introduction}
\label{sec:intro}

The discovery of the Higgs boson at the Large Hadron Collider along the absence of an observation of physics beyond the Standard Model (SM) has left open questions surrounding the nature of electroweak (EW) symmetry breaking.
One such question concerns the Higgs boson itself: Is the Higgs boson part of an $SU(2)_L$ doublet with hypercharge $Y = 1/2$, or is it an electroweak singlet?
Measurements at energies well below the electroweak scale interpreted in an effective field theory (EFT) framework provide one way of answering this question.
The effective field theory approach to analyzing experimental data has become widely popular due to its model-independence and systematic improvability. 

The Low-Energy Effective Field Theory below the Electroweak Scale (LEFT), as its name implies, is correct the EFT to use to interpret measurements below the EW scale~\cite{Jenkins:2017jig, Jenkins:2017dyc, Dekens:2019ept, Liao:2020zyx}.
The LEFT gauge group is the product of QCD and QED, $SU(3) \times U(1)_Q$, with field strengths $G_{\mu\nu}^A$ and $F_{\mu\nu}$, respectively.
There are $n_u = 2$ generations of up-type quarks, $u_L,\, u_R$, and $n_d = n_e = n_\nu = 3$ generations of down-type quarks, charged leptons, and left-handed neutrinos, $d_L,\, d_R, e_L,\, e_R,\, \nu_L$.
Through the lens of the LEFT the question concerning the Higgs boson can be restated as: What is the UV completion of the LEFT?
A leading candidate is the Standard Model Effective Field Theory (SMEFT) in which the Higgs boson forms part of an $SU(2)_L$ doublet with $Y = 1/2$~\cite{Grzadkowski:2010es, Jenkins:2013zja, Jenkins:2013wua, Alonso:2013hga, Alonso:2014zka, Henning:2015alf, Liao:2016hru, Liao:2019tep, Murphy:2020rsh, Li:2020gnx, Li:2020xlh, Liao:2020jmn}.
The SMEFT has a larger gauge group, $SU(3) \times SU(2)_L \times U(1)_Y$, than the LEFT leading to the seven fermion fields of the LEFT (ignoring flavor) being encoded into only five SMEFT fermion fields, $l,\, e,\, q,\, u,\, d$.
This leads to correlations among LEFT Wilson coefficients matching to the SMEFT.
For example, the dimension-6 SMEFT contribution to the LEFT operator $(\bar \tau_L \gamma_\mu \nu_{L\tau}) (\bar c_R \gamma^\mu b_R)$ cannot induce lepton universality violation~\cite{Bernard:2006gy, Cirigliano:2009wk, Alonso:2015sja}.
Ref.~\cite{Jenkins:2017jig} tabulated the full set of correlations between dimension-6 LEFT coefficients that would arise if it is UV-completed by the SMEFT. 

By measuring observables sensitive to these coefficients one can test whether the SMEFT is the correct completion of the LEFT.
Here correlated coefficients means a linear combination of Wilson coefficients that generally should be $O(1 / \Lambda^2)$, but is instead zero when matched to the SMEFT at dimension-6.
However, as any measurement has a finite precision, it is important to first determine the size of the corrections to these correlations resulting from higher-order effects to design an experiment with the appropriate level of sensitivity.
For correlated dimension-6 coefficients typically what will happen when the matching is extended to dimension-8 is that a correlated coefficient will no longer be zero, but instead will be of order $O(v^2 / \Lambda^4)$.

Furthermore, many dimension-8 operators are subject to positivity bounds arising from fundamental principles of quantum field theory including unitarity and analyticity, see~\cite{Remmen:2019cyz, Remmen:2020vts, Almeida:2020ylr, Zhang:2020jyn, Fuks:2020ujk, Yamashita:2020gtt, Remmen:2020uze, Bellazzini:2020cot, Gu:2020ldn, Bonnefoy:2020yee} for some recent applications.
The LEFT is a valid EFT in its own right, independent of whatever its UV completion may be.
As such the Wilson coefficients of the LEFT are subject to their own set of positivity bounds.
In general, the constraints on the underlying physics from LEFT positivity bounds will not be identical to those arising from its UV completion, say, the SMEFT.
This is due to the fact that the matching between the LEFT and the SMEFT is rich in nature, and is not simply a matter of setting the Higgs field to its vacuum expectation value (vev).
In addition to these contact contributions to the matching there are also exchanges of SM particles with masses comparable to the EW scale, and the number possible exchange contributions to the matching increases with mass increasing dimension~\cite{Jenkins:2017jig, Dekens:2019ept, Liao:2020zyx}.

The first step in tackling these matching problems is to construct a complete basis of dimension-8 LEFT operators, which is the goal of this work.
We find there are 35058 operators.
The overwhelming majority of these operators, 34721 to be exact, contain four fermions along with either a field strength or two derivatives.
In contrast the SMEFT has fewer operators belonging to classes $\psi^4 X$ or $\psi^4 D^2$, 21303~\cite{Henning:2015alf, Murphy:2020rsh, Li:2020gnx}, which will lead to correlations among LEFT dimension-8 operators when matched to the SMEFT, repeating the behavior first seen at dimension-6.


\section{LEFT Dimension-8 Operators}
\label{sec:left}

The LEFT consists of all the gauge invariant, non-redundant operators of dimension-3 and higher that can be formed out of the subset of SM fields whose masses are parametrically smaller than the electroweak scale.
Additionally, the LEFT possesses an expansion in the mass dimension, $d$, of its operators with expansion parameter $1 / v^{d-4}$ for $d > 4$ where $v$ is the vev of the Higgs field \textit{in the SM}.
LEFT operators with two or more fermions can violate lepton number, and LEFT operators with four fermions can violate both baryon and lepton number. 
Ref.~\cite{Jenkins:2017jig} constructed a LEFT operator basis for operators up to and including dimension-6.
At dimension-5 there is one operator class $\psi^2 X$, while at dimension-6 there are two classes, $X^3$ and $\psi^4$.
The one-loop anomalous dimension matrix for these operators was computed by the same authors in Ref.~\cite{Jenkins:2017dyc}.
Ref.~\cite{Liao:2020zyx} constructed a LEFT basis for dimension-7 operators.
There are two classes of dimension-7 LEFT operators: $\psi^2 X^2$ and $\psi^4 D$.

Much effort has been devoted to counting the number of operators in an EFT~\cite{Jenkins:2009dy, Hanany:2010vu, Lehman:2015via, Lehman:2015coa, Henning:2015alf, Fonseca:2017lem, Henning:2017fpj}.
Specifically, we use the Mathematica program $\mathtt{Sym2Int}$~\cite{Fonseca:2017lem} to count the number of LEFT dimension-8 operators along with their baryon and lepton numbers.
$\mathtt{Sym2Int}$ also provides the flavor structures of the Lagrangian terms up to ambiguities due to integration by parts relations between operators with repeated fields.
We find the number of LEFT dimension-8 operators is
\begin{equation}
\label{eq:Nops}
N_{\rm operators} = 35058 = 21144 |_{\Delta B = 0}^{\Delta L = 0} + 5442 |_{\Delta B = 0}^{\Delta L = 2} + 4536 |_{\Delta B = 1}^{\Delta L = 1} + 3888 |_{\Delta B = 1}^{\Delta L = -1} + 48 |_{\Delta B = 0}^{\Delta L = 4}
\end{equation}
for the SM case of $n_u = 2$ and $n_d = n_e = n_\nu = 3$.
The counting for arbitrary $n_{\nu, e, u, d}$ is given in Table~\ref{tab:count_sum} of Appendix~\ref{sec:count}.

In this work we construct a complete basis of dimension-8 LEFT operators.
There are four operators classes to consider: $X^4$, $\psi^2 X^2 D$, $\psi^4 X$, and $\psi^4 D^2$.
Each class of operators contains subclasses with definite transformation properties under the Lorentz group.
For example, $\{X_L^4,\, X_R^4,\, X_L^2 X_R^2\} \in X^4$ where $X_L \sim (1, 0)$ and $X_R \sim (0, 1)$ under $SU(2)_l \times SU(2)_r$ of the Lorentz group.
The existence of complete bases of dimension-8 SMEFT operators~\cite{Murphy:2020rsh, Li:2020gnx} aides in the construct of a LEFT dimension-8 basis as all of the LEFT operator classes are also present in the SMEFT at dimension-8.
However not all of the types of LEFT operators are present in the SMEFT.
For example, some LEFT operators involving four fermions can violate lepton number while preserving baryon number, but in the SMEFT even mass dimensions operators (below $d = 10$) conserve baryon minus lepton number, $\Delta B - \Delta L = 0$~\cite{Kobach:2016ami, Helset:2019eyc}.

The LEFT Lagrangian terms with two or less fermions are a subset of their SMEFT counterparts, and can be constructed by relabeling some of the fields of the appropriate operators.
For definiteness we use the SMEFT dimension-8 basis of Ref.~\cite{Murphy:2020rsh} when constructing these LEFT operators, and refer the reader there for more detail on the construction of the operator basis.
Each of these Lagrangian terms has trivial flavor structure with $X^4$ and $\psi^2 X^2 D$ terms containing one and $n_f^2$ operators, respectively.
In total there are 23 $X^4$ operators and $2 n_\nu^2 + 4 n_e^2 + 20 n_u^2 + 20 n_d^2 = 314$ $\psi^2 X^2 D$ operators.
Table~\ref{tab:left8_x4} of Appendix~\ref{sec:basis} lists the $X^4$ operators, while Tables~\ref{tab:left8_psi2x2d_L} and~\ref{tab:left8_psi2x2d_R} contain the $\psi^2 X^2 D$ operators.

As there are a large number of operators with four fermions it is convenient to break these operators classes into different cases.
For operators that conserve both baryon and lepton number we again start from the SMEFT operators of Ref.~\cite{Murphy:2020rsh} and relabel fields as appropriate.
However the counting of operators in Lagrangian terms with repeated field does not translate from the SMEFT to the LEFT as there are five fermion fields in the SMEFT, whereas there are seven fermion fields in the LEFT.
The flavor structure and operator counting for each Lagrangian term with four fermions is given in Tables~\ref{tab:count_4psi}, \ref{tab:count_deltaL}, and~\ref{tab:count_deltaB}.
We make one change from the approach of Ref.~\cite{Murphy:2020rsh}, namely that all of the baryon and lepton number conserving operators in this work have lepton bilinear and/or quark bilinears, but they do not contain leptoquark bilinears.
There is one baryon number conserving term in~\cite{Murphy:2020rsh} with a leptoquark bilinear, and we use a Fierz identity to eliminate it in favor of operators with the tensor Dirac matrix $\sigma^{\mu\nu}$.
A special case are the subclass $\psi_L^4 D^2$ operators with all charged-lepton fields.
The lack of color indicies allows for a symmetric and antisymmetric flavor representation to be unambiguously combined into a single Lagrangian term.
In equations, for up-quarks there is $D^\mu (\bar u_L u_R) D_\mu (\bar u_L u_R)$ \textit{and} $D^\mu (\bar u_L T^A u_R) D_\mu (\bar u_L T^A u_R)$, whereas for charged-leptons there can clearly only be $D^\mu (\bar e_L e_R) D_\mu (\bar e_L e_R)$.
Similarly, the lack of color indices on the lepton fields makes some operators of subclass $\psi_L^2 \psi_R^2 D^2$  redundant.
One such example is
\begin{equation}
(\bar e_{Rp} \gamma^\mu \overleftrightarrow{D}^\rho e_{Rr}) (\bar e_{Rs} \gamma_\mu \overleftrightarrow{D}_\rho e_{Rt}) = \O_{\substack{e^4D^2 \\ prst}}^{(1,RR)} + 2 \O_{\substack{e^4D^2 \\ ptsr}}^{(1,RR)} .
\end{equation}

For operators that violate baryon and/or lepton number we start from our operator counting and instead  use set of Lorentz structures for each subclass of operator from Ref.~\cite{Li:2020gnx}.
We then use Fierz identities to eliminate leptoquark bilinears from the $\Delta L = 2$ operators with two left-handed and two right-handed fermions.
On the other hand, for $\Delta L = 2$ operators with four left-handed fermions (and their Hermitian conjugates) we only eliminate leptoquark bilinears when there are no repeated fields.
That is, for operators with two left-handed neutrinos and two left-handed quarks we make the choice to have leptoquark bilinears to avoid writing Lagrangian terms as sum of two terms.
As an example, the neutrinos in $\O_{\nu^2d^2F}^{(2,ST)}$ can be in a symmetric or an antisymmetric flavor representation.
\begin{align}
L_{\substack{\nu^2d^2F \\ prst}}^{(2,ST)} \O_{\substack{\nu^2d^2F \\ prst}}^{(2,ST)} &\sim L_{\substack{\nu^2d^2F \\ [pt]sr}}^{(2a,ST)} \O_{\substack{\nu^2d^2F \\ ptsr}}^{(2a,ST)} + L_{\substack{\nu^2d^2F \\ (pt)sr}}^{(2s,ST)} \O_{\substack{\nu^2d^2F \\ ptsr}}^{(2s,ST)} , \nn
\O_{\nu^2d^2F}^{(2a,ST)} &= (\nu_{Lp}^\top C \nu_{Lr}) (\bar d_{Rs} \sigma^{\mu\rho} d_{Lt}) F_{\mu\rho} , \nn
 \O_{\nu^2d^2F}^{(2s,ST)} &= (\nu_{Lp}^\top C \sigma^{\mu\lambda} \nu_{Lr}) (\bar d_{Rs} \sigma_{\lambda\rho} d_{Lt}) F_\mu^\rho ,
\end{align}
where $[\cdot]$ and $(\cdot)$ indicate anti-symmetrization and symmetrization respectively.

There are 24108 $\psi^4 X$ LEFT operators and 10613 $\psi^4 D^2$ LEFT operators. 
These classes have more operators than their SMEFT counterparts despite the handicap of having one fewer generation of up-type quarks.
As such there are correlations between LEFT Wilson coefficients when matching from the SMEFT.
Measuring processes sensitive to these coefficients is way one to test whether the SMEFT is the correct UV completion of the LEFT.

The baryon and lepton number conserving operators of class $\psi^4 X$ and subclass $\psi_L^2 \psi_R^2 X$ are enumerated in Tables~\ref{tab:left8_psi4x_LL}, \ref{tab:left8_psi4x_RR_LRRL}, \ref{tab:left8_psi4x_LRG}, and~\ref{tab:left8_psi4x_LRF}, while Table~\ref{tab:left8_psi4x_LRLR} contains the $\psi_L^4 X_L + \hc$ operators.
The baryon and lepton number conserving operators of class $\psi^4 D^2$ and subclass $\psi_L^2 \psi_R^2 D^2$ are given in Tables~\ref{tab:left8_psi4d2_LL_RR} and \ref{tab:left8_psi4d2_LR}, whereas Table~\ref{tab:left8_psi4d2_LR_LR} lists the $\psi_L^4 D^2 + \hc$ operators.
The $\psi^4 X$ and $\psi^4 D^2$ operators that violate lepton number, but preserve baryon number are tabulated in Tables~\ref{tab:left8_psi4x_dL2} and~\ref{tab:left8_psi4d2_dL2}, respectively.
On the other hand, the $\psi^4 D^2$ and $\psi^4 X$ operators that violate baryon and lepton number, but preserve $\Delta B - \Delta L$ are listed in Tables~\ref{tab:left8_psi4d2_dB1dL1} and~\ref{tab:left8_psi4x_dB1dL1}, respectively.
Finally, Table~\ref{tab:left8_dB1dLm1} contains all the $\Delta B = - \Delta L = 1$ operators.


\section{Conclusions}
\label{sec:con}

In this work we constructed a complete basis of dimension-8 operators in the LEFT.
We found there are 35058 operators, see Eq.~\eqref{eq:Nops} or Table~\ref{tab:count_sum}.
In addition to the operator counting, the construction of the LEFT operator basis was aided by the existence of the dimension-8 bases of SMEFT operators as all of the LEFT dimension-8 operator classes are also found in the SMEFT.

The existence of an operator basis is a prerequisite for matching to the SMEFT at the dimension-8 level.
For a given class of LEFT operators involving four fermions there are fewer SMEFT operators of the same class leading to correlations between LEFT Wilson coefficients when matched to the SMEFT.
Measuring observables sensitive to these coefficients is one way to test whether the SMEFT is the correct EFT extension of the SM and the correct UV-completion of the LEFT.
Additionally, the LEFT has its own set of positivity bounds on dimension-8 operators from unitarity, analyticity, and crossing symmetry.
We encourage the study of these bounds as they are potentially complementary to those of the SMEFT.

\section*{Acknowledgements}
\addcontentsline{toc}{section}{Acknowledgements}

We thank Hakan Akdag for pointing out missing $SU(3)$ generators and unnecessary Dirac tensors in an earlier version of the paper.
As this work was being finalized Ref.~\cite{Li:2020tsi} appeared on the arXiv.
It constructs bases of LEFT operators through mass dimension 9 by expanding on the technology developed in Refs.~\cite{Li:2020gnx, Li:2020xlh}.


\begin{appendix}

\section{Dimension-8 Operator Counting}
\label{sec:count}

\input{sections/counting_v2}

\section{Dimension-8 Operator Basis}
\label{sec:basis}

\subsection{$X^4$}
\input{sections/basis_x4}

\subsection{$\psi^2X^2D$}
\input{sections/basis_psi2x2d_v3}

\subsection{$\psi^4X$}
\input{sections/basis_psi4x_llrr}

\input{sections/basis_psi4x_llll_v3}

\subsection{$\psi^4D^2$}
\input{sections/basis_psi4d2_llrr_v2}

\input{sections/basis_psi4d2_llll}

\subsection{$\Delta B = 0,\, \Delta L \neq 0$}
\input{sections/basis_dB0dL2L4}

\subsection{$\Delta B \neq 0,\, \Delta L \neq 0$}
\input{sections/basis_dB1dL1}

\input{sections/basis_dB1dLm1}

\end{appendix}

\clearpage


\bibliographystyle{JHEP}
\bibliography{left_dim8_basis}


\end{document}

%% file: sections/counting_v2.tex
\begin{table}[H]
\begin{center}
\begin{tabular}[t]{c | c | c}
\textbf{Class} & Number of Operators & SM \\ \hline \hline
\boldmath$X^4$ & 23 & 23 \\ \hline
\boldmath$\psi^2X^2D$ & $2 n_\nu^2 + 4 n_e^2 + 20 n_u^2 + 20 n_d^2$ & 314  \\ \hline
\boldmath$\psi^4X$ & $27 n_d^4 + 18 n_d^3 n_e + 2 n_d n_u (n_e + 27 n_e n_u + n_\nu + 36 n_e n_\nu) $ & 24108 \\ 
& $+ 
  n_d^2 (-3 + 2 n_e + 36 n_e^2 + 108 n_u^2 - 2 n_\nu + 54 n_u n_\nu + 
     18 n_\nu^2)$ & \\
     & $ + 
  \tfrac{1}{4} (-4 n_e^3 + 18 n_e^4 + 108 n_u^4 + 4 n_u^2 (-3 - 2 n_\nu + 18 n_\nu^2) $ & \\
  & $+ 
     2 n_e^2 (-3 + 72 n_u^2 - 2 n_\nu + 18 n_\nu^2) + 
     n_\nu (2 - 3 n_\nu - 2 n_\nu^2 + 3 n_\nu^3))$ & \\ \hline
\boldmath$\psi^4D^2$ & $9 n_d^4 + 6 n_d^3 n_e + n_e^3 + \tfrac{9}{2} n_e^4 + 
  n_e^2 (\tfrac{5}{2} + 18 n_u^2 + n_\nu+ 9 n_\nu^2) $ & 10613 \\ 
  & $+ 
  n_d^2 (5 - 2 n_e + 18 n_e^2 + 36 n_u^2 + n_\nu+ 18 n_u n_\nu+ 9 n_\nu^2) $ & \\
  & $+ 
  2 n_d n_u (-n_\nu+ n_e (-1 + 9 n_u + 18 n_\nu)) $ & \\  
  & $+ 
  \tfrac{1}{4} (36 n_u^4 + 4 n_u^2 (5 + n_\nu+ 9 n_\nu^2) + 
     n_\nu(2 + 5 n_\nu+ 2 n_\nu^2 + 3 n_\nu^3))$ & \\ \hline \hline
\boldmath$\Delta B = \Delta L = 0$ & $23 + 36 n_d^4 + 9 n_e^4 + 22 n_u^2 + 36 n_u^4 + 54 n_d n_e n_u n_\nu + 
  2 n_\nu^2 + 12 n_u^2 n_\nu^2 + n_\nu^4 $ & 21144 \\
  & $+ 
  2 n_d^2 (11 + 27 n_e^2 + 72 n_u^2 + 6 n_\nu^2) + 
  n_e^2 (5 + 54 n_u^2 + 8 n_\nu^2)$ & \\ \hline
\boldmath$ \Delta L = 2$ & $n_\nu (54 n_d n_e n_u + 10 n_e^2 n_\nu + n_d^2 (-1 + 15 n_\nu) + 
    n_u^2 (-1 + 15 n_\nu))$ & 5442 \\ \hline
\boldmath$\Delta B = \Delta L = 1$ & $36 n_d n_u (2 n_e n_u + n_d n_\nu)$ & 4536 \\ \hline
\boldmath$\Delta B = - \Delta L = 1$ & $12 n_d^2 (2 n_d n_e + 3 n_u n_\nu)$ & 3888 \\  \hline
\boldmath$ \Delta L = 4$ & $\tfrac{1}{2} n_\nu (2 + n_\nu + n_\nu^3)$ & 48 \\ \hline \hline
\textbf{Total} & $23 + 36 n_d^4 + 24 n_d^3 n_e + 9 n_e^4 + 22 n_u^2 + 36 n_u^4 + n_\nu - 
  n_u^2 n_\nu $ & 35058 \\
  & $+ \tfrac{5}{2} n_\nu^2 + 27 n_u^2 n_\nu^2 + \tfrac{3}{2} n_\nu^4 + 
  36 n_d n_e n_u (2 n_u + 3 n_\nu) $ & \\
  & $+ n_e^2 (5 + 54 n_u^2 + 18 n_\nu^2) + n_d^2 (22 + 54 n_e^2 + 144 n_u^2 - n_\nu + 72 n_u n_\nu + 27 n_\nu^2)$ &
\end{tabular}
\end{center}
\caption{The number of dimension-8 LEFT operators broken down by operator class and by baryon and lepton number. The $X^4$ Lagrangian terms each contain one operator whereas the $\psi^2X^2D$ Lagrangian terms each contain $n_f^2$ operators where in the SM $n_\nu = n_e = n_d = 3$ and $n_u = 2$. Tables~\ref{tab:count_4psi}, \ref{tab:count_deltaL}, and~\ref{tab:count_deltaB} and provide more detail on the counting of operators with four fermions.}
\label{tab:count_sum}
\end{table}


\begin{table}[H]
\begin{center}
\begin{adjustbox}{width=1.\textwidth,center}
\small
\begin{minipage}[t]{6.5cm}
\renewcommand{\arraystretch}{1.5}
\begin{tabular}[t]{c | c | c }
\multicolumn{3}{c}{\boldmath$\psi^4X$} \\ \hline
Term(s) & Number of Operators & SM \\ \hline
$\O_{\nu^4F}^{(1,\, \ldots\, 2, LL)}$ & $\tfrac{1}{4} n_\nu^2 (n_\nu^2 - 1)$ & 18 \\
$\O_{\nu^2e^2F}^{(\rm all)}$ & $n_\nu^2 n_e^2$ & 81 \\
$\O_{\nu^2u^2X}^{(\rm all)}$ & $n_\nu^2 n_u^2$ & 36 \\
$\O_{\nu^2d^2X}^{(\rm all)}$ & $n_\nu^2 n_d^2$ & 81 \\
$\O_{\nu eudX}^{(\rm all)}$ & $n_\nu n_e n_u n_d$ & 81 \\
$\O_{e^4F}^{(1,\, \ldots\, 2, LL)}$ & $\tfrac{1}{4} n_e^2 (n_e^2 - 1)$ & 18 \\
$\O_{e^4F}^{(1,\, \ldots\, 2, RR)}$ & $\tfrac{1}{4} n_e^2 (n_e^2 - 1)$ & 18 \\
$\O_{e^4F}^{(1, ST)}$ & $\tfrac{1}{2} n_e^2 (n_e^2 - 1) + \tfrac{1}{4} n_e^2 (n_e - 1)^2$ & 45 \\
$\O_{e^4F}^{(\rm all\, others)}$ & $n_e^4$ & 81 \\
$\O_{e^2u^2X}^{(\rm all)}$ & $n_e^2 n_u^2$ & 36 \\
$\O_{e^2d^2X}^{(\rm all)}$ & $n_e^2 n_d^2$ & 81 \\
$\O_{u^4F}^{(1,\, \ldots\, 2, LL)}$ & $\tfrac{1}{2} n_u^2 (n_u^2 - 1)$ & 6 \\
$\O_{u^4F}^{(1,\, \ldots\, 2, RR)}$ & $\tfrac{1}{2} n_u^2 (n_u^2 - 1)$ & 6 \\
$\O_{u^4F}^{(1, ST)}$ & $\tfrac{1}{2} n_u^2 (n_u^2 - 1) + \tfrac{1}{4} n_u^2 (n_u + 1)^2$ & 15 \\
$\O_{u^4F}^{(2, ST)}$ & $\tfrac{1}{2} n_u^2 (n_u^2 - 1) + \tfrac{1}{4} n_u^2 (n_u - 1)^2$ & 7 \\
$\O_{u^4X}^{(\rm all\, others)}$ & $n_u^4$ & 16 \\
$\O_{d^4F}^{(1,\, \ldots\, 2, LL)}$ & $\tfrac{1}{2} n_d^2 (n_d^2 - 1)$ & 36 \\
$\O_{d^4F}^{(1,\, \ldots\, 2, RR)}$ & $\tfrac{1}{2} n_d^2 (n_d^2 - 1)$ & 36 \\
$\O_{d^4F}^{(1, ST)}$ & $\tfrac{1}{2} n_d^2 (n_d^2 - 1) + \tfrac{1}{4} n_d^2 (n_d + 1)^2$ & 72 \\
$\O_{d^4F}^{(2, ST)}$ & $\tfrac{1}{2} n_d^2 (n_d^2 - 1) + \tfrac{1}{4} n_d^2 (n_d - 1)^2$ & 45 \\
$\O_{d^4X}^{(\rm all\, others)}$ & $n_d^4$ & 81
\end{tabular}
\end{minipage}
\hspace{1.4cm}
\begin{minipage}[t]{7.1cm}
\renewcommand{\arraystretch}{1.5}
\begin{tabular}[t]{c | c | c }
\multicolumn{3}{c}{\boldmath$\psi^4D^2$} \\ \hline
Term(s) & Number of Operators & SM \\ \hline
$\O_{\nu^4D^2}^{(1, LL)}$ & $\tfrac{1}{4} n_\nu^2 (n_\nu + 1)^2 + \tfrac{1}{4} n_\nu^2 (n_\nu - 1)^2$ & 45 \\
$\O_{\nu^2e^2D^2}^{(\rm all)}$ & $n_\nu^2 n_e^2$ & 81 \\
$\O_{\nu^2u^2D^2}^{(\rm all)}$ & $n_\nu^2 n_u^2$ & 36 \\
$\O_{\nu^2d^2D^2}^{(\rm all)}$ & $n_\nu^2 n_d^2$ & 81 \\
$\O_{\nu eudD^2}^{(\rm all)}$ & $n_\nu n_e n_u n_d$ & 81 \\
$\O_{e^4D^2}^{(1, LL, RR)}$ & $\tfrac{1}{4} n_e^2 (n_e + 1)^2 + \tfrac{1}{4} n_e^2 (n_e - 1)^2$ & 45 \\
$\O_{e^4D^2}^{(1, SS)}$ & $\tfrac{1}{4} n_e^2 (n_e + 1)^2 + \tfrac{1}{4} n_e^2 (n_e - 1)^2$ & 45 \\
$\O_{e^4D^2}^{(1, TT)}$ & $\tfrac{1}{4} n_e^2 (n_e + 1)^2$ & 36 \\
$\O_{e^4D^2}^{(\rm all\, others)}$ & $n_e^4$ & 81 \\
$\O_{e^2u^2D^2}^{(\rm all)}$ & $n_e^2 n_u^2$ & 36 \\
$\O_{e^2d^2D^2}^{(\rm all)}$ & $n_e^2 n_d^2$ & 81 \\
$\O_{u^4D^2}^{(1,\, \ldots\, 2, LL)}$ & $\tfrac{1}{4} n_u^2 (n_u + 1)^2 + \tfrac{1}{4} n_u^2 (n_u - 1)^2$ & 10 \\
$\O_{u^4D^2}^{(1,\, \ldots\, 2, RR)}$ & $\tfrac{1}{4} n_u^2 (n_u + 1)^2 + \tfrac{1}{4} n_u^2 (n_u - 1)^2$ & 10 \\
$\O_{u^4D^2}^{(1,\, \ldots\, 3, SS)}$ & $\tfrac{1}{4} n_u^2 (n_u + 1)^2 + \tfrac{1}{4} n_u^2 (n_u - 1)^2$ & 10 \\
$\O_{u^4D^2}^{(\rm all\, others)}$ & $n_u^4$ & 16 \\
$\O_{d^4D^2}^{(1,\, \ldots\, 2, LL)}$ & $\tfrac{1}{4} n_d^2 (n_d + 1)^2 + \tfrac{1}{4} n_d^2 (n_d - 1)^2$ & 45 \\
$\O_{d^4D^2}^{(1,\, \ldots\, 2, RR)}$ & $\tfrac{1}{4} n_d^2 (n_d + 1)^2 + \tfrac{1}{4} n_d^2 (n_d - 1)^2$ & 45 \\
$\O_{d^4D^2}^{(1,\, \ldots\, 3, SS)}$ & $\tfrac{1}{4} n_d^2 (n_d + 1)^2 + \tfrac{1}{4} n_d^2 (n_d - 1)^2$ & 45 \\
$\O_{d^4D^2}^{(\rm all\, others)}$ & $n_d^4$ & 81
\end{tabular}
\end{minipage}
\end{adjustbox}
\end{center}
\caption{Counting of operators in LEFT dimension-8 Lagrangian terms with four fermions that preserve both baryon and lepton number. Terms with repeated fields can have non-trivial flavor structure. In the SM $n_\nu = n_e = n_d = 3$ and $n_u = 2$.}
\label{tab:count_4psi}
\end{table}


\begin{table}[H]
\begin{center}
\begin{adjustbox}{width=0.75\textwidth,center}
\small
\begin{minipage}[t]{5.4cm}
\renewcommand{\arraystretch}{1.5}
\begin{tabular}[t]{c  | c | c}
\multicolumn{3}{c}{\boldmath$\Delta L = 2: \psi^4F$} \\ \hline
Term(s) & Number of Ops. & SM \\ \hline
$\O_{\nu^2e^2F}^{(1, ST)}$  &   $\tfrac{1}{2} n_e^2 n_\nu (n_\nu + 1)$ & 54 \\ 
$\O_{\nu^2e^2F}^{(2, ST)}$  &   $\tfrac{1}{2} n_e^2 n_\nu^2 $ & 81 \\ 
$\O_{\nu^2e^2F}^{(1,\, \ldots\, 2, TS)}$  &   $\tfrac{1}{2} n_e^2 n_\nu (n_\nu - 1)$ & 27 \\ 
$\O_{\nu^2u^2X}^{(1, ST)}$  &   $\tfrac{1}{2} n_u^2 n_\nu (n_\nu + 1)$ & 24 \\ 
$\O_{\nu^2u^2X}^{(2, ST)}$  &   $\tfrac{1}{2} n_u^2 n_\nu^2$ & 36 \\ 
$\O_{\nu^2u^2X}^{(1,\, \ldots\, 2, TS)}$  &   $\tfrac{1}{2} n_u^2 n_\nu (n_\nu - 1)$ & 12 \\ 
$\O_{\nu^2d^2X}^{(1, ST)}$  &   $\tfrac{1}{2} n_d^2 n_\nu (n_\nu + 1)$ & 54 \\ 
$\O_{\nu^2d^2X}^{(2, ST)}$  &   $\tfrac{1}{2} n_d^2 n_\nu^2$ & 81 \\ 
$\O_{\nu^2d^2X}^{(1,\, \ldots\, 2, TS)}$  &   $\tfrac{1}{2} n_d^2 n_\nu (n_\nu - 1)$ & 27 \\ 
$\O_{\nu eudX}^{(\rm all)}$  &   $n_\nu n_e n_u n_d$ & 81 
\end{tabular}
\end{minipage}
\hspace{1cm}
\begin{minipage}[t]{4.9cm}
\renewcommand{\arraystretch}{1.5}
\begin{tabular}[t]{c  | c | c}
\multicolumn{3}{c}{\boldmath$\Delta L = 2: \psi^4D^2$} \\ \hline
Term(s) & Number of Ops. & SM \\ \hline
$\O_{\nu^2e^2D^2}^{(1, 3, SS)}$  &   $\tfrac{1}{2} n_e^2 n_\nu (n_\nu + 1)$ & 54 \\ 
$\O_{\nu^2e^2D^2}^{(1, ST)}$  &   $\tfrac{1}{2} n_e^2 n_\nu (n_\nu + 1)$ & 54 \\ 
$\O_{\nu^2e^2D^2}^{(2, 4, SS)}$  &   $\tfrac{1}{2} n_e^2 n_\nu (n_\nu - 1)$ & 27 \\ 
$\O_{\nu^2u^2D^2}^{(1, 3, SS)}$  &   $\tfrac{1}{2} n_u^2 n_\nu (n_\nu + 1)$ & 24 \\ 
$\O_{\nu^2u^2D^2}^{(1, ST)}$  &   $\tfrac{1}{2} n_u^2 n_\nu (n_\nu + 1)$ & 24 \\ 
$\O_{\nu^2u^2D^2}^{(2, 4, SS)}$  &   $\tfrac{1}{2} n_u^2 n_\nu (n_\nu - 1)$ & 12 \\ 
$\O_{\nu^2d^2D^2}^{(1, 3, SS)}$  &   $\tfrac{1}{2} n_d^2 n_\nu (n_\nu + 1)$ & 54 \\ 
$\O_{\nu^2d^2D^2}^{(1, ST)}$  &   $\tfrac{1}{2} n_d^2 n_\nu (n_\nu + 1)$ & 54 \\ 
$\O_{\nu^2d^2D^2}^{(2, 4, SS)}$  &   $\tfrac{1}{2} n_d^2 n_\nu (n_\nu - 1)$ & 27 \\ 
$\O_{\nu eudD^2}^{(\rm all)}$  &   $n_\nu n_e n_u n_d$ & 81 
\end{tabular}
\end{minipage}
\end{adjustbox}
\begin{adjustbox}{width=0.59\textwidth,center}
\small
\begin{minipage}[t]{8.9cm}
\renewcommand{\arraystretch}{1.5}
\begin{tabular}[t]{c  | c | c}
\multicolumn{3}{c}{\boldmath$\Delta L = 4$} \\ \hline
Term & Number of Operators & SM \\ \hline
$\O_{\nu^4F}^{(1)}$  &   $\tfrac{1}{8} n_\nu (n_\nu^2 - 1) (n_\nu - 2)$ & 3 \\ 
$\O_{\nu^4D^2}^{(1)}$ & $\tfrac{1}{24} n_\nu (n_\nu + 1) (n_\nu + 2) (n_\nu + 3) + \tfrac{1}{12} n_\nu^2 (n_\nu^2 - 1)$ & 21
\end{tabular}
\end{minipage}
\end{adjustbox}
\end{center}
\caption{Counting of operators in LEFT dimension-8 Lagrangian terms that violate lepton number but preserve baryon number. Terms with repeated fields can have non-trivial flavor structure. In the SM $n_\nu = n_e = n_d = 3$ and $n_u = 2$.}
\label{tab:count_deltaL}
\end{table}


\begin{table}[H]
\begin{center}
\begin{adjustbox}{width=0.95\textwidth,center}
\small
\begin{minipage}[t]{7.8cm}
\renewcommand{\arraystretch}{1.5}
\begin{tabular}[t]{c  | c | c }
\multicolumn{3}{c }{\boldmath$\Delta B = - \Delta L = 1: \psi^4X$} \\ \hline
Term(s) & Number of Operators & SM \\ \hline
$\O_{\bar \nu ud^2F}^{(2, 5)}$  &   $\tfrac{1}{2} n_\nu n_u n_d (n_d + 1)$ & 36 \\ 
$\O_{\bar \nu ud^2F}^{(4)}$  &   $\tfrac{1}{2} n_\nu n_u n_d (n_d - 1)$ & 18 \\ 
$\O_{\bar \nu ud^2X}^{(\rm all\, others)}$  &   $n_\nu n_u n_d^2$ & 54 \\ 
$\O_{\bar e d^3F}^{(1, 2)}$  &   $\tfrac{1}{2} n_e n_d^2 (n_d + 1)$ & 54 \\ 
$\O_{\bar e d^3F}^{(3, 4)}$  &   $\tfrac{1}{2} n_e n_d^2 (n_d - 1)$ & 27 \\ 
$\O_{\bar e d^3X}^{(5, 6)}$  &   $\tfrac{1}{6} n_e n_d (n_d + 1) (n_d + 2) + \tfrac{1}{3} n_e n_d (n_d^2 - 1)$ & 54 \\ 
$\O_{\bar e d^3G}^{(7, 8)}$  &   $\tfrac{1}{6} n_e n_d (n_d - 1) (n_d - 2) + \tfrac{1}{3} n_e n_d (n_d^2 - 1)$ & 27 \\ 
$\O_{\bar e d^3G}^{(\rm all\, others)}$  &   $n_e n_u^2 n_d$ & 36
\end{tabular}
\end{minipage}
\hspace{1.5cm}
\begin{minipage}[t]{5.0cm}
\renewcommand{\arraystretch}{1.5}
\begin{tabular}[t]{c  | c | c}
\multicolumn{3}{c}{\boldmath$\Delta B = \Delta L = 1: \psi^4D^2$} \\ \hline
Term(s) & Number of Ops. & SM \\ \hline
$\O_{\nu ud^2D^2}^{(3, SS)}$  &   $\tfrac{1}{2} n_\nu n_u n_d (n_d + 1)$ & 36 \\ 
$\O_{\nu ud^2D^2}^{(2, ST)}$  &   $\tfrac{1}{2} n_\nu n_u n_d (n_d + 1)$ & 36 \\ 
$\O_{\nu ud^2D^2}^{(2, 4, SS)}$  &   $\tfrac{1}{2} n_\nu n_u n_d (n_d - 1)$ & 18 \\ 
$\O_{\nu ud^2D^2}^{(3, ST)}$  &   $\tfrac{1}{2} n_\nu n_u n_d (n_d - 1)$ & 18 \\ 
$\O_{\nu ud^2D^2}^{(\rm all\, others)}$  &   $n_\nu n_u n_d^2$ & 54 \\ 
$\O_{eu^2dD^2}^{(5, 6, SS)}$  &   $\tfrac{1}{2} n_e n_d n_u (n_u + 1)$ & 27 \\ 
$\O_{eu^2dD^2}^{(3, 4, ST)}$  &   $\tfrac{1}{2} n_e n_d n_u (n_u + 1)$ & 27 \\ 
$\O_{eu^2dD^2}^{(3, 4, 7, 8, SS)}$  &   $\tfrac{1}{2} n_e n_d n_u (n_u - 1)$ & 9 \\ 
$\O_{eu^2dD^2}^{(5, 6, ST)}$  &   $\tfrac{1}{2} n_e n_d n_u (n_u - 1)$ & 9 \\ 
$\O_{eu^2dD^2}^{(\rm all\, others)}$  &   $n_e n_u^2 n_d$ & 36
\end{tabular}
\end{minipage}
\end{adjustbox}
\begin{adjustbox}{width=0.73\textwidth,center}
\small
\begin{minipage}[t]{4.9cm}
\renewcommand{\arraystretch}{1.5}
\begin{tabular}[t]{c  | c | c}
\multicolumn{3}{c}{\boldmath$\Delta B = - \Delta L = 1: \psi^4D^2$} \\ \hline
Term(s) & Number of Ops. & SM \\ \hline
$\O_{\bar \nu ud^2D^2}^{(3, 5)}$  &   $\tfrac{1}{2} n_\nu n_u n_d (n_d + 1)$ & 36 \\ 
$\O_{\bar \nu ud^2D^2}^{(4, 6, 7)}$  &   $\tfrac{1}{2} n_\nu n_u n_d (n_d - 1)$ & 18 \\ 
$\O_{\bar \nu ud^2D^2}^{(\rm all\, others)}$  &   $n_\nu n_u n_d^2$ & 54 \\ 
$\O_{\bar e d^3D^2}^{(1, 2)}$  &   $\tfrac{1}{2} n_e n_d^2 (n_d + 1)$ & 54 \\ 
$\O_{\bar e d^3D^2}^{(3,\, \ldots\, 6)}$  &   $\tfrac{1}{2} n_e n_d^2 (n_d - 1)$ & 27 
\end{tabular}
\end{minipage}
\hspace{1cm}
\begin{minipage}[t]{5.0cm}
\renewcommand{\arraystretch}{1.5}
\begin{tabular}[t]{c  | c | c}
\multicolumn{3}{c}{\boldmath$\Delta B = \Delta L = 1: \psi^4X$} \\ \hline
Term(s) & Number of Ops. & SM \\ \hline
$\O_{\nu ud^2F}^{(2, 4, ST)}$  &   $\tfrac{1}{2} n_\nu n_u n_d (n_d + 1)$ & 36 \\ 
$\O_{\nu ud^2F}^{(2, TS)}$  &   $\tfrac{1}{2} n_\nu n_u n_d (n_d - 1)$ & 18 \\ 
$\O_{\nu ud^2X}^{(\rm all\, others)}$  &   $n_\nu n_u n_d^2$ & 54 \\ 
$\O_{eu^2dF}^{(3, 4, 7, 8, ST)}$  &   $\tfrac{1}{2} n_e n_d n_u (n_u + 1)$ & 27 \\ 
$\O_{eu^2dF}^{(3, 4, TS)}$  &   $\tfrac{1}{2} n_e n_d n_u (n_u - 1)$ & 9 \\ 
$\O_{eu^2dX}^{(\rm all\, others)}$  &   $n_e n_u^2 n_d$ & 36
\end{tabular}
\end{minipage}
\end{adjustbox}
\end{center}
\caption{Counting of operators in LEFT dimension-8 Lagrangian terms that violate baryon and lepton number. Terms with repeated fields can have non-trivial flavor structure. In the SM $n_\nu = n_e = n_d = 3$ and $n_u = 2$.}
\label{tab:count_deltaB}
\end{table}

%% file: sections/basis_x4.tex
\begin{table}[H]
\begin{center}
\begin{adjustbox}{width=0.4\textwidth,center}
\small
\begin{minipage}[t]{6cm}
\renewcommand{\arraystretch}{1.5}
\begin{tabular}[t]{c|c}
\multicolumn{2}{c}{\boldmath$X^4$} \\
\hline
$\O_{G^4}^{(1)}$  &  $(G_{\mu\nu}^A G^{A\mu\nu}) (G_{\rho\sigma}^B G^{B\rho\sigma})$ \\
$\O_{G^4}^{(2)}$  &  $(G_{\mu\nu}^A \widetilde{G}^{A\mu\nu}) (G_{\rho\sigma}^B \widetilde{G}^{B\rho\sigma})$ \\
$\O_{G^4}^{(3)}$  &  $(G_{\mu\nu}^A G^{B\mu\nu}) (G_{\rho\sigma}^A G^{B\rho\sigma})$ \\
$\O_{G^4}^{(4)}$  &  $(G_{\mu\nu}^A \widetilde{G}^{B\mu\nu}) (G_{\rho\sigma}^A \widetilde{G}^{B\rho\sigma})$ \\
$\O_{G^4}^{(5)}$  &  $(G_{\mu\nu}^A G^{A\mu\nu}) (G_{\rho\sigma}^B \widetilde{G}^{B\rho\sigma})$ \\
$\O_{G^4}^{(6)}$  &  $(G_{\mu\nu}^A G^{B\mu\nu}) (G_{\rho\sigma}^A \widetilde{G}^{B\rho\sigma})$ \\
$\O_{G^4}^{(7)}$  &  $d^{ABE} d^{CDE} (G_{\mu\nu}^A G^{B\mu\nu}) (G_{\rho\sigma}^C G^{D\rho\sigma})$ \\
$\O_{G^4}^{(8)}$  &  $d^{ABE} d^{CDE} (G_{\mu\nu}^A \widetilde{G}^{B\mu\nu}) (G_{\rho\sigma}^C \widetilde{G}^{D\rho\sigma})$ \\
$\O_{G^4}^{(9)}$  &  $d^{ABE} d^{CDE} (G_{\mu\nu}^A G^{B\mu\nu}) (G_{\rho\sigma}^C \widetilde{G}^{D\rho\sigma})$ \\
$\O_{F^4}^{(1)}$  &  $(F_{\mu\nu} F^{\mu\nu}) (F_{\rho\sigma} F^{\rho\sigma})$ \\
$\O_{F^4}^{(2)}$  &  $(F_{\mu\nu} \widetilde{F}^{\mu\nu}) (F_{\rho\sigma} \widetilde{F}^{\rho\sigma})$ \\
$\O_{F^4}^{(3)}$  &  $(F_{\mu\nu} F^{\mu\nu}) (F_{\rho\sigma} \widetilde{F}^{\rho\sigma})$ \\
$\O_{G^3F}^{(1)}$  &  $d^{ABC} (F_{\mu\nu} G^{A\mu\nu}) (G_{\rho\sigma}^B G^{C\rho\sigma})$ \\
$\O_{G^3F}^{(2)}$  &  $d^{ABC} (F_{\mu\nu} \widetilde{G}^{A\mu\nu}) (G_{\rho\sigma}^B \widetilde{G}^{C\rho\sigma})$ \\
$\O_{G^3F}^{(3)}$  &  $d^{ABC} (F_{\mu\nu} \widetilde{G}^{A\mu\nu}) (G_{\rho\sigma}^B G^{C\rho\sigma})$ \\
$\O_{G^3F}^{(4)}$  &  $d^{ABC} (F_{\mu\nu} G^{A\mu\nu}) (G_{\rho\sigma}^B \widetilde{G}^{C\rho\sigma})$ \\
$\O_{G^2F^2}^{(1)}$  &  $(F_{\mu\nu} F^{\mu\nu}) (G_{\rho\sigma}^A G^{A\rho\sigma})$ \\
$\O_{G^2F^2}^{(2)}$  &  $(F_{\mu\nu} \widetilde{F}^{\mu\nu}) (G_{\rho\sigma}^A \widetilde{G}^{A\rho\sigma})$ \\
$\O_{G^2F^2}^{(3)}$  &  $(F_{\mu\nu} G^{A\mu\nu}) (F_{\rho\sigma} G^{A\rho\sigma})$ \\
$\O_{G^2F^2}^{(4)}$  &  $(F_{\mu\nu} \widetilde{G}^{A\mu\nu}) (F_{\rho\sigma} \widetilde{G}^{A\rho\sigma})$ \\
$\O_{G^2F^2}^{(5)}$  &  $(F_{\mu\nu} \widetilde{F}^{\mu\nu}) (G_{\rho\sigma}^A G^{A\rho\sigma})$ \\
$\O_{G^2F^2}^{(6)}$  &  $(F_{\mu\nu} F^{\mu\nu}) (G_{\rho\sigma}^A \widetilde{G}^{A\rho\sigma})$ \\
$\O_{G^2F^2}^{(7)}$  &  $(F_{\mu\nu} G^{A\mu\nu}) (F_{\rho\sigma} \widetilde{G}^{A\rho\sigma})$ \\
\end{tabular}
\end{minipage}
\end{adjustbox}
\end{center}
\caption{The dimension-eight operators in the LEFT whose field content is entirely gauge field strengths.}
\label{tab:left8_x4}
\end{table}

%% file: sections/basis_psi2x2d_v3.tex
\begin{table}[H]
\begin{center}
\begin{adjustbox}{width=0.53\textwidth,center}
\small
\begin{minipage}[t]{7.9cm}
\renewcommand{\arraystretch}{1.5}
\begin{tabular}[t]{c|c}
\multicolumn{2}{c}{\boldmath$(\bar L L)X^2D$} \\
\hline
$\O_{\nu^2G^2D}^{(1, L)}$  &  $i (\bar \nu_{Lp} \gamma^\mu \overleftrightarrow{D}^\lambda \nu_{Lr}) G_{\mu\rho}^A G_\lambda^{A\rho}$ \\
$\O_{\nu^2F^2D}^{(1, L)}$  &  $i (\bar \nu_{Lp} \gamma^\mu \overleftrightarrow{D}^\lambda \nu_{Lr}) F_{\mu\rho} F_\lambda^{\,\,\,\rho}$ \\
$\O_{e^2G^2D}^{(1, L)}$  &  $i (\bar e_{Lp} \gamma^\mu \overleftrightarrow{D}^\lambda e_{Lr}) G_{\mu\rho}^A G_\lambda^{A\rho}$ \\
$\O_{e^2F^2D}^{(1, L)}$  &  $i (\bar e_{Lp} \gamma^\mu \overleftrightarrow{D}^\lambda e_{Lr}) F_{\mu\rho} F_\lambda^{\,\,\,\rho}$ \\
$\O_{u^2G^2D}^{(1, L)}$  &  $i (\bar u_{Lp} \gamma^\mu \overleftrightarrow{D}^\lambda u_{Lr}) G_{\mu\rho}^A G_\lambda^{A\rho}$ \\
$\O_{u^2G^2D}^{(2, L)}$  &  $f^{ABC} (\bar u_{Lp} \gamma^\mu T^A \overleftrightarrow{D}^\lambda u_{Lr}) G_{\mu\rho}^B G_\lambda^{C\rho}$ \\
$\O_{u^2G^2D}^{(3, L)}$  &  $i d^{ABC} (\bar u_{Lp} \gamma^\mu T^A \overleftrightarrow{D}^\lambda u_{Lr}) G_{\mu\rho}^B G_\lambda^{C\rho}$ \\
$\O_{u^2G^2D}^{(4, L)}$  &  $i f^{ABC} (\bar u_{Lp} \gamma^\mu T^A \overleftrightarrow{D}^\lambda u_{Lr}) (G_{\mu\rho}^B \widetilde{G}_\lambda^{C\rho} - \widetilde{G}_{\mu\rho}^B G_\lambda^{C\rho})$ \\
$\O_{u^2G^2D}^{(5, L)}$  &  $f^{ABC} (\bar u_{Lp} \gamma^\mu T^A \overleftrightarrow{D}^\lambda u_{Lr}) (G_{\mu\rho}^B \widetilde{G}_\lambda^{C\rho} + \widetilde{G}_{\mu\rho}^B G_\lambda^{C\rho})$ \\
$\O_{u^2F^2D}^{(1, L)}$  &  $i (\bar u_{Lp} \gamma^\mu \overleftrightarrow{D}^\lambda u_{Lr}) F_{\mu\rho} F_\lambda^{\,\,\,\rho}$ \\
$\O_{u^2GFD}^{(1, L)}$  &  $(\bar u_{Lp} \gamma^\mu T^A \overleftrightarrow{D}^\lambda u_{Lr}) (F_{\mu\rho} G_\lambda^{A\rho} - F_{\lambda\rho} G_{\mu}^{A\rho})$ \\
$\O_{u^2GFD}^{(2, L)}$  &  $i (\bar u_{Lp} \gamma^\mu T^A \overleftrightarrow{D}^\lambda u_{Lr}) (F_{\mu\rho} G_\lambda^{A\rho} + F_{\lambda\rho} G_{\mu}^{A\rho})$ \\
$\O_{u^2GFD}^{(3, L)}$  &  $(\bar u_{Lp} \gamma^\mu T^A \overleftrightarrow{D}^\lambda u_{Lr}) (F_{\mu\rho} \widetilde{G}_\lambda^{A \rho} - F_{\lambda\rho} \widetilde{G}_\mu^{A \rho})$ \\
$\O_{u^2GFD}^{(4, L)}$  &  $i (\bar u_{Lp} \gamma^\mu T^A \overleftrightarrow{D}^\lambda u_{Lr}) (F_{\mu\rho} \widetilde{G}_\lambda^{A \rho} + F_{\lambda\rho} \widetilde{G}_\mu^{A \rho})$ \\
$\O_{d^2G^2D}^{(1, L)}$  &  $i (\bar d_{Lp} \gamma^\mu \overleftrightarrow{D}^\lambda d_{Lr}) G_{\mu\rho}^A G_\lambda^{A\rho}$ \\
$\O_{d^2G^2D}^{(2, L)}$  &  $f^{ABC} (\bar d_{Lp} \gamma^\mu T^A \overleftrightarrow{D}^\lambda d_{Lr}) G_{\mu\rho}^B G_\lambda^{C\rho}$ \\
$\O_{d^2G^2D}^{(3, L)}$  &  $i d^{ABC} (\bar d_{Lp} \gamma^\mu T^A \overleftrightarrow{D}^\lambda d_{Lr}) G_{\mu\rho}^B G_\lambda^{C\rho}$ \\
$\O_{d^2G^2D}^{(4, L)}$  &  $i f^{ABC} (\bar d_{Lp} \gamma^\mu T^A \overleftrightarrow{D}^\lambda d_{Lr}) (G_{\mu\rho}^B \widetilde{G}_\lambda^{C\rho} - \widetilde{G}_{\mu\rho}^B G_\lambda^{C\rho})$ \\
$\O_{d^2G^2D}^{(5, L)}$  &  $f^{ABC} (\bar d_{Lp} \gamma^\mu T^A \overleftrightarrow{D}^\lambda d_{Lr}) (G_{\mu\rho}^B \widetilde{G}_\lambda^{C\rho} + \widetilde{G}_{\mu\rho}^B G_\lambda^{C\rho})$ \\
$\O_{d^2F^2D}^{(1, L)}$  &  $i (\bar d_{Lp} \gamma^\mu \overleftrightarrow{D}^\lambda d_{Lr}) F_{\mu\rho} F_\lambda^{\,\,\,\rho}$ \\
$\O_{d^2GFD}^{(1, L)}$  &  $(\bar d_{Lp} \gamma^\mu T^A \overleftrightarrow{D}^\lambda d_{Lr}) (F_{\mu\rho} G_\lambda^{A\rho} - F_{\lambda\rho} G_{\mu}^{A\rho})$ \\
$\O_{d^2GFD}^{(2, L)}$  &  $i (\bar d_{Lp} \gamma^\mu T^A \overleftrightarrow{D}^\lambda d_{Lr}) (F_{\mu\rho} G_\lambda^{A\rho} + F_{\lambda\rho} G_{\mu}^{A\rho})$ \\
$\O_{d^2GFD}^{(3, L)}$  &  $(\bar d_{Lp} \gamma^\mu T^A \overleftrightarrow{D}^\lambda d_{Lr}) (F_{\mu\rho} \widetilde{G}_\lambda^{A \rho} - F_{\lambda\rho} \widetilde{G}_\mu^{A \rho})$ \\
$\O_{d^2GFD}^{(4, L)}$  &  $i (\bar d_{Lp} \gamma^\mu T^A \overleftrightarrow{D}^\lambda d_{Lr}) (F_{\mu\rho} \widetilde{G}_\lambda^{A \rho} + F_{\lambda\rho} \widetilde{G}_\mu^{A \rho})$ 
\end{tabular}
\end{minipage}
\end{adjustbox}
\end{center}
\caption{The baryon and lepton number conserving operators of class $\psi^2 X^2 D$ with left-handed chiral projectors in the fermion bilinear. The subscripts $p, r$ are weak-eigenstate indices.}
\label{tab:left8_psi2x2d_L}
\end{table}

\begin{table}[H]
\begin{center}
\begin{adjustbox}{width=0.53\textwidth,center}
\small
\begin{minipage}[t]{7.9cm}
\renewcommand{\arraystretch}{1.5}
\begin{tabular}[t]{c|c}
\multicolumn{2}{c}{\boldmath$(\bar R R)X^2D$} \\
\hline
$\O_{e^2G^2D}^{(1, R)}$  &  $i (\bar e_{Rp} \gamma^\mu \overleftrightarrow{D}^\lambda e_{Rr}) G_{\mu\rho}^A G_\lambda^{A\rho}$ \\
$\O_{e^2F^2D}^{(1, R)}$  &  $i (\bar e_{Rp} \gamma^\mu \overleftrightarrow{D}^\lambda e_{Rr}) F_{\mu\rho} F_\lambda^{\,\,\,\rho}$ \\
$\O_{u^2G^2D}^{(1, R)}$  &  $i (\bar u_{Rp} \gamma^\mu \overleftrightarrow{D}^\lambda u_{Rr}) G_{\mu\rho}^A G_\lambda^{A\rho}$ \\
$\O_{u^2G^2D}^{(2, R)}$  &  $f^{ABC} (\bar u_{Rp} \gamma^\mu T^A \overleftrightarrow{D}^\lambda u_{Rr}) G_{\mu\rho}^B G_\lambda^{C\rho}$ \\
$\O_{u^2G^2D}^{(3, R)}$  &  $i d^{ABC} (\bar u_{Rp} \gamma^\mu T^A \overleftrightarrow{D}^\lambda u_{Rr}) G_{\mu\rho}^B G_\lambda^{C\rho}$ \\
$\O_{u^2G^2D}^{(4, R)}$  &  $i f^{ABC} (\bar u_{Rp} \gamma^\mu T^A \overleftrightarrow{D}^\lambda u_{Rr}) (G_{\mu\rho}^B \widetilde{G}_\lambda^{C\rho} - \widetilde{G}_{\mu\rho}^B G_\lambda^{C\rho})$ \\
$\O_{u^2G^2D}^{(5, R)}$  &  $f^{ABC} (\bar u_{Rp} \gamma^\mu T^A \overleftrightarrow{D}^\lambda u_{Rr}) (G_{\mu\rho}^B \widetilde{G}_\lambda^{C\rho} + \widetilde{G}_{\mu\rho}^B G_\lambda^{C\rho})$ \\
$\O_{u^2F^2D}^{(1, R)}$  &  $i (\bar u_{Rp} \gamma^\mu \overleftrightarrow{D}^\lambda u_{Rr}) F_{\mu\rho} F_\lambda^{\,\,\,\rho}$ \\
$\O_{u^2GFD}^{(1, R)}$  &  $(\bar u_{Rp} \gamma^\mu T^A \overleftrightarrow{D}^\lambda u_{Rr}) (F_{\mu\rho} G_\lambda^{A\rho} - F_{\lambda\rho} G_{\mu}^{A\rho})$ \\
$\O_{u^2GFD}^{(2, R)}$  &  $i (\bar u_{Rp} \gamma^\mu T^A \overleftrightarrow{D}^\lambda u_{Rr}) (F_{\mu\rho} G_\lambda^{A\rho} + F_{\lambda\rho} G_{\mu}^{A\rho})$ \\
$\O_{u^2GFD}^{(3, R)}$  &  $(\bar u_{Rp} \gamma^\mu T^A \overleftrightarrow{D}^\lambda u_{Rr}) (F_{\mu\rho} \widetilde{G}_\lambda^{A \rho} - F_{\lambda\rho} \widetilde{G}_\mu^{A \rho})$ \\
$\O_{u^2GFD}^{(4, R)}$  &  $i (\bar u_{Rp} \gamma^\mu T^A \overleftrightarrow{D}^\lambda u_{Rr}) (F_{\mu\rho} \widetilde{G}_\lambda^{A \rho} + F_{\lambda\rho} \widetilde{G}_\mu^{A \rho})$ \\
$\O_{d^2G^2D}^{(1, R)}$  &  $i (\bar d_{Rp} \gamma^\mu \overleftrightarrow{D}^\lambda d_{Rr}) G_{\mu\rho}^A G_\lambda^{A\rho}$ \\
$\O_{d^2G^2D}^{(2, R)}$  &  $f^{ABC} (\bar d_{Rp} \gamma^\mu T^A \overleftrightarrow{D}^\lambda d_{Rr}) G_{\mu\rho}^B G_\lambda^{C\rho}$ \\
$\O_{d^2G^2D}^{(3, R)}$  &  $i d^{ABC} (\bar d_{Rp} \gamma^\mu T^A \overleftrightarrow{D}^\lambda d_{Rr}) G_{\mu\rho}^B G_\lambda^{C\rho}$ \\
$\O_{d^2G^2D}^{(4, R)}$  &  $i f^{ABC} (\bar d_{Rp} \gamma^\mu T^A \overleftrightarrow{D}^\lambda d_{Rr}) (G_{\mu\rho}^B \widetilde{G}_\lambda^{C\rho} - \widetilde{G}_{\mu\rho}^B G_\lambda^{C\rho})$ \\
$\O_{d^2G^2D}^{(5, R)}$  &  $f^{ABC} (\bar d_{Rp} \gamma^\mu T^A \overleftrightarrow{D}^\lambda d_{Rr}) (G_{\mu\rho}^B \widetilde{G}_\lambda^{C\rho} + \widetilde{G}_{\mu\rho}^B G_\lambda^{C\rho})$ \\
$\O_{d^2F^2D}^{(1, R)}$  &  $i (\bar d_{Rp} \gamma^\mu \overleftrightarrow{D}^\lambda d_{Rr}) F_{\mu\rho} F_\lambda^{\,\,\,\rho}$ \\
$\O_{d^2GFD}^{(1, R)}$  &  $(\bar d_{Rp} \gamma^\mu T^A \overleftrightarrow{D}^\lambda d_{Rr}) (F_{\mu\rho} G_\lambda^{A\rho} - F_{\lambda\rho} G_{\mu}^{A\rho})$ \\
$\O_{d^2GFD}^{(2, R)}$  &  $i (\bar d_{Rp} \gamma^\mu T^A \overleftrightarrow{D}^\lambda d_{Rr}) (F_{\mu\rho} G_\lambda^{A\rho} + F_{\lambda\rho} G_{\mu}^{A\rho})$ \\
$\O_{d^2GFD}^{(3, R)}$  &  $(\bar d_{Rp} \gamma^\mu T^A \overleftrightarrow{D}^\lambda d_{Rr}) (F_{\mu\rho} \widetilde{G}_\lambda^{A \rho} - F_{\lambda\rho} \widetilde{G}_\mu^{A \rho})$ \\
$\O_{d^2GFD}^{(4, R)}$  &  $i (\bar d_{Rp} \gamma^\mu T^A \overleftrightarrow{D}^\lambda d_{Rr}) (F_{\mu\rho} \widetilde{G}_\lambda^{A \rho} + F_{\lambda\rho} \widetilde{G}_\mu^{A \rho})$ 
\end{tabular}
\end{minipage}
\end{adjustbox}
\end{center}
\caption{The baryon and lepton number conserving operators of class $\psi^2 X^2 D$ with right-handed chiral projectors in the fermion bilinear. The subscripts $p, r$ are weak-eigenstate indices.}
\label{tab:left8_psi2x2d_R}
\end{table}

%% file: sections/basis_psi4x_llrr.tex

\begin{table}[H]
\begin{center}
\begin{adjustbox}{width=0.94\textwidth,center}
\small
\begin{minipage}[t]{6.4cm}
\renewcommand{\arraystretch}{1.5}
\begin{tabular}[t]{c|c}
\multicolumn{2}{c}{\boldmath$(\bar L L)(\bar L L)X$} \\
\hline
$\O_{\nu^2e^2F}^{(1, LL)}$  &  $(\bar \nu_{Lp} \gamma^\mu \nu_{Lr}) (\bar e_{Ls} \gamma^\rho e_{Lt})  F_{\mu\rho}$ \\
$\O_{\nu^2e^2F}^{(2, LL)}$  &  $(\bar \nu_{Lp} \gamma^\mu \nu_{Lr}) (\bar e_{Ls} \gamma^\rho e_{Lt})  \widetilde F_{\mu\rho}$ \\
$\O_{\nu^2u^2G}^{(1, LL)}$  &  $(\bar \nu_{Lp} \gamma^\mu \nu_{Lr}) (\bar u_{Ls} \gamma^\rho T^A u_{Lt}) G_{\mu\rho}^A$ \\
$\O_{\nu^2u^2G}^{(2, LL)}$  &  $(\bar \nu_{Lp} \gamma^\mu \nu_{Lr}) (\bar u_{Ls} \gamma^\rho T^A u_{Lt}) \widetilde G_{\mu\rho}^A$ \\
$\O_{\nu^2u^2F}^{(1, LL)}$  &  $(\bar \nu_{Lp} \gamma^\mu \nu_{Lr}) (\bar u_{Ls} \gamma^\rho u_{Lt}) F_{\mu\rho}$ \\
$\O_{\nu^2u^2F}^{(2, LL)}$  &  $(\bar \nu_{Lp} \gamma^\mu \nu_{Lr}) (\bar u_{Ls} \gamma^\rho u_{Lt}) \widetilde F_{\mu\rho}$ \\
$\O_{\nu^2d^2G}^{(1, LL)}$  &  $(\bar \nu_{Lp} \gamma^\mu \nu_{Lr}) (\bar d_{Ls} \gamma^\rho T^A d_{Lt}) G_{\mu\rho}^A$ \\
$\O_{\nu^2d^2G}^{(2, LL)}$  &  $(\bar \nu_{Lp} \gamma^\mu \nu_{Lr}) (\bar d_{Ls} \gamma^\rho T^A d_{Lt}) \widetilde G_{\mu\rho}^A$ \\
$\O_{\nu^2d^2F}^{(1, LL)}$  &  $(\bar \nu_{Lp} \gamma^\mu \nu_{Lr}) (\bar d_{Ls} \gamma^\rho d_{Lt}) F_{\mu\rho}$ \\
$\O_{\nu^2d^2F}^{(2, LL)}$  &  $(\bar \nu_{Lp} \gamma^\mu \nu_{Lr}) (\bar d_{Ls} \gamma^\rho d_{Lt}) \widetilde F_{\mu\rho}$ \\ 
$\O_{\nu eudG}^{(1, LL)}$  &  $(\bar \nu_{Lp} \gamma^\mu e_{Lr}) (\bar d_{Ls} \gamma^\rho T^A u_{Lt}) G_{\mu\rho}^A + \hc$ \\
$\O_{\nu eudG}^{(2, LL)}$  &  $(\bar \nu_{Lp} \gamma^\mu e_{Lr}) (\bar d_{Ls} \gamma^\rho T^A u_{Lt}) \widetilde G_{\mu\rho}^A + \hc$ \\
$\O_{\nu eudF}^{(1, LL)}$  &  $(\bar \nu_{Lp} \gamma^\mu e_{Lr}) (\bar d_{Ls} \gamma^\rho u_{Lt}) F_{\mu\rho} + \hc$ \\
$\O_{\nu eudF}^{(2, LL)}$  &  $(\bar \nu_{Lp} \gamma^\mu e_{Lr}) (\bar d_{Ls} \gamma^\rho u_{Lt}) \widetilde F_{\mu\rho} + \hc$ \\ 
$\O_{e^2u^2G}^{(1, LL)}$  &  $(\bar e_{Lp} \gamma^\mu e_{Lr})(\bar u_{Ls} \gamma^\rho T^A u_{Lt}) G_{\mu\rho}^A$ \\
$\O_{e^2u^2G}^{(2, LL)}$  &  $(\bar e_{Lp} \gamma^\mu e_{Lr})(\bar u_{Ls} \gamma^\rho T^A u_{Lt}) \widetilde G_{\mu\rho}^A$ \\
$\O_{e^2u^2F}^{(1, LL)}$  &  $(\bar e_{Lp} \gamma^\mu e_{Lr})(\bar u_{Ls} \gamma^\rho u_{Lt}) F_{\mu\rho}$ \\
$\O_{e^2u^2F}^{(2, LL)}$  &  $(\bar e_{Lp} \gamma^\mu e_{Lr})(\bar u_{Ls} \gamma^\rho u_{Lt}) \widetilde F_{\mu\rho}$ \\
$\O_{e^2d^2G}^{(1, LL)}$  &  $(\bar e_{Lp} \gamma^\mu e_{Lr})(\bar d_{Ls} \gamma^\rho T^A d_{Lt}) G_{\mu\rho}^A$ \\
$\O_{e^2d^2G}^{(2, LL)}$  &  $(\bar e_{Lp} \gamma^\mu e_{Lr})(\bar d_{Ls} \gamma^\rho T^A d_{Lt}) \widetilde G_{\mu\rho}^A$ \\
$\O_{e^2d^2F}^{(1, LL)}$  &  $(\bar e_{Lp} \gamma^\mu e_{Lr})(\bar d_{Ls} \gamma^\rho d_{Lt}) F_{\mu\rho}$ \\
$\O_{e^2d^2F}^{(2, LL)}$  &  $(\bar e_{Lp} \gamma^\mu e_{Lr})(\bar d_{Ls} \gamma^\rho d_{Lt}) \widetilde F_{\mu\rho}$ \\ \hdashline
$\O_{\nu^4F}^{(1, LL)}$  &  $(\bar \nu_{Lp} \gamma^\mu \nu_{Lr}) (\bar \nu_{Ls} \gamma^\rho \nu_{Lt}) F_{\mu\rho}$ \\
$\O_{\nu^4F}^{(2, LL)}$  &  $(\bar \nu_{Lp} \gamma^\mu \nu_{Lr}) (\bar \nu_{Ls} \gamma^\rho \nu_{Lt}) \widetilde F_{\mu\rho}$ \\
$\O_{e^4F}^{(1, LL)}$  &  $(\bar e_{Lp} \gamma^\mu e_{Lr}) (\bar e_{Ls} \gamma^\rho e_{Lt}) F_{\mu\rho}$ \\
$\O_{e^4F}^{(2, LL)}$  &  $(\bar e_{Lp} \gamma^\mu e_{Lr}) (\bar e_{Ls} \gamma^\rho e_{Lt}) \widetilde F_{\mu\rho}$
\end{tabular}
\end{minipage}
\hspace{1cm}
\begin{minipage}[t]{6.7cm}
\renewcommand{\arraystretch}{1.5}
\begin{tabular}[t]{c|c}
\multicolumn{2}{c}{\boldmath$(\bar L L)(\bar L L)X$} \\
\hline
$\O_{u^4G}^{(1, LL)}$  &  $(\bar u_{Lp} \gamma^\mu u_{Lr}) (\bar u_{Ls} \gamma^\rho T^A u_{Lt}) G^A_{\mu\rho}$ \\
$\O_{u^4G}^{(2, LL)}$  &  $(\bar u_{Lp} \gamma^\mu u_{Lr}) (\bar u_{Ls} \gamma^\rho T^A u_{Lt}) \widetilde G^A_{\mu\rho}$ \\
$\O_{d^4G}^{(1, LL)}$  &  $(\bar d_{Lp} \gamma^\mu d_{Lr}) (\bar d_{Ls} \gamma^\rho T^A d_{Lt}) G^A_{\mu\rho}$ \\
$\O_{d^4G}^{(2, LL)}$  &  $(\bar d_{Lp} \gamma^\mu d_{Lr}) (\bar d_{Ls} \gamma^\rho T^A d_{Lt}) \widetilde G^A_{\mu\rho}$ \\
$\O_{u^2d^2G}^{(1, LL)}$  &  $(\bar u_{Lp} \gamma^\mu u_{Lr})(\bar d_{Ls} \gamma^\rho T^A d_{Lt}) G_{\mu\rho}^A$ \\
$\O_{u^2d^2G}^{(2, LL)}$  &  $(\bar u_{Lp} \gamma^\mu u_{Lr})(\bar d_{Ls} \gamma^\rho T^A d_{Lt}) \widetilde G_{\mu\rho}^A$ \\
$\O_{u^2d^2G}^{(3, LL)}$  &  $(\bar u_{Lp} \gamma^\mu T^A u_{Lr})(\bar d_{Ls} \gamma^\rho d_{Lt}) G_{\mu\rho}^A$ \\
$\O_{u^2d^2G}^{(4, LL)}$  &  $(\bar u_{Lp} \gamma^\mu T^A u_{Lr})(\bar d_{Ls} \gamma^\rho d_{Lt}) \widetilde G_{\mu\rho}^A$ \\
$\O_{u^2d^2G}^{(5, LL)}$  &  $f^{ABC} (\bar u_{Lp} \gamma^\mu T^A u_{Lr})(\bar d_{Ls} \gamma^\rho T^B d_{Lt}) G_{\mu\rho}^C$ \\
$\O_{u^2d^2G}^{(6, LL)}$  &  $f^{ABC} (\bar u_{Lp} \gamma^\mu T^A u_{Lr})(\bar d_{Ls} \gamma^\rho T^B d_{Lt}) \widetilde G_{\mu\rho}^C$ \\
$\O_{u^2d^2G}^{(7, LL)}$  &  $d^{ABC} (\bar u_{Lp} \gamma^\mu T^A u_{Lr})(\bar d_{Ls} \gamma^\rho T^B d_{Lt}) G_{\mu\rho}^C$ \\
$\O_{u^2d^2G}^{(8, LL)}$  &  $d^{ABC} (\bar u_{Lp} \gamma^\mu T^A u_{Lr})(\bar d_{Ls} \gamma^\rho T^B d_{Lt}) \widetilde G_{\mu\rho}^C$ \\
$\O_{u^2d^2F}^{(1, LL)}$  &  $(\bar u_{Lp} \gamma^\mu u_{Lr})(\bar d_{Ls} \gamma^\rho d_{Lt}) F_{\mu\rho}$ \\
$\O_{u^2d^2F}^{(2, LL)}$  &  $(\bar u_{Lp} \gamma^\mu u_{Lr})(\bar d_{Ls} \gamma^\rho d_{Lt}) \widetilde F_{\mu\rho}$ \\
$\O_{u^2d^2F}^{(3, LL)}$  &  $(\bar u_{Lp} \gamma^\mu T^A u_{Lr})(\bar d_{Ls} \gamma^\rho T^A d_{Lt}) F_{\mu\rho}$ \\
$\O_{u^2d^2F}^{(4, LL)}$  &  $(\bar u_{Lp} \gamma^\mu T^A u_{Lr})(\bar d_{Ls} \gamma^\rho T^A d_{Lt}) \widetilde F_{\mu\rho}$ \\ \hdashline
$\O_{u^4F}^{(1, LL)}$  &  $(\bar u_{Lp} \gamma^\mu u_{Lr}) (\bar u_{Ls} \gamma^\rho u_{Lt}) F_{\mu\rho}$ \\
$\O_{u^4F}^{(2, LL)}$  &  $(\bar u_{Lp} \gamma^\mu u_{Lr}) (\bar u_{Ls} \gamma^\rho u_{Lt}) \widetilde F_{\mu\rho}$ \\
$\O_{d^4F}^{(1, LL)}$  & $(\bar d_{Lp} \gamma^\mu d_{Lr}) (\bar d_{Ls} \gamma^\rho d_{Lt}) F_{\mu\rho}$ \\
$\O_{d^4F}^{(2, LL)}$  & $(\bar d_{Lp} \gamma^\mu d_{Lr}) (\bar d_{Ls} \gamma^\rho d_{Lt}) \widetilde F_{\mu\rho}$
\end{tabular}
\end{minipage}
\end{adjustbox}
\end{center}
\caption{The baryon and lepton number conserving operators of class $\psi^4 X$ where both fermion bilinears involve left-handed chiral projectors. 
Operators with $\hc$ have distinct Hermitian conjugates.
The subscripts $p, r, s, t$ are weak-eigenstate indices.
Operators below the dashed lines vanish when there is only one generation of fermions.}
\label{tab:left8_psi4x_LL}
\end{table}

%

\begin{table}[H]
\begin{center}
\begin{adjustbox}{width=0.89\textwidth,center}
\small
\begin{minipage}[t]{5.6cm}
\renewcommand{\arraystretch}{1.5}
\begin{tabular}[t]{c|c}
\multicolumn{2}{c}{\boldmath$(\bar R R)(\bar R R)X$} \\
\hline
$\O_{u^4G}^{(1, RR)}$  &  $(\bar u_{Rp} \gamma^\mu u_{Rr}) (\bar u_{Rs} \gamma^\rho T^A u_{Rt}) G^A_{\mu\rho}$ \\
$\O_{u^4G}^{(2, RR)}$  &  $(\bar u_{Rp} \gamma^\mu u_{Rr}) (\bar u_{Rs} \gamma^\rho T^A u_{Rt}) \widetilde G^A_{\mu\rho}$ \\
$\O_{d^4G}^{(1, RR)}$  &  $(\bar d_{Rp} \gamma^\mu d_{Rr}) (\bar d_{Rs} \gamma^\rho T^A d_{Rt}) G^A_{\mu\rho}$ \\
$\O_{d^4G}^{(2, RR)}$  &  $(\bar d_{Rp} \gamma^\mu d_{Rr}) (\bar d_{Rs} \gamma^\rho T^A d_{Rt}) \widetilde G^A_{\mu\rho}$ \\
$\O_{e^2u^2G}^{(1, RR)}$  &  $(\bar e_{Rp} \gamma^\mu e_{Rr})(\bar u_{Rs} \gamma^\rho T^A u_{Rt}) G_{\mu\rho}^A$ \\
$\O_{e^2u^2G}^{(2, RR)}$  &  $(\bar e_{Rp} \gamma^\mu e_{Rr})(\bar u_{Rs} \gamma^\rho T^A u_{Rt}) \widetilde G_{\mu\rho}^A$ \\
$\O_{e^2u^2F}^{(1, RR)}$  &  $(\bar e_{Rp} \gamma^\mu e_{Rr})(\bar u_{Rs} \gamma^\rho u_{Rt}) F_{\mu\rho}$ \\
$\O_{e^2u^2F}^{(2, RR)}$  &  $(\bar e_{Rp} \gamma^\mu e_{Rr})(\bar u_{Rs} \gamma^\rho u_{Rt}) \widetilde F_{\mu\rho}$ \\
$\O_{e^2d^2G}^{(1, RR)}$  &  $(\bar e_{Rp} \gamma^\mu e_{Rr})(\bar d_{Rs} \gamma^\rho T^A d_{Rt}) G_{\mu\rho}^A$ \\
$\O_{e^2d^2G}^{(2, RR)}$  &  $(\bar e_{Rp} \gamma^\mu e_{Rr})(\bar d_{Rs} \gamma^\rho T^A d_{Rt}) \widetilde G_{\mu\rho}^A$ \\
$\O_{e^2d^2F}^{(1, RR)}$  &  $(\bar e_{Rp} \gamma^\mu e_{Rr})(\bar d_{Rs} \gamma^\rho d_{Rt}) F_{\mu\rho}$ \\
$\O_{e^2d^2F}^{(2, RR)}$  &  $(\bar e_{Rp} \gamma^\mu e_{Rr})(\bar d_{Rs} \gamma^\rho d_{Rt}) \widetilde F_{\mu\rho}$ \\ \hdashline
$\O_{e^4F}^{(1, RR)}$  &  $(\bar e_{Rp} \gamma^\mu e_{Rr}) (\bar e_{Rs} \gamma^\rho e_{Rt}) F_{\mu\rho}$ \\
$\O_{e^4F}^{(2, RR)}$  &  $(\bar e_{Rp} \gamma^\mu e_{Rr}) (\bar e_{Rs} \gamma^\rho e_{Rt}) \widetilde F_{\mu\rho}$ \\
\end{tabular}
\end{minipage}
\hspace{1cm}
\begin{minipage}[t]{6.7cm}
\renewcommand{\arraystretch}{1.5}
\begin{tabular}[t]{c|c}
\multicolumn{2}{c}{\boldmath$(\bar R R)(\bar R R)X$} \\
\hline
$\O_{u^2d^2G}^{(1, RR)}$  &  $(\bar u_{Rp} \gamma^\mu u_{Rr})(\bar d_{Rs} \gamma^\rho T^A d_{Rt}) G_{\mu\rho}^A$ \\
$\O_{u^2d^2G}^{(2, RR)}$  &  $(\bar u_{Rp} \gamma^\mu u_{Rr})(\bar d_{Rs} \gamma^\rho T^A d_{Rt}) \widetilde G_{\mu\rho}^A$ \\
$\O_{u^2d^2G}^{(3, RR)}$  &  $(\bar u_{Rp} \gamma^\mu T^A u_{Rr})(\bar d_{Rs} \gamma^\rho d_{Rt}) G_{\mu\rho}^A$ \\
$\O_{u^2d^2G}^{(4, RR)}$  &  $(\bar u_{Rp} \gamma^\mu T^A u_{Rr})(\bar d_{Rs} \gamma^\rho d_{Rt}) \widetilde G_{\mu\rho}^A$ \\
$\O_{u^2d^2G}^{(5, RR)}$  &  $f^{ABC} (\bar u_{Rp} \gamma^\mu T^A u_{Rr})(\bar d_{Rs} \gamma^\rho T^B d_{Rt}) G_{\mu\rho}^C$ \\
$\O_{u^2d^2G}^{(6, RR)}$  &  $f^{ABC} (\bar u_{Rp} \gamma^\mu T^A u_{Rr})(\bar d_{Rs} \gamma^\rho T^B d_{Rt}) \widetilde G_{\mu\rho}^C$ \\
$\O_{u^2d^2G}^{(7, RR)}$  &  $d^{ABC} (\bar u_{Rp} \gamma^\mu T^A u_{Rr})(\bar d_{Rs} \gamma^\rho T^B d_{Rt}) G_{\mu\rho}^C$ \\
$\O_{u^2d^2G}^{(8, RR)}$  &  $d^{ABC} (\bar u_{Rp} \gamma^\mu T^A u_{Rr})(\bar d_{Rs} \gamma^\rho T^B d_{Rt}) \widetilde G_{\mu\rho}^C$ \\
$\O_{u^2d^2F}^{(1, RR)}$  &  $(\bar u_{Rp} \gamma^\mu u_{Rr})(\bar d_{Rs} \gamma^\rho d_{Rt}) F_{\mu\rho}$ \\
$\O_{u^2d^2F}^{(2, RR)}$  &  $(\bar u_{Rp} \gamma^\mu u_{Rr})(\bar d_{Rs} \gamma^\rho d_{Rt}) \widetilde F_{\mu\rho}$ \\
$\O_{u^2d^2F}^{(3, RR)}$  &  $(\bar u_{Rp} \gamma^\mu T^A u_{Rr})(\bar d_{Rs} \gamma^\rho T^A d_{Rt}) F_{\mu\rho}$ \\
$\O_{u^2d^2F}^{(4, RR)}$  &  $(\bar u_{Rp} \gamma^\mu T^A u_{Rr})(\bar d_{Rs} \gamma^\rho T^A d_{Rt}) \widetilde F_{\mu\rho}$ \\ \hdashline
$\O_{u^4F}^{(1, RR)}$  &  $(\bar u_{Rp} \gamma^\mu u_{Rr}) (\bar u_{Rs} \gamma^\rho u_{Rt}) F_{\mu\rho}$ \\
$\O_{u^4F}^{(2, RR)}$  &  $(\bar u_{Rp} \gamma^\mu u_{Rr}) (\bar u_{Rs} \gamma^\rho u_{Rt}) \widetilde F_{\mu\rho}$ \\
$\O_{d^4F}^{(1, RR)}$  & $(\bar d_{Rp} \gamma^\mu d_{Rr}) (\bar d_{Rs} \gamma^\rho d_{Rt}) F_{\mu\rho}$ \\
$\O_{d^4F}^{(2, RR)}$  & $(\bar d_{Rp} \gamma^\mu d_{Rr}) (\bar d_{Rs} \gamma^\rho d_{Rt}) \widetilde F_{\mu\rho}$
\end{tabular}
\end{minipage}
\end{adjustbox}
\begin{adjustbox}{width=0.79\textwidth,center}
\small
\begin{minipage}[t]{5.4cm}
\renewcommand{\arraystretch}{1.5}
\begin{tabular}[t]{c|c}
\multicolumn{2}{c}{\boldmath$(\bar L R)(\bar R L)X + \hc$} \\
\hline
$\O_{\nu eudG}^{(1, ST)}$  &  $(\bar \nu_{Lp} e_{Rr}) (\bar d_{Rs} \sigma^{\mu\rho} T^A u_{Lt}) G_{\mu\rho}^A$ \\
$\O_{\nu eudG}^{(1, TS)}$  &  $(\bar \nu_{Lp} \sigma^{\mu\rho} e_{Rr}) (\bar d_{Rs} T^A u_{Lt}) G_{\mu\rho}^A$ \\
$\O_{\nu eudF}^{(1, ST)}$  &  $(\bar \nu_{Lp} e_{Rr}) (\bar d_{Rs} \sigma^{\mu\rho} u_{Lt}) F_{\mu\rho}$ \\
$\O_{\nu eudF}^{(1, TS)}$  &  $(\bar \nu_{Lp} \sigma^{\mu\rho} e_{Rr}) (\bar d_{Rs} u_{Lt}) F_{\mu\rho}$ \\
$\O_{e^2u^2G}^{(1, ST)}$  &  $(\bar e_{Lp} e_{Rr}) (\bar u_{Rs} \sigma^{\mu\rho} T^A u_{Lt}) G_{\mu\rho}^A$ \\
$\O_{e^2u^2G}^{(1, TS)}$  &  $(\bar e_{Lp} \sigma^{\mu\rho} e_{Rr}) (\bar u_{Rs} T^A u_{Lt}) G_{\mu\rho}^A$ 
\end{tabular}
\end{minipage}
\hspace{1cm}
\begin{minipage}[t]{5.4cm}
\renewcommand{\arraystretch}{1.5}
\begin{tabular}[t]{c|c}
\multicolumn{2}{c}{\boldmath$(\bar L R)(\bar R L)X + \hc$} \\
\hline
$\O_{e^2u^2F}^{(1, ST)}$  &  $(\bar e_{Lp} e_{Rr}) (\bar u_{Rs} \sigma^{\mu\rho} u_{Lt}) F_{\mu\rho}$ \\
$\O_{e^2u^2F}^{(1, TS)}$  &  $(\bar e_{Lp} \sigma^{\mu\rho} e_{Rr}) (\bar u_{Rs} u_{Lt}) F_{\mu\rho}$ \\
$\O_{e^2d^2G}^{(1, ST)}$  &  $(\bar e_{Lp} e_{Rr}) (\bar d_{Rs} \sigma^{\mu\rho} T^A d_{Lt}) G_{\mu\rho}^A$ \\
$\O_{e^2d^2G}^{(1, TS)}$  &  $(\bar e_{Lp} \sigma^{\mu\rho} e_{Rr}) (\bar d_{Rs} T^A d_{Lt}) G_{\mu\rho}^A$ \\
$\O_{e^2d^2F}^{(1, ST)}$  &  $(\bar e_{Lp} e_{Rr}) (\bar d_{Rs} \sigma^{\mu\rho} d_{Lt}) F_{\mu\rho}$ \\
$\O_{e^2d^2F}^{(1, TS)}$  &  $(\bar e_{Lp} \sigma^{\mu\rho} e_{Rr}) (\bar d_{Rs} d_{Lt}) F_{\mu\rho}$ 
\end{tabular}
\end{minipage}
\end{adjustbox}
\end{center}
\caption{The baryon and lepton number conserving operators of class $\psi^4 X$ and subclass $\psi_L^2 \psi_R^2 X$ involving a scalar and a tensor fermion bilinear or where both fermion bilinears involve right-handed chiral projectors. 
Operators with $+ \hc$ have distinct Hermitian conjugates.
Operators below the dashed lines vanish when there is only one generation of fermions.
The subscripts $p, r, s, t$ are weak-eigenstate indices.}
\label{tab:left8_psi4x_RR_LRRL}
\end{table}

%

\begin{table}[H]
\begin{center}
\begin{adjustbox}{width=1.04\textwidth,center}
\small
\begin{minipage}[t]{6.9cm}
\renewcommand{\arraystretch}{1.5}
\begin{tabular}[t]{c|c}
\multicolumn{2}{c}{\boldmath$(\bar L L)(\bar R R)G$} \\
\hline
$\O_{\nu^2u^2G}^{(1, LR)}$  &  $(\bar \nu_{Lp} \gamma^\mu \nu_{Lr}) (\bar u_{Rs} \gamma^\rho T^A u_{Rt}) G_{\mu\rho}^A$ \\
$\O_{\nu^2u^2G}^{(2, LR)}$  &  $(\bar \nu_{Lp} \gamma^\mu \nu_{Lr}) (\bar u_{Rs} \gamma^\rho T^A u_{Rt}) \widetilde G_{\mu\rho}^A$ \\
$\O_{\nu^2d^2G}^{(1, LR)}$  &  $(\bar \nu_{Lp} \gamma^\mu \nu_{Lr}) (\bar d_{Rs} \gamma^\rho T^A d_{Rt}) G_{\mu\rho}^A$ \\
$\O_{\nu^2d^2G}^{(2, LR)}$  &  $(\bar \nu_{Lp} \gamma^\mu \nu_{Lr}) (\bar d_{Rs} \gamma^\rho T^A d_{Rt}) \widetilde G_{\mu\rho}^A$ \\
$\O_{\nu eudG}^{(1, LR)}$  &  $(\bar \nu_{Lp} \gamma^\mu e_{Lr}) (\bar d_{Rs} \gamma^\rho T^A u_{Rt}) G_{\mu\rho}^A + \hc$ \\
$\O_{\nu eudG}^{(2, LR)}$  &  $(\bar \nu_{Lp} \gamma^\mu e_{Lr}) (\bar d_{Rs} \gamma^\rho T^A u_{Rt}) \widetilde G_{\mu\rho}^A + \hc$ \\
$\O_{e^2u^2G}^{(1, LR)}$  &  $(\bar e_{Lp} \gamma^\mu e_{Lr})(\bar u_{Rs} \gamma^\rho T^A u_{Rt}) G_{\mu\rho}^A$ \\
$\O_{e^2u^2G}^{(2, LR)}$  &  $(\bar e_{Lp} \gamma^\mu e_{Lr})(\bar u_{Rs} \gamma^\rho T^A u_{Rt}) \widetilde G_{\mu\rho}^A$ \\
$\O_{u^2e^2G}^{(1, LR)}$  &  $(\bar u_{Lp} \gamma^\mu T^A u_{Lr})(\bar e_{Rs} \gamma^\rho e_{Rt}) G_{\mu\rho}^A$ \\
$\O_{u^2e^2G}^{(2, LR)}$  &  $(\bar u_{Lp} \gamma^\mu T^A u_{Lr})(\bar e_{Rs} \gamma^\rho e_{Rt}) \widetilde G_{\mu\rho}^A$ \\
$\O_{e^2d^2G}^{(1, LR)}$  &  $(\bar e_{Lp} \gamma^\mu e_{Lr})(\bar d_{Rs} \gamma^\rho T^A d_{Rt}) G_{\mu\rho}^A$ \\
$\O_{e^2d^2G}^{(2, LR)}$  &  $(\bar e_{Lp} \gamma^\mu e_{Lr})(\bar d_{Rs} \gamma^\rho T^A d_{Rt}) \widetilde G_{\mu\rho}^A$ \\
$\O_{d^2e^2G}^{(1, LR)}$  &  $(\bar d_{Lp} \gamma^\mu T^A d_{Lr})(\bar e_{Rs} \gamma^\rho d_{Rt}) G_{\mu\rho}^A$ \\
$\O_{d^2e^2G}^{(2, LR)}$  &  $(\bar d_{Lp} \gamma^\mu T^A d_{Lr})(\bar e_{Rs} \gamma^\rho d_{Rt}) \widetilde G_{\mu\rho}^A$ \\
$\O_{u^4G}^{(1, LR)}$  &  $(\bar u_{Lp} \gamma^\mu u_{Lr})(\bar u_{Rs} \gamma^\rho T^A u_{Rt}) G_{\mu\rho}^A$ \\
$\O_{u^4G}^{(2, LR)}$  &  $(\bar u_{Lp} \gamma^\mu u_{Lr})(\bar u_{Rs} \gamma^\rho T^A u_{Rt}) \widetilde G_{\mu\rho}^A$ \\
$\O_{u^4G}^{(3, LR)}$  &  $(\bar u_{Lp} \gamma^\mu T^A u_{Lr})(\bar u_{Rs} \gamma^\rho u_{Rt}) G_{\mu\rho}^A$ \\
$\O_{u^4G}^{(4, LR)}$  &  $(\bar u_{Lp} \gamma^\mu T^A u_{Lr})(\bar u_{Rs} \gamma^\rho u_{Rt}) \widetilde G_{\mu\rho}^A$ \\
$\O_{u^4G}^{(5, LR)}$  &  $f^{ABC} (\bar u_{Lp} \gamma^\mu T^A u_{Lr})(\bar u_{Rs} \gamma^\rho T^B u_{Rt}) G_{\mu\rho}^C$ \\
$\O_{u^4G}^{(6, LR)}$  &  $f^{ABC} (\bar u_{Lp} \gamma^\mu T^A u_{Lr})(\bar u_{Rs} \gamma^\rho T^B u_{Rt}) \widetilde G_{\mu\rho}^C$ \\
$\O_{u^4G}^{(7, LR)}$  &  $d^{ABC} (\bar u_{Lp} \gamma^\mu T^A u_{Lr})(\bar u_{Rs} \gamma^\rho T^B u_{Rt}) G_{\mu\rho}^C$ \\
$\O_{u^4G}^{(8, LR)}$  &  $d^{ABC} (\bar u_{Lp} \gamma^\mu T^A u_{Lr})(\bar u_{Rs} \gamma^\rho T^B u_{Rt}) \widetilde G_{\mu\rho}^C$ \\
$\O_{d^4G}^{(1, LR)}$  &  $(\bar d_{Lp} \gamma^\mu d_{Lr})(\bar d_{Rs} \gamma^\rho T^A d_{Rt}) G_{\mu\rho}^A$ \\
$\O_{d^4G}^{(2, LR)}$  &  $(\bar d_{Lp} \gamma^\mu d_{Lr})(\bar d_{Rs} \gamma^\rho T^A d_{Rt}) \widetilde G_{\mu\rho}^A$ \\
$\O_{d^4G}^{(3, LR)}$  &  $(\bar d_{Lp} \gamma^\mu T^A d_{Lr})(\bar d_{Rs} \gamma^\rho d_{Rt}) G_{\mu\rho}^A$ \\
$\O_{d^4G}^{(4, LR)}$  &  $(\bar d_{Lp} \gamma^\mu T^A d_{Lr})(\bar d_{Rs} \gamma^\rho d_{Rt}) \widetilde G_{\mu\rho}^A$ \\
$\O_{d^4G}^{(5, LR)}$  &  $f^{ABC} (\bar d_{Lp} \gamma^\mu T^A d_{Lr})(\bar d_{Rs} \gamma^\rho T^B d_{Rt}) G_{\mu\rho}^C$ \\
$\O_{d^4G}^{(6, LR)}$  &  $f^{ABC} (\bar d_{Lp} \gamma^\mu T^A d_{Lr})(\bar d_{Rs} \gamma^\rho T^B d_{Rt}) \widetilde G_{\mu\rho}^C$ \\
$\O_{d^4G}^{(7, LR)}$  &  $d^{ABC} (\bar d_{Lp} \gamma^\mu T^A d_{Lr})(\bar d_{Rs} \gamma^\rho T^B d_{Rt}) G_{\mu\rho}^C$ \\
$\O_{d^4G}^{(8, LR)}$  &  $d^{ABC} (\bar d_{Lp} \gamma^\mu T^A d_{Lr})(\bar d_{Rs} \gamma^\rho T^B d_{Rt}) \widetilde G_{\mu\rho}^C$
\end{tabular}
\end{minipage}
\hspace{1cm}
\begin{minipage}[t]{7.7cm}
\renewcommand{\arraystretch}{1.5}
\begin{tabular}[t]{c|c}
\multicolumn{2}{c}{\boldmath$(\bar L L)(\bar R R)G$} \\
\hline
$\O_{u^2d^2G}^{(1, LR)}$  &  $(\bar u_{Lp} \gamma^\mu u_{Lr})(\bar d_{Rs} \gamma^\rho T^A d_{Rt}) G_{\mu\rho}^A$ \\
$\O_{u^2d^2G}^{(2, LR)}$  &  $(\bar u_{Lp} \gamma^\mu u_{Lr})(\bar d_{Rs} \gamma^\rho T^A d_{Rt}) \widetilde G_{\mu\rho}^A$ \\
$\O_{u^2d^2G}^{(3, LR)}$  &  $(\bar u_{Lp} \gamma^\mu T^A u_{Lr})(\bar d_{Rs} \gamma^\rho d_{Rt}) G_{\mu\rho}^A$ \\
$\O_{u^2d^2G}^{(4, LR)}$  &  $(\bar u_{Lp} \gamma^\mu T^A u_{Lr})(\bar d_{Rs} \gamma^\rho d_{Rt}) \widetilde G_{\mu\rho}^A$ \\
$\O_{u^2d^2G}^{(5, LR)}$  &  $f^{ABC} (\bar u_{Lp} \gamma^\mu T^A u_{Lr})(\bar d_{Rs} \gamma^\rho T^B d_{Rt}) G_{\mu\rho}^C$ \\
$\O_{u^2d^2G}^{(6, LR)}$  &  $f^{ABC} (\bar u_{Lp} \gamma^\mu T^A u_{Lr})(\bar d_{Rs} \gamma^\rho T^B d_{Rt}) \widetilde G_{\mu\rho}^C$ \\
$\O_{u^2d^2G}^{(7, LR)}$  &  $d^{ABC} (\bar u_{Lp} \gamma^\mu T^A u_{Lr})(\bar d_{Rs} \gamma^\rho T^B d_{Rt}) G_{\mu\rho}^C$ \\
$\O_{u^2d^2G}^{(8, LR)}$  &  $d^{ABC} (\bar u_{Lp} \gamma^\mu T^A u_{Lr})(\bar d_{Rs} \gamma^\rho T^B d_{Rt}) \widetilde G_{\mu\rho}^C$ \\
$\O_{d^2u^2G}^{(1, LR)}$  &  $(\bar d_{Lp} \gamma^\mu T^A d_{Lr}) (\bar u_{Rs} \gamma^\rho u_{Rt}) G_{\mu\rho}^A$ \\
$\O_{d^2u^2G}^{(2, LR)}$  &  $(\bar d_{Lp} \gamma^\mu T^A d_{Lr})  (\bar u_{Rs} \gamma^\rho u_{Rt}) \widetilde G_{\mu\rho}^A$ \\
$\O_{d^2u^2G}^{(3, LR)}$  &  $(\bar d_{Lp} \gamma^\mu d_{Lr}) (\bar u_{Rs} \gamma^\rho T^A u_{Rt}) G_{\mu\rho}^A$ \\
$\O_{d^2u^2G}^{(4, LR)}$  &  $(\bar d_{Lp} \gamma^\mu d_{Lr})  (\bar u_{Rs} \gamma^\rho T^A u_{Rt}) \widetilde G_{\mu\rho}^A$ \\
$\O_{d^2u^2G}^{(5, LR)}$  &  $f^{ABC} (\bar d_{Lp} \gamma^\mu T^A d_{Lr}) (\bar u_{Rs} \gamma^\rho T^B u_{Rt}) G_{\mu\rho}^C$ \\
$\O_{d^2u^2G}^{(6, LR)}$  &  $f^{ABC} (\bar d_{Lp} \gamma^\mu T^A d_{Lr}) (\bar u_{Rs} \gamma^\rho T^B u_{Rt}) \widetilde G_{\mu\rho}^C$ \\
$\O_{d^2u^2G}^{(7, LR)}$  &  $d^{ABC} (\bar d_{Lp} \gamma^\mu T^A d_{Lr}) (\bar u_{Rs} \gamma^\rho T^B u_{Rt}) G_{\mu\rho}^C$ \\
$\O_{d^2u^2G}^{(8, LR)}$  &  $d^{ABC} (\bar d_{Lp} \gamma^\mu T^A d_{Lr}) (\bar u_{Rs} \gamma^\rho T^B u_{Rt}) \widetilde G_{\mu\rho}^C$ \\
$\O_{(ud)^2G}^{(1, LR)}$  &  $(\bar d_{Lp} \gamma^\mu u_{Lr})(\bar u_{Rs} \gamma^\rho T^A d_{Rt}) G_{\mu\rho}^A + \hc$ \\
$\O_{(ud)^2G}^{(2, LR)}$  &  $(\bar d_{Lp} \gamma^\mu u_{Lr})(\bar u_{Rs} \gamma^\rho T^A d_{Rt}) \widetilde G_{\mu\rho}^A + \hc$ \\
$\O_{(ud)^2G}^{(3, LR)}$  &  $(\bar d_{Lp} \gamma^\mu T^A u_{Lr})(\bar u_{Rs} \gamma^\rho d_{Rt}) G_{\mu\rho}^A + \hc$ \\
$\O_{(ud)^2G}^{(4, LR)}$  &  $(\bar d_{Lp} \gamma^\mu T^A u_{Lr})(\bar u_{Rs} \gamma^\rho d_{Rt}) \widetilde G_{\mu\rho}^A + \hc$ \\
$\O_{(ud)^2G}^{(5, LR)}$  &  $f^{ABC} (\bar d_{Lp} \gamma^\mu T^A u_{Lr})(\bar u_{Rs} \gamma^\rho T^B d_{Rt}) G_{\mu\rho}^C + \hc$ \\
$\O_{(ud)^2G}^{(6, LR)}$  &  $f^{ABC} (\bar d_{Lp} \gamma^\mu T^A u_{Lr})(\bar u_{Rs} \gamma^\rho T^B d_{Rt}) \widetilde G_{\mu\rho}^C + \hc$ \\
$\O_{(ud)^2G}^{(7, LR)}$  &  $d^{ABC} (\bar d_{Lp} \gamma^\mu T^A u_{Lr})(\bar u_{Rs} \gamma^\rho T^B d_{Rt}) G_{\mu\rho}^C + \hc$ \\
$\O_{(ud)^2G}^{(8, LR)}$  &  $d^{ABC} (\bar d_{Lp} \gamma^\mu T^A u_{Lr})(\bar u_{Rs} \gamma^\rho T^B d_{Rt}) \widetilde G_{\mu\rho}^C + \hc$ 
\end{tabular}
\end{minipage}
\end{adjustbox}
\end{center}
\caption{The baryon and lepton number conserving operators of class $\psi^4 X$ involving a gluon field strength where one fermion bilinear has left-handed chiral projectors and the other has right-handed chiral projectors. 
Operators with $\hc$ have distinct Hermitian conjugates.
The subscripts $p, r, s, t$ are weak-eigenstate indices}
\label{tab:left8_psi4x_LRG}
\end{table}

%

\begin{table}[H]
\begin{center}
\begin{adjustbox}{width=0.93\textwidth,center}
\small
\begin{minipage}[t]{6.0cm}
\renewcommand{\arraystretch}{1.5}
\begin{tabular}[t]{c|c}
\multicolumn{2}{c}{\boldmath$(\bar L L)(\bar R R)F$} \\
\hline
$\O_{\nu eudF}^{(1, LR)}$  &  $(\bar \nu_{Lp} \gamma^\mu e_{Lr}) (\bar d_{Rs} \gamma^\rho u_{Rt}) F_{\mu\rho} + \hc$ \\
$\O_{\nu eudF}^{(2, LR)}$  &  $(\bar \nu_{Lp} \gamma^\mu e_{Lr}) (\bar d_{Rs} \gamma^\rho u_{Rt}) \widetilde F_{\mu\rho} + \hc$ \\ 
$\O_{e^4F}^{(1, LR)}$  &  $(\bar e_{Lp} \gamma^\mu e_{Lr}) (\bar e_{Rs} \gamma^\rho e_{Rt})  F_{\mu\rho}$ \\
$\O_{e^4F}^{(2, LR)}$  &  $(\bar e_{Lp} \gamma^\mu e_{Lr}) (\bar e_{Rs} \gamma^\rho e_{Rt})  \widetilde F_{\mu\rho}$ \\
$\O_{\nu^2e^2F}^{(1, LR)}$  &  $(\bar \nu_{Lp} \gamma^\mu \nu_{Lr}) (\bar e_{Rs} \gamma^\rho e_{Rt})  F_{\mu\rho}$ \\
$\O_{\nu^2e^2F}^{(2, LR)}$  &  $(\bar \nu_{Lp} \gamma^\mu \nu_{Lr}) (\bar e_{Rs} \gamma^\rho e_{Rt})  \widetilde F_{\mu\rho}$ \\
$\O_{\nu^2u^2F}^{(1, LR)}$  &  $(\bar \nu_{Lp} \gamma^\mu \nu_{Lr}) (\bar u_{Rs} \gamma^\rho u_{rt}) F_{\mu\rho}$ \\
$\O_{\nu^2u^2F}^{(2, LR)}$  &  $(\bar \nu_{Lp} \gamma^\mu \nu_{Lr}) (\bar u_{Rs} \gamma^\rho u_{Rt}) \widetilde F_{\mu\rho}$ \\
$\O_{\nu^2d^2F}^{(1, LR)}$  &  $(\bar \nu_{Lp} \gamma^\mu \nu_{Lr}) (\bar d_{Rs} \gamma^\rho d_{Rt}) F_{\mu\rho}$ \\
$\O_{\nu^2d^2F}^{(2, LR)}$  &  $(\bar \nu_{Lp} \gamma^\mu \nu_{Lr}) (\bar d_{Rs} \gamma^\rho d_{Rt}) \widetilde F_{\mu\rho}$ \\ 
$\O_{e^2u^2F}^{(1, LR)}$  &  $(\bar e_{Lp} \gamma^\mu e_{Lr})(\bar u_{Rs} \gamma^\rho u_{Rt}) F_{\mu\rho}$ \\
$\O_{e^2u^2F}^{(2, LR)}$  &  $(\bar e_{Lp} \gamma^\mu e_{Lr})(\bar u_{Rs} \gamma^\rho u_{Rt}) \widetilde F_{\mu\rho}$ \\
$\O_{u^2e^2F}^{(1, LR)}$  &  $(\bar u_{Lp} \gamma^\mu u_{Lr})(\bar e_{Rs} \gamma^\rho e_{Rt}) F_{\mu\rho}$ \\
$\O_{u^2e^2F}^{(2, LR)}$  &  $(\bar u_{Lp} \gamma^\mu u_{Lr})(\bar e_{Rs} \gamma^\rho e_{Rt})  \widetilde F_{\mu\rho}$ \\
$\O_{e^2d^2F}^{(1, LR)}$  &  $(\bar e_{Lp} \gamma^\mu e_{Lr})(\bar d_{Rs} \gamma^\rho d_{Rt}) F_{\mu\rho}$ \\
$\O_{e^2d^2F}^{(2, LR)}$  &  $(\bar e_{Lp} \gamma^\mu e_{Lr})(\bar d_{Rs} \gamma^\rho d_{Rt}) \widetilde F_{\mu\rho}$ \\
$\O_{d^2e^2F}^{(3, LR)}$  &  $(\bar d_{Lp} \gamma^\mu d_{Lr})(\bar e_{Rs} \gamma^\rho e_{Rt})  F_{\mu\rho}$ \\
$\O_{d^2e^2F}^{(4, LR)}$  &  $(\bar d_{Lp} \gamma^\mu d_{Lr})(\bar e_{Rs} \gamma^\rho e_{Rt}) \widetilde F_{\mu\rho}$ 
\end{tabular}
\end{minipage}
\hspace{1cm}
\begin{minipage}[t]{6.9cm}
\renewcommand{\arraystretch}{1.5}
\begin{tabular}[t]{c|c}
\multicolumn{2}{c}{\boldmath$(\bar L L)(\bar R R)F$} \\
\hline
$\O_{u^4F}^{(1, LR)}$  &  $(\bar u_{Lp} \gamma^\mu u_{Lr}) (\bar u_{Rs} \gamma^\rho u_{Rt}) F_{\mu\rho}$ \\
$\O_{u^4F}^{(2, LR)}$  &  $(\bar u_{Lp} \gamma^\mu u_{Lr}) (\bar u_{Rs} \gamma^\rho u_{Rt}) \widetilde F_{\mu\rho}$ \\
$\O_{u^4F}^{(3, LR)}$  &  $(\bar u_{Lp} \gamma^\mu T^A u_{Lr}) (\bar u_{Rs} \gamma^\rho T^A u_{Rt}) F_{\mu\rho}$ \\
$\O_{u^4F}^{(4, LR)}$  &  $(\bar u_{Lp} \gamma^\mu T^A u_{Lr}) (\bar u_{Rs} \gamma^\rho T^A u_{Rt}) \widetilde F_{\mu\rho}$ \\
$\O_{d^4F}^{(1, LR)}$  & $(\bar d_{Lp} \gamma^\mu d_{Lr}) (\bar d_{Rs} \gamma^\rho d_{Rt}) F_{\mu\rho}$ \\
$\O_{d^4F}^{(2, LR)}$  & $(\bar d_{Lp} \gamma^\mu d_{Lr}) (\bar d_{Rs} \gamma^\rho d_{Rt}) \widetilde F_{\mu\rho}$ \\
$\O_{d^4F}^{(3, LR)}$  & $(\bar d_{Lp} \gamma^\mu T^A d_{Lr}) (\bar d_{Rs} \gamma^\rho T^A d_{Rt}) F_{\mu\rho}$ \\
$\O_{d^4F}^{(4, LR)}$  & $(\bar d_{Lp} \gamma^\mu T^A d_{Lr}) (\bar d_{Rs} \gamma^\rho T^A d_{Rt}) \widetilde F_{\mu\rho}$ \\
$\O_{u^2d^2F}^{(1, LR)}$  &  $(\bar u_{Lp} \gamma^\mu u_{Lr})(\bar d_{Rs} \gamma^\rho d_{Rt}) F_{\mu\rho}$ \\
$\O_{u^2d^2F}^{(2, LR)}$  &  $(\bar u_{Lp} \gamma^\mu u_{Lr})(\bar d_{Rs} \gamma^\rho d_{Rt}) \widetilde F_{\mu\rho}$ \\
$\O_{u^2d^2F}^{(3, LR)}$  &  $(\bar u_{Lp} \gamma^\mu T^A u_{Lr})(\bar d_{Rs} \gamma^\rho T^A d_{Rt}) F_{\mu\rho}$ \\
$\O_{u^2d^2F}^{(4, LR)}$  &  $(\bar u_{Lp} \gamma^\mu T^A u_{Lr})(\bar d_{Rs} \gamma^\rho T^A d_{Rt}) \widetilde F_{\mu\rho}$ \\
$\O_{d^2u^2F}^{(1, LR)}$  &  $(\bar d_{Lp} \gamma^\mu d_{Lr})(\bar u_{Rs} \gamma^\rho u_{Rt}) F_{\mu\rho}$ \\
$\O_{d^2u^2F}^{(2, LR)}$  &  $(\bar d_{Lp} \gamma^\mu d_{Lr})(\bar u_{Rs} \gamma^\rho u_{Rt})\widetilde F_{\mu\rho}$ \\
$\O_{d^2u^2F}^{(3, LR)}$  &  $(\bar d_{Lp} \gamma^\mu T^A d_{Lr})(\bar u_{Rs} \gamma^\rho T^A u_{Rt}) F_{\mu\rho}$ \\
$\O_{d^2u^2F}^{(4, LR)}$  &  $(\bar d_{Lp} \gamma^\mu T^A d_{Lr})(\bar u_{Rs} \gamma^\rho T^A u_{Rt}) \widetilde F_{\mu\rho}$ \\
$\O_{(ud)^2F}^{(1, LR)}$  &  $(\bar d_{Lp} \gamma^\mu u_{Lr})(\bar u_{Rs} \gamma^\rho d_{Rt}) F_{\mu\rho} + \hc$ \\
$\O_{(ud)^2F}^{(2, LR)}$  &  $(\bar d_{Lp} \gamma^\mu u_{Lr})(\bar u_{Rs} \gamma^\rho d_{Rt}) \widetilde F_{\mu\rho} + \hc$ \\
$\O_{(ud)^2F}^{(3, LR)}$  &  $(\bar d_{Lp} \gamma^\mu T^A u_{Lr})(\bar u_{Rs} \gamma^\rho T^A d_{Rt}) F_{\mu\rho} + \hc$ \\
$\O_{(ud)^2F}^{(4, LR)}$  &  $(\bar d_{Lp} \gamma^\mu T^A u_{Lr})(\bar u_{Rs} \gamma^\rho T^A d_{Rt}) \widetilde F_{\mu\rho} + \hc$ 
\end{tabular}
\end{minipage}
\end{adjustbox}
\end{center}
\caption{The baryon and lepton number conserving operators of class $\psi^4 X$ involving a photon field strength where one fermion bilinear has left-handed chiral projectors and the other has right-handed chiral projectors. 
Operators with $\hc$ have distinct Hermitian conjugates.
The subscripts $p, r, s, t$ are weak-eigenstate indices.}
\label{tab:left8_psi4x_LRF}
\end{table}

%% file: sections/basis_psi4x_llll_v3.tex

\begin{table}[H]
\begin{center}
\begin{adjustbox}{width=0.94\textwidth,center}
\small
\begin{minipage}[t]{6.0cm}
\renewcommand{\arraystretch}{1.5}
\begin{tabular}[t]{c|c}
\multicolumn{2}{c}{\boldmath$(\bar L R)(\bar L R)X + \hc$} \\
\hline
$\O_{\nu eudG}^{(2, ST)}$  &  $(\bar \nu_{Lp} e_{Rr}) (\bar d_{Ls} \sigma^{\mu\rho} T^A u_{Rt}) G_{\mu\rho}^A $ \\
$\O_{\nu eudG}^{(2, TS)}$  &  $(\bar \nu_{Lp} \sigma^{\mu\rho} e_{Rr}) (\bar d_{Ls} T^A u_{Rt}) G_{\mu\rho}^A $ \\
$\O_{\nu eudG}^{(1, TT)}$  &  $(\bar \nu_{Lp} \sigma^{\mu\lambda} e_{Rr}) (\bar d_{Ls} \sigma_{\lambda\rho} T^A u_{Rt}) G_\mu^{A \rho} $ \\
$\O_{\nu eudF}^{(2, ST)}$  &  $(\bar \nu_{Lp} e_{Rr}) (\bar d_{Ls} \sigma^{\mu\rho} u_{Rt}) F_{\mu\rho} $ \\
$\O_{\nu eudF}^{(2, TS)}$  &  $(\bar \nu_{Lp} \sigma^{\mu\rho} e_{Rr}) (\bar d_{Ls} u_{Rt}) F_{\mu\rho} $ \\
$\O_{\nu eudF}^{(1, TT)}$  &  $(\bar \nu_{Lp} \sigma^{\mu\lambda} e_{Rr}) (\bar d_{Ls} \sigma_{\lambda\rho} u_{Rt}) F_\mu^{\,\, \rho} $ \\
$\O_{e^2u^2G}^{(2, ST)}$  &  $(\bar e_{Lp} e_{Rr}) (\bar u_{Ls} \sigma^{\mu\rho} T^A u_{Rt}) G_{\mu\rho}^A $ \\
$\O_{e^2u^2G}^{(2, TS)}$  &  $(\bar e_{Lp} \sigma^{\mu\rho}  e_{Rr}) (\bar u_{Ls} T^A u_{Rt}) G_{\mu\rho}^A $ \\
$\O_{e^2u^2G}^{(1, TT)}$  &  $(\bar e_{Lp} \sigma^{\mu\lambda} e_{Rr}) (\bar u_{Ls} \sigma_{\lambda\rho} T^A u_{Rt}) G_\mu^{A \rho} $ \\
$\O_{e^2u^2F}^{(2, ST)}$  &  $(\bar e_{Lp} e_{Rr}) (\bar u_{Ls} \sigma^{\mu\rho} u_{Rt}) F_{\mu\rho} $ \\
$\O_{e^2u^2F}^{(2, TS)}$  &  $(\bar e_{Lp} \sigma^{\mu\rho} e_{Rr}) (\bar u_{Ls} u_{Rt}) F_{\mu\rho} $ \\
$\O_{e^2u^2F}^{(1, TT)}$  &  $(\bar e_{Lp} \sigma^{\mu\lambda} e_{Rr}) (\bar u_{Ls} \sigma_{\lambda\rho} u_{Rt}) F_\mu^{\,\, \rho}$ \\
$\O_{e^2d^2G}^{(2, ST)}$  &  $(\bar e_{Lp} e_{Rr}) (\bar d_{Ls} \sigma^{\mu\rho} T^A d_{Rt}) G_{\mu\rho}^A $ \\
$\O_{e^2d^2G}^{(2, TS)}$  &  $(\bar e_{Lp} \sigma^{\mu\rho}  e_{Rr}) (\bar d_{Ls} T^A d_{Rt}) G_{\mu\rho}^A $ \\
$\O_{e^2d^2G}^{(1, TT)}$  &  $(\bar e_{Lp} \sigma^{\mu\lambda} e_{Rr}) (\bar d_{Ls} \sigma_{\lambda\rho} T^A d_{Rt}) G_\mu^{A \rho} $ \\
$\O_{e^2d^2F}^{(2, ST)}$  &  $(\bar e_{Lp} e_{Rr}) (\bar d_{Ls} \sigma^{\mu\rho} d_{Rt}) F_{\mu\rho} $ \\
$\O_{e^2d^2F}^{(2, TS)}$  &  $(\bar e_{Lp} \sigma^{\mu\rho} e_{Rr}) (\bar d_{Ls} d_{Rt}) F_{\mu\rho} $ \\
$\O_{e^2d^2F}^{(1, TT)}$  &  $(\bar e_{Lp} \sigma^{\mu\lambda} e_{Rr}) (\bar d_{Ls} \sigma_{\lambda\rho} d_{Rt}) F_\mu^{\,\, \rho}$ \\ \hdashline
$\O_{e^4F}^{(1, ST)}$  &  $(\bar e_{Lp} e_{Rr}) (\bar e_{Ls} \sigma^{\mu\rho} e_{Rt}) F_{\mu\rho} $
\end{tabular}
\end{minipage}
\hspace{1cm}
\begin{minipage}[t]{7.1cm}
\renewcommand{\arraystretch}{1.5}
\begin{tabular}[t]{c|c}
\multicolumn{2}{c}{\boldmath$(\bar L R)(\bar L R)X + \hc$} \\
\hline
$\O_{u^4G}^{(1, ST)}$  &  $(\bar u_{Lp} u_{Rr}) (\bar u_{Ls} \sigma^{\mu\rho} T^A u_{Rt}) G_{\mu\rho}^A $ \\
$\O_{u^4G}^{(1, TS)}$  &  $(\bar u_{Lp} \sigma^{\mu\rho} u_{Rr}) (\bar u_{Ls} T^A u_{Rt}) G_{\mu\rho}^A $ \\
$\O_{u^4G}^{(1, TT)}$  &  $(\bar u_{Lp} \sigma^{\mu\lambda} u_{Rr}) (\bar u_{Ls} \sigma_{\lambda\rho}T^A u_{Rt}) G_{\mu}^{A \rho} $ \\
$\O_{u^4F}^{(1, ST)}$  &  $(\bar u_{Lp} T^A u_{Rr}) (\bar u_{Ls} \sigma^{\mu\rho} T^A u_{Rt}) F_{\mu\rho} $ \\
$\O_{d^4G}^{(1, ST)}$  &  $(\bar d_{Lp} d_{Rr}) (\bar d_{Ls} \sigma^{\mu\rho} T^A d_{Rt}) G_{\mu\rho}^A $ \\
$\O_{d^4G}^{(1, TS)}$  &  $(\bar d_{Lp} \sigma^{\mu\rho} d_{Rr}) (\bar d_{Ls} T^A d_{Rt}) G_{\mu\rho}^A $ \\
$\O_{d^4G}^{(1, TT)}$  &  $(\bar d_{Lp} \sigma^{\mu\lambda} d_{Rr}) (\bar d_{Ls} \sigma_{\lambda\rho}T^A d_{Rt}) G_{\mu}^{A \rho} $ \\
$\O_{d^4F}^{(1, ST)}$  &  $(\bar d_{Lp} T^A d_{Rr}) (\bar d_{Ls} \sigma^{\mu\rho} T^A d_{Rt}) F_{\mu\rho} $ \\
$\O_{u^2d^2G}^{(1, ST)}$  &  $(\bar u_{Lp} d_{Rr}) (\bar d_{Ls} \sigma_{\mu\rho} T^A u_{Rt}) G_{\mu\rho}^A$ \\ 
$\O_{u^2d^2G}^{(2, ST)}$  &  $(\bar u_{Lp} T^A d_{Rr}) (\bar d_{Ls} \sigma_{\mu\rho} u_{Rt}) G_{\mu\rho}^A$ \\ 
$\O_{u^2d^2G}^{(3, ST)}$  &  $f^{ABC} (\bar u_{Lp} T^A d_{Rr}) (\bar d_{Ls} \sigma_{\mu\rho} T^B u_{Rt}) G_{\mu\rho}^C$ \\ 
$\O_{u^2d^2G}^{(4, ST)}$  &  $d^{ABC} (\bar u_{Lp} T^A d_{Rr}) (\bar d_{Ls} \sigma_{\mu\rho} T^B u_{Rt}) G_{\mu\rho}^C$ \\ 
$\O_{u^2d^2G}^{(1, TS)}$  &  $(\bar u_{Lp} \sigma^{\mu\rho} d_{Rr}) (\bar d_{Ls} T^A u_{Rt}) G_{\mu\rho}^A$ \\ 
$\O_{u^2d^2G}^{(2, TS)}$  &  $(\bar u_{Lp} \sigma^{\mu\rho} T^A d_{Rr}) (\bar d_{Ls} u_{Rt}) G_{\mu\rho}^A$ \\ 
$\O_{u^2d^2G}^{(3, TS)}$  &  $f^{ABC} (\bar u_{Lp} \sigma^{\mu\rho} T^A d_{Rr}) (\bar d_{Ls} T^B u_{Rt}) G_{\mu\rho}^C$ \\ 
$\O_{u^2d^2G}^{(4, TS)}$  &  $d^{ABC} (\bar u_{Lp} \sigma^{\mu\rho} T^A d_{Rr}) (\bar d_{Ls} T^B u_{Rt}) G_{\mu\rho}^C$ \\ 
$\O_{u^2d^2G}^{(1, TT)}$  &  $(\bar u_{Lp} \sigma^{\mu\lambda} d_{Rr}) (\bar d_{Ls} \sigma_{\lambda\rho} T^A u_{Rt}) G_\mu^{A \rho}$ \\
$\O_{u^2d^2G}^{(2, TT)}$  &  $(\bar u_{Lp} \sigma^{\mu\lambda} T^A d_{Rr}) (\bar d_{Ls} \sigma_{\lambda\rho} u_{Rt}) G_\mu^{A \rho}$ \\
$\O_{u^2d^2G}^{(3, TT)}$  &  $f^{ABC} (\bar u_{Lp} \sigma^{\mu\lambda} T^A d_{Rr}) (\bar d_{Ls} \sigma_{\lambda\rho} T^B u_{Rt}) G_\mu^{C \rho}$ \\ 
$\O_{u^2d^2G}^{(4, TT)}$  &  $d^{ABC} (\bar u_{Lp} \sigma^{\mu\lambda} T^A d_{Rr}) (\bar d_{Ls} \sigma_{\lambda\rho} T^B u_{Rt}) G_\mu^{C \rho}$ \\ 
$\O_{u^2d^2F}^{(1, ST)}$  &  $(\bar u_{Lp} d_{Rr}) (\bar d_{Ls} \sigma_{\mu\rho} u_{Rt}) F_{\mu\rho}$ \\ 
$\O_{u^2d^2F}^{(2, ST)}$  &  $(\bar u_{Lp} T^A d_{Rr}) (\bar d_{Ls} \sigma_{\mu\rho} T^A u_{Rt}) F_{\mu\rho}$ \\ 
$\O_{u^2d^2F}^{(1, TS)}$  &  $(\bar u_{Lp} \sigma^{\mu\rho} d_{Rr}) (\bar d_{Ls}  u_{Rt}) F_{\mu\rho}$ \\ 
$\O_{u^2d^2F}^{(2, TS)}$  &  $(\bar u_{Lp} \sigma^{\mu\rho} T^A d_{Rr}) (\bar d_{Ls}  T^A u_{Rt}) F_{\mu\rho}$ \\ 
$\O_{u^2d^2F}^{(1, TT)}$  &  $(\bar u_{Lp} \sigma^{\mu\lambda} d_{Rr}) (\bar d_{Ls} \sigma_{\lambda\rho} u_{Rt}) F_\mu^{\,\, \rho}$ \\
$\O_{u^2d^2F}^{(2, TT)}$  &  $(\bar u_{Lp} \sigma^{\mu\lambda} T^A d_{Rr}) (\bar d_{Ls} \sigma_{\lambda\rho} T^A u_{Rt}) F_\mu^{\,\, \rho}$ \\ \hdashline
$\O_{u^4F}^{(2, ST)}$  &  $(\bar u_{Lp} u_{Rr}) (\bar u_{Ls} \sigma^{\mu\rho} u_{Rt}) F_{\mu\rho} $ \\
$\O_{d^4F}^{(2, ST)}$  &  $(\bar d_{Lp} d_{Rr}) (\bar d_{Ls} \sigma^{\mu\rho} d_{Rt}) F_{\mu\rho} $
\end{tabular}
\end{minipage}
\end{adjustbox}
\end{center}
\caption{The baryon and lepton number conserving operators of class $\psi^4 X$ and subclass $\psi_R^4 X_R$. 
All of the operators in this table have distinct Hermitian conjugates.
The subscripts $p, r, s, t$ are weak-eigenstate indices.
Operators below the dashed lines vanish when there is only one generation of fermions.}
\label{tab:left8_psi4x_LRLR}
\end{table}

%% file: sections/basis_psi4d2_llrr_v2.tex

\begin{table}[H]
\begin{center}
\begin{adjustbox}{width=0.93\textwidth,center}
\small
\begin{minipage}[t]{6.5cm}
\renewcommand{\arraystretch}{1.5}
\begin{tabular}[t]{c|c}
\multicolumn{2}{c}{\boldmath$(\bar L L)(\bar L L)D^2$} \\
\hline
$\O_{\nu^4D^2}^{(1, LL)}$  &  $D^\rho (\bar \nu_{Lp} \gamma^\mu \nu_{Lr}) D_\rho (\bar \nu_{Ls} \gamma_\mu \nu_{Lt}) $ \\
$\O_{e^4D^2}^{(1, LL)}$  &  $D^\rho (\bar e_{Lp} \gamma^\mu e_{Lr}) D_\rho (\bar e_{Ls} \gamma_\mu e_{Lt}) $ \\
$\O_{u^4D^2}^{(1, LL)}$  &  $D^\rho (\bar u_{Lp} \gamma^\mu u_{Lr}) D_\rho (\bar u_{Ls} \gamma_\mu u_{Lt}) $ \\
$\O_{u^4D^2}^{(2, LL)}$  &  $(\bar u_{Lp} \gamma^\mu \overleftrightarrow{D}^\rho u_{Lr}) (\bar u_{Ls} \gamma_\mu \overleftrightarrow{D}_\rho u_{Lt}) $ \\
$\O_{d^4D^2}^{(1, LL)}$  &  $D^\rho (\bar d_{Lp} \gamma^\mu d_{Lr}) D_\rho (\bar d_{Ls} \gamma_\mu d_{Lt}) $ \\
$\O_{d^4D^2}^{(2, LL)}$  &  $(\bar d_{Lp} \gamma^\mu \overleftrightarrow{D}^\rho d_{Lr}) (\bar d_{Ls} \gamma_\mu \overleftrightarrow{D}_\rho d_{Lt}) $ \\
$\O_{\nu^2e^2D^2}^{(1, LL)}$  &  $D^\rho (\bar \nu_{Lp} \gamma^\mu \nu_{Lr}) D_\rho (\bar e_{Ls} \gamma_\mu e_{Lt}) $ \\
$\O_{\nu^2e^2D^2}^{(2, LL)}$  &  $(\bar \nu_{Lp} \gamma^\mu \overleftrightarrow{D}^\rho \nu_{Lr}) (\bar e_{Ls} \gamma_\mu \overleftrightarrow{D}_\rho e_{Lt}) $ \\
$\O_{\nu^2u^2D^2}^{(1, LL)}$  &  $D^\rho (\bar \nu_{Lp} \gamma^\mu \nu_{Lr}) D_\rho (\bar u_{Ls} \gamma_\mu u_{Lt}) $ \\
$\O_{\nu^2u^2D^2}^{(2, LL)}$  &  $(\bar \nu_{Lp} \gamma^\mu \overleftrightarrow{D}^\rho \nu_{Lr}) (\bar u_{Ls} \gamma_\mu \overleftrightarrow{D}_\rho u_{Lt}) $ \\
$\O_{\nu^2d^2D^2}^{(1, LL)}$  &  $D^\rho (\bar \nu_{Lp} \gamma^\mu \nu_{Lr}) D_\rho (\bar d_{Ls} \gamma_\mu d_{Lt}) $ \\
$\O_{\nu^2d^2D^2}^{(2, LL)}$  &  $(\bar \nu_{Lp} \gamma^\mu \overleftrightarrow{D}^\rho \nu_{Lr}) (\bar d_{Ls} \gamma_\mu \overleftrightarrow{D}_\rho d_{Lt}) $ \\
$\O_{e^2u^2D^2}^{(1, LL)}$  &  $D^\rho (\bar e_{Lp} \gamma^\mu e_{Lr}) D_\rho (\bar u_{Ls} \gamma_\mu u_{Lt}) $ \\
$\O_{e^2u^2D^2}^{(2, LL)}$  &  $(\bar e_{Lp} \gamma^\mu \overleftrightarrow{D}^\rho e_{Lr}) (\bar u_{Ls} \gamma_\mu \overleftrightarrow{D}_\rho u_{Lt}) $ \\
$\O_{e^2d^2D^2}^{(1, LL)}$  &  $D^\rho (\bar e_{Lp} \gamma^\mu e_{Lr}) D_\rho (\bar d_{Ls} \gamma_\mu d_{Lt}) $ \\
$\O_{e^2d^2D^2}^{(2, LL)}$  &  $(\bar e_{Lp} \gamma^\mu \overleftrightarrow{D}^\rho e_{Lr}) (\bar d_{Ls} \gamma_\mu \overleftrightarrow{D}_\rho d_{Lt}) $ \\ 
$\O_{u^2d^2D^2}^{(1, LL)}$  &  $D^\rho (\bar u_{Lp} \gamma^\mu u_{Lr}) D_\rho (\bar d_{Ls} \gamma_\mu d_{Lt}) $ \\
$\O_{u^2d^2D^2}^{(2, LL)}$  &  $(\bar u_{Lp} \gamma^\mu \overleftrightarrow{D}^\rho u_{Lr}) (\bar d_{Ls} \gamma_\mu \overleftrightarrow{D}_\rho d_{Lt}) $ \\
$\O_{u^2d^2D^2}^{(3, LL)}$  &  $D^\rho (\bar u_{Lp} \gamma^\mu T^A u_{Lr}) D_\rho (\bar d_{Ls} \gamma_\mu T^A d_{Lt}) $ \\
$\O_{u^2d^2D^2}^{(4, LL)}$  &  $(\bar u_{Lp} \gamma^\mu T^A \overleftrightarrow{D}^\rho u_{Lr}) (\bar d_{Ls} \gamma_\mu T^A \overleftrightarrow{D}_\rho d_{Lt}) $ \\ 
$\O_{\nu eudD^2}^{(1, LL)}$  &  $D^\rho (\bar \nu_{Lp} \gamma^\mu e_{Lr}) D_\rho (\bar d_{Ls} \gamma^\rho u_{Lt})  + \hc$ \\
$\O_{\nu eudD^2}^{(2, LL)}$  &  $(\bar \nu_{Lp} \gamma^\mu \overleftrightarrow{D}^\rho e_{Lr}) (\bar d_{Ls} \gamma_\mu \overleftrightarrow{D}_\rho u_{Lt})  + \hc$ 
\end{tabular}
\end{minipage}
\hspace{1cm}
\begin{minipage}[t]{6.5cm}
\renewcommand{\arraystretch}{1.5}
\begin{tabular}[t]{c|c}
\multicolumn{2}{c}{\boldmath$(\bar R R)(\bar R R)D^2$} \\
\hline
$\O_{e^4D^2}^{(1, RR)}$  &  $D^\rho (\bar e_{Rp} \gamma^\mu e_{Rr}) D_\rho (\bar e_{Rs} \gamma_\mu e_{Rt}) $ \\
$\O_{u^4D^2}^{(1, RR)}$  &  $D^\rho (\bar u_{Rp} \gamma^\mu u_{Rr}) D_\rho (\bar u_{Rs} \gamma_\mu u_{Rt}) $ \\
$\O_{u^4D^2}^{(2, RR)}$  &  $(\bar u_{Rp} \gamma^\mu \overleftrightarrow{D}^\rho u_{Rr}) (\bar u_{Rs} \gamma_\mu \overleftrightarrow{D}_\rho u_{Rt}) $ \\
$\O_{d^4D^2}^{(1, RR)}$  &  $D^\rho (\bar d_{Rp} \gamma^\mu d_{Rr}) D_\rho (\bar d_{Rs} \gamma_\mu d_{Rt}) $ \\
$\O_{d^4D^2}^{(2, RR)}$  &  $(\bar d_{Rp} \gamma^\mu \overleftrightarrow{D}^\rho d_{Rr}) (\bar d_{Rs} \gamma_\mu \overleftrightarrow{D}_\rho d_{Rt}) $ \\
$\O_{e^2u^2D^2}^{(1, RR)}$  &  $D^\rho (\bar e_{Rp} \gamma^\mu e_{Rr}) D_\rho (\bar u_{Rs} \gamma_\mu u_{Rt}) $ \\
$\O_{e^2u^2D^2}^{(2, RR)}$  &  $(\bar e_{Rp} \gamma^\mu \overleftrightarrow{D}^\rho e_{Rr}) (\bar u_{Rs} \gamma_\mu \overleftrightarrow{D}_\rho u_{Rt}) $ \\
$\O_{e^2d^2D^2}^{(1, RR)}$  &  $D^\rho (\bar e_{Rp} \gamma^\mu e_{Rr}) D_\rho (\bar d_{Rs} \gamma_\mu d_{Rt}) $ \\
$\O_{e^2d^2D^2}^{(2, RR)}$  &  $(\bar e_{Rp} \gamma^\mu \overleftrightarrow{D}^\rho e_{Rr}) (\bar d_{Rs} \gamma_\mu \overleftrightarrow{D}_\rho d_{Rt}) $ \\ 
$\O_{u^2d^2D^2}^{(1, RR)}$  &  $D^\rho (\bar u_{Rp} \gamma^\mu u_{Rr}) D_\rho (\bar d_{Rs} \gamma_\mu d_{Rt}) $ \\
$\O_{u^2d^2D^2}^{(2, RR)}$  &  $(\bar u_{Rp} \gamma^\mu \overleftrightarrow{D}^\rho u_{Rr}) (\bar d_{Rs} \gamma_\mu \overleftrightarrow{D}_\rho d_{Rt}) $ \\
$\O_{u^2d^2D^2}^{(3, RR)}$  &  $D^\rho (\bar u_{Rp} \gamma^\mu T^A u_{Rr}) D_\rho (\bar d_{Rs} \gamma_\mu T^A d_{Rt}) $ \\
$\O_{u^2d^2D^2}^{(4, RR)}$  &  $(\bar u_{Rp} \gamma^\mu T^A \overleftrightarrow{D}^\rho u_{Rr}) (\bar d_{Rs} \gamma_\mu T^A \overleftrightarrow{D}_\rho d_{Rt}) $ \\ 
\end{tabular}
\end{minipage}
\end{adjustbox}
\end{center}
\caption{The baryon and lepton number conserving operators of class $\psi^4 D^2$ where both fermion bilinears involve either left-handed or right-handed chiral projectors. 
Operators with $\hc$ have distinct Hermitian conjugates.
The subscripts $p, r, s, t$ are weak-eigenstate indices.}
\label{tab:left8_psi4d2_LL_RR}
\end{table}

%

\begin{table}[H]
\begin{center}
\begin{adjustbox}{width=1.0\textwidth,center}
\small
\begin{minipage}[t]{6.5cm}
\renewcommand{\arraystretch}{1.5}
\begin{tabular}[t]{c|c}
\multicolumn{2}{c}{\boldmath$(\bar L L)(\bar R R)D^2$} \\
\hline
$\O_{e^4D^2}^{(1, LR)}$  &  $D^\rho (\bar e_{Lp} \gamma^\mu e_{Lr}) D_\rho (\bar e_{Rs} \gamma_\mu e_{Rt}) $ \\
$\O_{e^4D^2}^{(2, LR)}$  &  $(\bar e_{Lp} \gamma^\mu \overleftrightarrow{D}^\rho e_{Lr}) (\bar e_{Rs} \gamma_\mu \overleftrightarrow{D}_\rho e_{Rt}) $ \\
$\O_{\nu^2e^2D^2}^{(1, LR)}$  &  $D^\rho (\bar \nu_{Lp} \gamma^\mu \nu_{Lr}) D_\rho (\bar e_{Rs} \gamma_\mu e_{Rt}) $ \\
$\O_{\nu^2e^2D^2}^{(2, LR)}$  &  $(\bar \nu_{Lp} \gamma^\mu \overleftrightarrow{D}^\rho \nu_{Lr}) (\bar e_{Rs} \gamma_\mu \overleftrightarrow{D}_\rho e_{Rt}) $ \\
$\O_{\nu^2u^2D^2}^{(1, LR)}$  &  $D^\rho (\bar \nu_{Lp} \gamma^\mu \nu_{Lr}) D_\rho (\bar u_{Rs} \gamma_\mu u_{Rt}) $ \\
$\O_{\nu^2u^2D^2}^{(2, LR)}$  &  $(\bar \nu_{Lp} \gamma^\mu \overleftrightarrow{D}^\rho \nu_{Lr}) (\bar u_{Rs} \gamma_\mu \overleftrightarrow{D}_\rho u_{Rt}) $ \\
$\O_{\nu^2d^2D^2}^{(1, LR)}$  &  $D^\rho (\bar \nu_{Lp} \gamma^\mu \nu_{Lr}) D_\rho (\bar d_{Rs} \gamma_\mu d_{Rt}) $ \\
$\O_{\nu^2d^2D^2}^{(2, LR)}$  &  $(\bar \nu_{Lp} \gamma^\mu \overleftrightarrow{D}^\rho \nu_{Lr}) (\bar d_{Rs} \gamma_\mu \overleftrightarrow{D}_\rho d_{Rt}) $ \\
$\O_{e^2u^2D^2}^{(1, LR)}$  &  $D^\rho (\bar e_{Lp} \gamma^\mu e_{Lr}) D_\rho (\bar u_{Rs} \gamma_\mu u_{Rt}) $ \\
$\O_{e^2u^2D^2}^{(2, LR)}$  &  $(\bar e_{Lp} \gamma^\mu \overleftrightarrow{D}^\rho e_{Lr}) (\bar u_{Rs} \gamma_\mu \overleftrightarrow{D}_\rho u_{Rt}) $ \\
$\O_{u^2e^2D^2}^{(1, LR)}$  &  $D^\rho (\bar u_{Lp} \gamma^\mu u_{Lr}) D_\rho (\bar e_{Rs} \gamma_\mu e_{Rt}) $ \\
$\O_{u^2e^2D^2}^{(2, LR)}$  &  $(\bar u_{Lp} \gamma^\mu \overleftrightarrow{D}^\rho u_{Lr}) (\bar e_{Rs} \gamma_\mu \overleftrightarrow{D}_\rho e_{Rt}) $ \\
$\O_{e^2d^2D^2}^{(1, LR)}$  &  $D^\rho (\bar e_{Lp} \gamma^\mu e_{Lr}) D_\rho (\bar d_{Rs} \gamma_\mu d_{Rt}) $ \\
$\O_{e^2d^2D^2}^{(2, LR)}$  &  $(\bar e_{Lp} \gamma^\mu \overleftrightarrow{D}^\rho e_{Lr}) (\bar d_{Rs} \gamma_\mu \overleftrightarrow{D}_\rho d_{Rt}) $ \\
$\O_{d^2e^2D^2}^{(1, LR)}$  &  $D^\rho (\bar d_{Lp} \gamma^\mu d_{Lr}) D_\rho (\bar e_{Rs} \gamma_\mu e_{Rt}) $ \\
$\O_{d^2e^2D^2}^{(2, LR)}$  &  $(\bar d_{Lp} \gamma^\mu \overleftrightarrow{D}^\rho d_{Lr}) (\bar e_{Rs} \gamma_\mu \overleftrightarrow{D}_\rho e_{Rt}) $ \\
$\O_{\nu eudD^2}^{(1, LR)}$  &  $D^\rho (\bar \nu_{Lp} \gamma^\mu e_{Lr}) D_\rho (\bar d_{Rs} \gamma^\rho u_{Rt})  + \hc$ \\
$\O_{\nu eudD^2}^{(2, LR)}$  &  $(\bar \nu_{Lp} \gamma^\mu \overleftrightarrow{D}^\rho e_{Lr}) (\bar d_{Rs} \gamma_\mu \overleftrightarrow{D}_\rho u_{Rt})  + \hc$ \\ 
\end{tabular}
\end{minipage}
\hspace{1cm}
\begin{minipage}[t]{7.5cm}
\renewcommand{\arraystretch}{1.5}
\begin{tabular}[t]{c|c}
\multicolumn{2}{c}{\boldmath$(\bar L L)(\bar R R)D^2$} \\
\hline
$\O_{u^4D^2}^{(1, LR)}$  &  $D^\rho (\bar u_{Lp} \gamma^\mu u_{Lr}) D_\rho (\bar u_{Rs} \gamma_\mu u_{Rt}) $ \\
$\O_{u^4D^2}^{(2, LR)}$  &  $(\bar u_{Lp} \gamma^\mu \overleftrightarrow{D}^\rho u_{Lr}) (\bar u_{Rs} \gamma_\mu \overleftrightarrow{D}_\rho u_{Rt}) $ \\
$\O_{u^4D^2}^{(3, LR)}$  &  $D^\rho (\bar u_{Lp} \gamma^\mu T^A u_{Lr}) D_\rho (\bar u_{Rs} \gamma_\mu T^A u_{Rt}) $ \\
$\O_{u^4D^2}^{(4, LR)}$  &  $(\bar u_{Lp} \gamma^\mu T^A \overleftrightarrow{D}^\rho u_{Lr}) (\bar u_{Rs} \gamma_\mu T^A \overleftrightarrow{D}_\rho u_{Rt}) $ \\ 
$\O_{d^4D^2}^{(1, LR)}$  &  $D^\rho (\bar d_{Lp} \gamma^\mu d_{Lr}) D_\rho (\bar d_{Rs} \gamma_\mu d_{Rt}) $ \\
$\O_{d^4D^2}^{(2, LR)}$  &  $(\bar d_{Lp} \gamma^\mu \overleftrightarrow{D}^\rho d_{Lr}) (\bar d_{Rs} \gamma_\mu \overleftrightarrow{D}_\rho d_{Rt}) $ \\
$\O_{d^4D^2}^{(3, LR)}$  &  $D^\rho (\bar d_{Lp} \gamma^\mu T^A d_{Lr}) D_\rho (\bar d_{Rs} \gamma_\mu T^A d_{Rt}) $ \\
$\O_{d^4D^2}^{(4, LR)}$  &  $(\bar d_{Lp} \gamma^\mu T^A \overleftrightarrow{D}^\rho d_{Lr}) (\bar d_{Rs} \gamma_\mu T^A \overleftrightarrow{D}_\rho d_{Rt}) $ \\ 
$\O_{u^2d^2D^2}^{(1, LR)}$  &  $D^\rho (\bar u_{Lp} \gamma^\mu u_{Lr}) D_\rho (\bar d_{Rs} \gamma_\mu d_{Rt}) $ \\
$\O_{u^2d^2D^2}^{(2, LR)}$  &  $(\bar u_{Lp} \gamma^\mu \overleftrightarrow{D}^\rho u_{Lr}) (\bar d_{Rs} \gamma_\mu \overleftrightarrow{D}_\rho d_{Rt}) $ \\
$\O_{u^2d^2D^2}^{(3, LR)}$  &  $D^\rho (\bar u_{Lp} \gamma^\mu T^A u_{Lr}) D_\rho (\bar d_{Rs} \gamma_\mu T^A d_{Rt}) $ \\
$\O_{u^2d^2D^2}^{(4, LR)}$  &  $(\bar u_{Lp} \gamma^\mu T^A \overleftrightarrow{D}^\rho u_{Lr}) (\bar d_{Rs} \gamma_\mu T^A \overleftrightarrow{D}_\rho d_{Rt}) $ \\ 
$\O_{d^2u^2D^2}^{(1, LR)}$  &  $D^\rho (\bar d_{Lp} \gamma^\mu d_{Lr}) D_\rho (\bar u_{Rs} \gamma_\mu u_{Rt}) $ \\
$\O_{d^2u^2D^2}^{(2, LR)}$  &  $(\bar d_{Lp} \gamma^\mu \overleftrightarrow{D}^\rho d_{Lr}) (\bar u_{Rs} \gamma_\mu \overleftrightarrow{D}_\rho u_{Rt}) $ \\
$\O_{d^2u^2D^2}^{(3, LR)}$  &  $D^\rho (\bar d_{Lp} \gamma^\mu T^A d_{Lr}) D_\rho (\bar u_{Rs} \gamma_\mu T^A u_{Rt}) $ \\
$\O_{d^2u^2D^2}^{(4, LR)}$  &  $(\bar d_{Lp} \gamma^\mu T^A \overleftrightarrow{D}^\rho d_{Lr}) (\bar u_{Rs} \gamma_\mu T^A \overleftrightarrow{D}_\rho u_{Rt}) $ \\ 
$\O_{(ud)^2D^2}^{(1, LR)}$  &  $D^\rho (\bar d_{Lp} \gamma^\mu u_{Lr}) D_\rho (\bar u_{Rs} \gamma_\mu d_{Rt}) + \hc$ \\
$\O_{(ud)^2D^2}^{(2, LR)}$  &  $(\bar d_{Lp} \gamma^\mu \overleftrightarrow{D}^\rho u_{Lr}) (\bar u_{Rs} \gamma_\mu \overleftrightarrow{D}_\rho d_{Rt}) + \hc$ \\
$\O_{(ud)^2D^2}^{(3, LR)}$  &  $D^\rho (\bar d_{Lp} \gamma^\mu T^A u_{Lr}) D_\rho (\bar u_{Rs} \gamma_\mu T^A d_{Rt}) + \hc$ \\
$\O_{(ud)^2D^2}^{(4, LR)}$  &  $(\bar d_{Lp} \gamma^\mu T^A \overleftrightarrow{D}^\rho u_{Lr}) (\bar u_{Rs} \gamma_\mu T^A \overleftrightarrow{D}_\rho d_{Rt}) + \hc$ 
\end{tabular}
\end{minipage}
\end{adjustbox}
\begin{adjustbox}{width=0.34\textwidth,center}
\small
\begin{minipage}[t]{5.1cm}
\renewcommand{\arraystretch}{1.5}
\begin{tabular}[t]{c|c}
\multicolumn{2}{c}{\boldmath$(\bar L R)(\bar R L)D^2 + \hc$} \\
\hline
$\O_{\nu eudD^2}^{(1, SS)}$  &   $D^\mu (\bar \nu_{Lp} e_{Rr}) D_\mu (\bar d_{Rs}  u_{Lt}) $ \\
$\O_{\nu eudD^2}^{(2, SS)}$  &  $(\bar \nu_{Lp}  \overleftrightarrow{D}^\mu e_{Rr}) (\bar d_{Rs} \overleftrightarrow{D}_\mu u_{Lt}) $ \\
$\O_{e^2u^2D^2}^{(1, SS)}$  &   $D^\mu (\bar e_{Lp} e_{Rr}) D_\mu (\bar u_{Rs}  u_{Lt}) $ \\
$\O_{e^2u^2D^2}^{(2, SS)}$  &  $(\bar e_{Lp} \overleftrightarrow{D}^\mu e_{Rr}) (\bar u_{Rs} \overleftrightarrow{D}_\mu u_{Lt}) $ \\
$\O_{e^2d^2D^2}^{(1, SS)}$  &   $D^\mu (\bar e_{Lp} e_{Rr}) D_\mu (\bar d_{Rs}  d_{Lt}) $ \\
$\O_{e^2d^2D^2}^{(2, SS)}$  &  $(\bar e_{Lp} \overleftrightarrow{D}^\mu e_{Rr}) (\bar d_{Rs} \overleftrightarrow{D}_\mu d_{Lt}) $ 
\end{tabular}
\end{minipage}
\end{adjustbox}
\end{center}
\caption{The baryon and lepton number conserving operators of class $\psi^4 D^2$ with two left-handed and two right-handed fermions. 
Operators with $\hc$ have distinct Hermitian conjugates.
The subscripts $p, r, s, t$ are weak-eigenstate indices.}
\label{tab:left8_psi4d2_LR}
\end{table}

%% file: sections/basis_psi4d2_llll.tex

\begin{table}[H]
\begin{center}
\begin{adjustbox}{width=0.84\textwidth,center}
\small
\begin{minipage}[t]{5.8cm}
\renewcommand{\arraystretch}{1.5}
\begin{tabular}[t]{c|c}
\multicolumn{2}{c}{\boldmath$(\bar L R)(\bar L R)D^2 + \hc$} \\
\hline
$\O_{\nu eudD^2}^{(3, SS)}$  &  $D^\mu (\bar \nu_{Lp} e_{Rr}) D_\mu (\bar d_{Ls} u_{Rt}) $ \\
$\O_{\nu eudD^2}^{(4, SS)}$  &  $(\bar \nu_{Lp} \overleftrightarrow{D}^\mu e_{Rr}) (\bar d_{Ls} \overleftrightarrow{D}_\mu u_{Rt}) $ \\
$\O_{\nu eudD^2}^{(1, TT)}$  &  $D^\lambda (\bar \nu_{Lp} \sigma^{\mu\rho} e_{Rr}) D_\lambda (\bar d_{Ls} \sigma_{\mu\rho} u_{Rt}) $ \\
$\O_{e^4D^2}^{(1, SS)}$  &  $D^\mu (\bar e_{Lp} e_{Rr}) D_\mu (\bar e_{Ls} e_{Rt}) $ \\
$\O_{e^4D^2}^{(1, TT)}$  &  $D^\lambda (\bar e_{Lp} \sigma^{\mu\rho} e_{Rr}) D_\lambda (\bar e_{Ls} \sigma_{\mu\rho}  e_{Rt}) $ \\
$\O_{e^2u^2D^2}^{(3, SS)}$  &  $D^\mu (\bar e_{Lp} e_{Rr}) D_\mu (\bar u_{Ls} u_{Rt}) $ \\
$\O_{e^2u^2D^2}^{(4, SS)}$  &  $(\bar e_{Lp} \overleftrightarrow{D}^\mu e_{Rr}) (\bar u_{Ls} \overleftrightarrow{D}_\mu u_{Rt}) $ \\
$\O_{e^2u^2D^2}^{(1, TT)}$  &  $D^\lambda (\bar e_{Lp} \sigma^{\mu\rho} e_{Rr}) D_\lambda (\bar u_{Ls} \sigma_{\mu\rho} u_{Rt}) $ \\
$\O_{e^2d^2D^2}^{(3, SS)}$  &  $D^\mu (\bar e_{Lp} e_{Rr}) D_\mu (\bar d_{Ls} d_{Rt}) $ \\
$\O_{e^2d^2D^2}^{(4, SS)}$  &  $(\bar e_{Lp} \overleftrightarrow{D}^\mu e_{Rr}) (\bar d_{Ls} \overleftrightarrow{D}_\mu d_{Rt}) $ \\
$\O_{e^2d^2D^2}^{(1, TT)}$  &  $D^\lambda (\bar e_{Lp} \sigma^{\mu\rho} e_{Rr}) D_\lambda (\bar d_{Ls} \sigma_{\mu\rho} d_{Rt}) $ 
\end{tabular}
\end{minipage}
\hspace{1cm}
\begin{minipage}[t]{5.8cm}
\renewcommand{\arraystretch}{1.5}
\begin{tabular}[t]{c|c}
\multicolumn{2}{c}{\boldmath$(\bar L R)(\bar L R)D^2 + \hc$} \\
\hline
$\O_{u^4D^2}^{(1, SS)}$  &  $D^\mu (\bar u_{Lp} u_{Rr}) D_\mu (\bar u_{Ls} u_{Rt}) $ \\
$\O_{u^4D^2}^{(2, SS)}$  &  $(\bar u_{Lp} \overleftrightarrow{D}^\mu u_{Rr}) (\bar u_{Ls} \overleftrightarrow{D}_\mu u_{Rt}) $ \\
$\O_{u^4D^2}^{(3, SS)}$  &  $D^\mu (\bar u_{Lp} T^A u_{Rr}) D_\mu (\bar u_{Ls} T^A u_{Rt}) $ \\
$\O_{d^4D^2}^{(1, SS)}$  &  $D^\mu (\bar d_{Lp} d_{Rr}) D_\mu (\bar d_{Ls} d_{Rt}) $ \\
$\O_{d^4D^2}^{(2, SS)}$  &  $(\bar d_{Lp} \overleftrightarrow{D}^\mu d_{Rr}) (\bar d_{Ls} \overleftrightarrow{D}_\mu d_{Rt}) $ \\
$\O_{d^4D^2}^{(3, SS)}$  &  $D^\mu (\bar d_{Lp} T^A d_{Rr}) D_\mu (\bar d_{Ls} T^A d_{Rt}) $ \\
$\O_{u^2d^2D^2}^{(1, SS)}$  &  $D^\mu (\bar u_{Lp} u_{Rr}) D_\mu (\bar d_{Ls} d_{Rt}) $ \\
$\O_{u^2d^2D^2}^{(2, SS)}$  &  $(\bar u_{Lp} \overleftrightarrow{D}^\mu u_{Rr}) (\bar d_{Ls} \overleftrightarrow{D}_\mu d_{Rt}) $ \\
$\O_{u^2d^2D^2}^{(3, SS)}$  &  $D^\mu (\bar u_{Lp} T^A u_{Rr}) D_\mu (\bar d_{Ls} T^A d_{Rt}) $ \\
$\O_{(ud)^2D^2}^{(1, SS)}$  &  $D^\mu (\bar d_{Lp} u_{Rr}) D_\mu (\bar u_{Ls} d_{Rt}) $ \\
$\O_{(ud)^2D^2}^{(2, SS)}$  &  $(\bar d_{Lp} \overleftrightarrow{D}^\mu u_{Rr}) (\bar u_{Ls} \overleftrightarrow{D}_\mu d_{Rt}) $ \\
$\O_{(ud)^2D^2}^{(3, SS)}$  &  $D^\mu (\bar d_{Lp} T^A u_{Rr}) D_\mu (\bar u_{Ls} T^A d_{Rt}) $ 
\end{tabular}
\end{minipage}
\end{adjustbox}
\end{center}
\caption{The baryon and lepton number conserving operators of class $\psi^4 D^2$ and subclass $\psi_R^4 D^2$. 
All of the operators in this table have distinct Hermitian conjugates.
The subscripts $p, r, s, t$ are weak-eigenstate indices.
Operators below the dashed lines vanish when there is only one generation of fermions.}
\label{tab:left8_psi4d2_LR_LR}
\end{table}

%% file: sections/basis_dB0dL2L4.tex

\begin{table}[H]
\begin{center}
\begin{adjustbox}{width=0.89\textwidth,center}
\small
\begin{minipage}[t]{5.8cm}
\renewcommand{\arraystretch}{1.5}
\begin{tabular}[t]{c|c}
\multicolumn{2}{c}{\boldmath$\Delta L = 2: \psi_L^2 \psi_R^2 X + \hc$} \\
\hline
$\O_{\nu^2e^2F}^{(1, ST)}$  &  $(\nu_{Lp}^\top C  \nu_{Lr}) (\bar e_{Ls} \sigma^{\mu\rho} e_{Rt}) F_{\mu\rho} $  \\ 
$\O_{\nu^2u^2G}^{(1, ST)}$  &  $(\nu_{Lp}^\top C  \nu_{Lr}) (\bar u_{Ls} \sigma^{\mu\rho} T^A u_{Rt}) G_{\mu\rho}^A $  \\ 
$\O_{\nu^2u^2F}^{(1, ST)}$  &  $(\nu_{Lp}^\top C  \nu_{Lr}) (\bar u_{Ls} \sigma^{\mu\rho} u_{Rt}) F_{\mu\rho} $  \\ 
$\O_{\nu^2d^2G}^{(1, ST)}$  &  $(\nu_{Lp}^\top C  \nu_{Lr}) (\bar d_{Ls} \sigma^{\mu\rho} T^A d_{Rt}) G_{\mu\rho}^A $  \\ 
$\O_{\nu^2d^2F}^{(1, ST)}$  &  $(\nu_{Lp}^\top C  \nu_{Lr}) (\bar d_{Ls} \sigma^{\mu\rho} d_{Rt}) F_{\mu\rho} $  \\ 
$\O_{\nu eudG}^{(1, RL)}$  &  $(\nu_{Lp}^\top C  \gamma^\mu e_{Rr}) (\bar d_{Ls} \gamma^\rho T^A u_{Lt}) G_{\mu\rho}^A $  \\ 
$\O_{\nu eudG}^{(2, RL)}$  &  $(\nu_{Lp}^\top C  \gamma^\mu e_{Rr}) (\bar d_{Ls} \gamma^\rho T^A u_{Lt}) \widetilde G_{\mu\rho}^A $  \\ 
$\O_{\nu eudG}^{(1, RR)}$  &  $(\nu_{Lp}^\top C  \gamma^\mu e_{Rr}) (\bar d_{Rs} \gamma^\rho T^A u_{Rt}) G_{\mu\rho}^A $  \\ 
$\O_{\nu eudG}^{(2, RR)}$  &  $(\nu_{Lp}^\top C  \gamma^\mu e_{Rr}) (\bar d_{Rs} \gamma^\rho T^A u_{Rt}) \widetilde G_{\mu\rho}^A $  \\ 
$\O_{\nu eudG}^{(3, ST)}$  &  $(\nu_{Lp}^\top C  e_{Lr}) (\bar d_{Ls} \sigma^{\mu\rho} T^A u_{Rt}) G_{\mu\rho}^A $  \\ 
$\O_{\nu eudG}^{(3, TS)}$  &  $(\nu_{Lp}^\top C  \sigma^{\mu\rho} e_{Lr}) (\bar d_{Ls} T^A u_{Rt}) G_{\mu\rho}^A $  \\ 
$\O_{\nu eudF}^{(1, RL)}$  &  $(\nu_{Lp}^\top C  \gamma^\mu e_{Rr}) (\bar d_{Ls} \gamma^\rho u_{Lt}) F_{\mu\rho}$  \\ 
$\O_{\nu eudF}^{(2, RL)}$  &  $(\nu_{Lp}^\top C  \gamma^\mu e_{Rr}) (\bar d_{Ls} \gamma^\rho u_{Lt}) \widetilde F_{\mu\rho} $  \\ 
$\O_{\nu eudF}^{(1, RR)}$  &  $(\nu_{Lp}^\top C  \gamma^\mu e_{Rr}) (\bar d_{Rs} \gamma^\rho u_{Rt}) F_{\mu\rho} $  \\ 
$\O_{\nu eudF}^{(2, RR)}$  &  $(\nu_{Lp}^\top C  \gamma^\mu e_{Rr}) (\bar d_{Rs} \gamma^\rho u_{Rt}) \widetilde F_{\mu\rho} $  \\ 
$\O_{\nu eudF}^{(3, ST)}$  &  $(\nu_{Lp}^\top C  e_{Lr}) (\bar d_{Ls} \sigma^{\mu\rho} u_{Rt}) F_{\mu\rho} $  \\ 
$\O_{\nu eudF}^{(3, TS)}$  &  $(\nu_{Lp}^\top C  \sigma^{\mu\rho} e_{Lr}) (\bar d_{Ls} u_{Rt}) F_{\mu\rho} $  \\ \hdashline
$\O_{\nu^2e^2F}^{(1, TS)}$  &  $(\nu_{Lp}^\top C  \sigma^{\mu\rho} \nu_{Lr}) (\bar e_{Ls} e_{Rt}) F_{\mu\rho} $ \\
$\O_{\nu^2u^2G}^{(1, TS)}$  &  $(\nu_{Lp}^\top C  \sigma^{\mu\rho} \nu_{Lr}) (\bar u_{Ls}  T^A u_{Rt}) G_{\mu\rho}^A $ \\
$\O_{\nu^2u^2F}^{(1, TS)}$  &  $(\nu_{Lp}^\top C  \sigma^{\mu\rho} \nu_{Lr}) (\bar u_{Ls}  u_{Rt}) F_{\mu\rho} $ \\
$\O_{\nu^2d^2G}^{(1, TS)}$  &  $(\nu_{Lp}^\top C  \sigma^{\mu\rho} \nu_{Lr}) (\bar d_{Ls}  T^A d_{Rt}) G_{\mu\rho}^A $ \\
$\O_{\nu^2d^2F}^{(1, TS)}$  &  $(\nu_{Lp}^\top C  \sigma^{\mu\rho} \nu_{Lr}) (\bar d_{Ls}  d_{Rt}) F_{\mu\rho} $
\end{tabular}
\end{minipage}
\hspace{1cm}
\begin{minipage}[t]{6.6cm}
\renewcommand{\arraystretch}{1.5}
\begin{tabular}[t]{c|c}
\multicolumn{2}{c}{\boldmath$\Delta L = 2: \psi_L^4 X_L + \hc$} \\
\hline
$\O_{\nu^2e^2F}^{(2, ST)}$ & $(\nu_{Lp}^\top C e_{Lr}) (\bar e_{Rs}  \sigma^{\mu\rho} \nu_{Lt}) F_{\mu\rho} $ \\
$\O_{\nu^2u^2G}^{(2, ST)}$ & $(T^A)_\alpha^\beta (\nu_{Lp}^\top C u_{Lr}^\alpha) (\bar u_{R\beta s}  \sigma^{\mu\rho} \nu_{Lt}) G_{\mu\rho}^A$ \\
$\O_{\nu^2u^2F}^{(2, ST)}$ & $(\nu_{Lp}^\top C u_{Lr}) (\bar u_{Rs}  \sigma^{\mu\rho} \nu_{Lt}) F_{\mu\rho} $ \\
$\O_{\nu^2d^2G}^{(2, ST)}$ & $(T^A)_\alpha^\beta (\nu_{Lp}^\top C d_{Lr}^\alpha) (\bar d_{R\beta s} \sigma^{\mu\rho}  \nu_{Lt}) G_{\mu\rho}^A$ \\
$\O_{\nu^2d^2F}^{(2, ST)}$ & $(\nu_{Lp}^\top C d_{Lr}) (\bar d_{Rs}  \sigma^{\mu\rho} \nu_{Lt}) F_{\mu\rho} $ \\
$\O_{\nu eudG}^{(4, ST)}$  &  $(\nu_{Lp}^\top C e_{Lr}) (\bar d_{Rs} \sigma^{\mu\rho} T^A u_{Lt}) G_{\mu\rho}^A $ \\
$\O_{\nu eudG}^{(4, TS)}$  &  $(\nu_{Lp}^\top C \sigma^{\mu\rho}e_{Lr}) (\bar d_{Rs}  T^A u_{Lt}) G_{\mu\rho}^A $ \\
$\O_{\nu eudG}^{(2, TT)}$  &  $(\nu_{Lp}^\top C \sigma^{\mu\lambda} e_{Lr}) (\bar d_{Rs} \sigma_{\lambda\rho} T^A u_{Lt}) G_\mu^{A \rho} $ \\
$\O_{\nu eudF}^{(4, ST)}$  &  $(\nu_{Lp}^\top C e_{Lr}) (\bar d_{Rs} \sigma^{\mu\rho} u_{Lt}) F_{\mu\rho} $ \\
$\O_{\nu eudF}^{(4, TS)}$  &  $(\nu_{Lp}^\top C \sigma^{\mu\rho}e_{Lr}) (\bar d_{Rs}  u_{Lt}) F_{\mu\rho} $ \\
$\O_{\nu eudF}^{(2, TT)}$  &  $(\nu_{Lp}^\top C \sigma^{\mu\lambda} e_{Lr}) (\bar d_{Rs} \sigma_{\lambda\rho}  u_{Lt}) F_\mu^{\,\, \rho} $ \\ \hdashline
$\O_{\nu^2e^2F}^{(2, TS)}$ & $(\nu_{Lp}^\top C \sigma^{\mu\rho} \nu_{Lr}) (\bar e_{Rs}  e_{Lt}) F_{\mu\rho} $ \\
$\O_{\nu^2u^2G}^{(2, TS)}$ & $(\nu_{Lp}^\top C \sigma^{\mu\rho} \nu_{Lr}) (\bar u_{Rs}  T^A u_{Lt}) G_{\mu\rho}^A$ \\
$\O_{\nu^2u^2F}^{(2, TS)}$ & $(\nu_{Lp}^\top C \sigma^{\mu\rho} \nu_{Lr}) (\bar u_{Rs}  u_{Lt}) F_{\mu\rho} $ \\
$\O_{\nu^2d^2G}^{(2, TS)}$ & $(\nu_{Lp}^\top C \sigma^{\mu\rho} \nu_{Lr}) (\bar d_{Rs}  T^A d_{Lt}) G_{\mu\rho}^A$ \\
$\O_{\nu^2d^2F}^{(2, TS)}$ & $(\nu_{Lp}^\top C \sigma^{\mu\rho} \nu_{Lr}) (\bar d_{Rs}  d_{Lt}) F_{\mu\rho} $ 
\end{tabular}
\end{minipage}
\end{adjustbox}
\begin{adjustbox}{width=0.33\textwidth,center}
\small
\begin{minipage}[t]{4.9cm}
\renewcommand{\arraystretch}{1.5}
\begin{tabular}[t]{c|c}
\multicolumn{2}{c}{\boldmath$\Delta L = 4: \psi_L^4 X_L + \hc$} \\ \hdashline \hdashline
$\O_{\nu^4F}^{(1)}$  &  $(\nu_{Lp}^\top C \nu_{Lr}) (\nu_{Ls}^\top C \sigma^{\mu\rho} \nu_{Lt}) F_{\mu\rho} $ 
\end{tabular}
\end{minipage}
\end{adjustbox}
\end{center}
\caption{The dimension-8 LEFT operators of class $\psi^4 X$ with $\Delta B = 0$ and $\Delta L \neq 0$.
All of the operators in this table have distinct Hermitian conjugates.
The subscripts $p, r, s, t$ are weak-eigenstate indices.
Operators below the (double) dashed lines vanish when there is only one (or two) generation(s) of neutrinos.}
\label{tab:left8_psi4x_dL2}
\end{table}


\begin{table}[H]
\begin{center}
\begin{adjustbox}{width=0.91\textwidth,center}
\small
\begin{minipage}[t]{6.0cm}
\renewcommand{\arraystretch}{1.5}
\begin{tabular}[t]{c|c}
\boldmath$\Delta L = 2$ & \boldmath$\psi_L^2 \psi_R^2 D^2 + \hc$ \\
\hline
$\O_{\nu^2e^2D^2}^{(1, SS)}$  &  $(D^\mu \nu_{Lp}^\top C D_\mu \nu_{Lr}) (\bar e_{Ls} e_{Rt}) $ \\
$\O_{\nu^2u^2D^2}^{(1, SS)}$  &  $(D^\mu \nu_{Lp}^\top C D_\mu \nu_{Lr}) (\bar u_{Ls} u_{Rt}) $ \\
$\O_{\nu^2d^2D^2}^{(1, SS)}$  &  $(D^\mu \nu_{Lp}^\top C D_\mu \nu_{Lr}) (\bar d_{Ls} d_{Rt}) $ \\
$\O_{\nu eudD^2}^{(1, RL)}$  &  $D^\rho (\nu_{Lp}^\top C \gamma^\mu e_{Rr}) D_\rho (\bar d_{Ls} \gamma_\mu u_{Lt}) $ \\
$\O_{\nu eudD^2}^{(2, RL)}$  &  $(\nu_{Lp}^\top C \gamma^\mu \overleftrightarrow{D}^\rho e_{Rr}) (\bar d_{Ls} \gamma_\mu \overleftrightarrow{D}_\rho u_{Lt}) $ \\
$\O_{\nu eudD^2}^{(1, RR)}$  &  $D^\rho (\nu_{Lp}^\top C \gamma^\mu e_{Rr}) D_\rho (\bar d_{Rs} \gamma_\mu u_{Rt}) $ \\
$\O_{\nu eudD^2}^{(2, RR)}$  &  $(\nu_{Lp}^\top C \gamma^\mu \overleftrightarrow{D}^\rho e_{Rr})  (\bar d_{Rs} \gamma_\mu \overleftrightarrow{D}_\rho u_{Rt}) $ \\
$\O_{\nu eudD^2}^{(5, SS)}$  &  $(D^\mu \nu_{Lp}^\top C D_\mu e_{Lr}) (\bar d_{Ls} u_{Rt}) $ \\
$\O_{\nu eudD^2}^{(6, SS)}$  &  $(D_\mu \nu_{Lp}^\top C D_\rho e_{Lr}) (\bar d_{Ls} \sigma^{\mu\rho} u_{Rt}) $ \\ \hdashline
$\O_{\nu^2e^2D^2}^{(2, SS)}$  &  $(D_\mu \nu_{Lp}^\top C D_\rho \nu_{Lr}) (\bar e_{Ls} \sigma^{\mu\rho} e_{Rt}) $ \\
$\O_{\nu^2u^2D^2}^{(2, SS)}$  &  $(D_\mu \nu_{Lp}^\top C D_\rho \nu_{Lr}) (\bar u_{Ls} \sigma^{\mu\rho} u_{Rt}) $ \\
$\O_{\nu^2d^2D^2}^{(2, SS)}$  &  $(D_\mu \nu_{Lp}^\top C D_\rho \nu_{Lr}) (\bar d_{Ls} \sigma^{\mu\rho} d_{Rt}) $ 
\end{tabular}
\end{minipage}
\hspace{1cm}
\begin{minipage}[t]{6.6cm}
\renewcommand{\arraystretch}{1.5}
\begin{tabular}[t]{c|c}
\boldmath$\Delta L = 2$ & \boldmath$\psi_L^4 D^2 + \hc$ \\
\hline
$\O_{\nu eudD^2}^{(7, SS)}$  &  $(D^\mu \nu_{Lp}^\top C D_\mu e_{Lr})  (\bar d_{Rs} u_{Lt}) $ \\
$\O_{\nu eudD^2}^{(8, SS)}$  &  $(D_\mu \nu_{Lp}^\top C D_\rho e_{Lr}) (\bar d_{Rs} \sigma^{\mu\rho}u_{Lt}) $ \\
$\O_{\nu eudD^2}^{(2, TT)}$  &  $(\nu_{Lp}^\top C  \sigma^{\mu\rho} D^\lambda e_{Lr}) (D_\lambda \bar d_{Rs} \sigma_{\mu\rho} u_{Lt}) $ \\
$\O_{\nu^2e^2D^2}^{(3, SS)}$  &  $(D^\mu \nu_{Lp}^\top C D_\mu \nu_{Lr})  (\bar e_{Rs} e_{Lt}) $ \\
$\O_{\nu^2e^2D^2}^{(1, ST)}$  &  $(D_\mu \bar e_{Rp} C D_\rho \nu_{Lr})  (\nu_{Ls}^\top \sigma^{\mu\rho}e_{Lt}) $ \\
$\O_{\nu^2u^2D^2}^{(3, SS)}$  &  $(D^\mu \nu_{Lp}^\top C D_\mu \nu_{Lr})  (\bar u_{Rs} u_{Lt}) $ \\
$\O_{\nu^2u^2D^2}^{(1, ST)}$  &  $(D_\mu \bar u_{R\alpha p} C D_\rho \nu_{Lr})  (\nu_{Ls}^\top  \sigma^{\mu\rho} u_{Lt}^\alpha) $ \\
$\O_{\nu^2d^2D^2}^{(3, SS)}$  &  $(D^\mu \nu_{Lp}^\top C D_\mu \nu_{Lr})  (\bar d_{Rs} d_{Lt}) $ \\
$\O_{\nu^2d^2D^2}^{(1, ST)}$  &  $(D_\mu \bar d_{R\alpha p} C D_\rho \nu_{Lr})  (\nu_{Ls}^\top \sigma^{\mu\rho} d_{Lt}^\alpha) $ \\ \hdashline
$\O_{\nu^2e^2D^2}^{(4, SS)}$  &  $(D^\mu \bar e_{Rp} C D_\mu \nu_{Lr})  (\nu_{Ls}^\top e_{Lt}) $ \\
$\O_{\nu^2u^2D^2}^{(4, SS)}$  &  $(D^\mu \bar u_{R\alpha p} C D_\mu \nu_{Lr})  (\nu_{Ls}^\top  u_{Lt}^\alpha) $ \\
$\O_{\nu^2d^2D^2}^{(4, SS)}$  &  $(D^\mu \bar d_{R\alpha p} C D_\mu \nu_{Lr})  (\nu_{Ls}^\top d_{Lt}^\alpha) $ 
\end{tabular}
\end{minipage}
\end{adjustbox}
\begin{adjustbox}{width=0.33\textwidth,center}
\small
\begin{minipage}[t]{4.9cm}
\renewcommand{\arraystretch}{1.5}
\begin{tabular}[t]{c|c}
\boldmath$\Delta L = 4$ & \boldmath$\psi_L^4 D^2 + \hc$ \\ \hline
$\O_{\nu^4D^2}^{(1)}$  & $(\partial^\mu \nu_{Lp}^\top C \partial_\mu \nu_{Lr}) (\nu_{Ls}^\top C \nu_{Lt})$
\end{tabular}
\end{minipage}
\end{adjustbox}
\end{center}
\caption{The dimension-8 LEFT operators of class $\psi^4 D^2$ with $\Delta B = 0$, $\Delta L \neq  2$. 
All of the operators in this table have distinct Hermitian conjugates.
The subscripts $p, r, s, t$ are weak-eigenstate indices.
Operators below the dashed lines vanish when there is only one generation of neutrinos.}
\label{tab:left8_psi4d2_dL2}
\end{table}

%% file: sections/basis_dB1dL1.tex

\begin{table}[H]
\begin{center}
\begin{adjustbox}{width=0.96\textwidth,center}
\small
\begin{minipage}[t]{6.4cm}
\renewcommand{\arraystretch}{1.5}
\begin{tabular}[t]{c|c}
\multicolumn{2}{c}{\boldmath$\Delta B =  \Delta L = 1: \psi_L^2 \psi_R^2 D^2 + \hc$} \\
\hline
$\O_{\nu ud^2D^2}^{(1, SS)}$  &  $\epsilon_{\alpha\beta\gamma} (D^\mu \nu_{Lp}^\top C D_\mu d_{Lr}^\alpha)  (d_{Rs}^{\beta \top} C u_{Rt}^\gamma) $ \\
$\O_{\nu ud^2D^2}^{(1, ST)}$  &  $\epsilon_{\alpha\beta\gamma} (D_\mu \nu_{Lp}^\top C D_\rho d_{Lr}^\alpha)  (d_{Rs}^{\beta \top} C \sigma^{\mu\rho} u_{Rt}^\gamma) $ \\
$\O_{\nu ud^2D^2}^{(2, ST)}$  &  $\epsilon_{\alpha\beta\gamma} (D_\mu \nu_{Lp}^\top C D_\rho u_{Lr}^\alpha)  (d_{Rs}^{\beta \top} C \sigma^{\mu\rho} d_{Rt}^\gamma) $ \\
$\O_{eu^2dD^2}^{(1, SS)}$  &  $\epsilon_{\alpha\beta\gamma} (D^\mu e_{Lp}^\top C D_\mu u_{Lr}^\alpha)  (u_{Rs}^{\beta \top} C d_{Rt}^\gamma) $ \\
$\O_{eu^2dD^2}^{(2, SS)}$  &  $\epsilon_{\alpha\beta\gamma} (D^\mu e_{Rp}^\top C D_\mu u_{Rr}^\alpha)  (u_{Ls}^{\beta \top} C d_{Lt}^\gamma) $ \\
$\O_{eu^2dD^2}^{(1, ST)}$  &  $\epsilon_{\alpha\beta\gamma} (D_\mu e_{Lp}^\top C D_\rho u_{Lr}^\alpha)  (u_{Rs}^{\beta \top} C \sigma^{\mu\rho} d_{Rt}^\gamma) $ \\
$\O_{eu^2dD^2}^{(2, ST)}$  &  $\epsilon_{\alpha\beta\gamma} (D_\mu e_{Rp}^\top C D_\rho u_{Rr}^\alpha)  (u_{Ls}^{\beta \top} C \sigma^{\mu\rho} d_{Lt}^\gamma) $ \\ 
$\O_{eu^2dD^2}^{(3, ST)}$  &  $\epsilon_{\alpha\beta\gamma} (D_\mu e_{Lp}^\top C D_\rho d_{Lr}^\alpha)  (u_{Rs}^{\beta \top} C \sigma^{\mu\rho} u_{Rt}^\gamma) $ \\
$\O_{eu^2dD^2}^{(4, ST)}$  &  $\epsilon_{\alpha\beta\gamma} (D_\mu e_{Rp}^\top C D_\rho d_{Rr}^\alpha)  (u_{Ls}^{\beta \top} C \sigma^{\mu\rho} u_{Lt}^\gamma) $ \\ \hdashline
$\O_{\nu ud^2D^2}^{(2, SS)}$  &  $\epsilon_{\alpha\beta\gamma} (D^\mu \nu_{Lp}^\top C D_\mu u_{Lr}^\alpha)  (d_{Rs}^{\beta \top} C d_{Rt}^\gamma) $ \\
$\O_{eu^2dD^2}^{(3, SS)}$  &  $\epsilon_{\alpha\beta\gamma} (D^\mu e_{Lp}^\top C D_\mu d_{Lr}^\alpha)  (u_{Rs}^{\beta \top} C u_{Rt}^\gamma) $ \\
$\O_{eu^2dD^2}^{(4, SS)}$  &  $\epsilon_{\alpha\beta\gamma} (D^\mu e_{Rp}^\top C D_\mu d_{Rr}^\alpha)  (u_{Ls}^{\beta \top} C u_{Lt}^\gamma) $ 
\end{tabular}
\end{minipage}
\hspace{1cm}
\begin{minipage}[t]{7.0cm}
\renewcommand{\arraystretch}{1.5}
\begin{tabular}[t]{c|c}
\multicolumn{2}{c}{\boldmath$\Delta B =  \Delta L = 1: \psi_L^4 D^2 + \hc$} \\
\hline
$\O_{\nu ud^2D^2}^{(3, SS)}$  &  $\epsilon_{\alpha\beta\gamma} (D^\mu \nu_{Lp}^\top C D_\mu d_{Lr}^\alpha)  (d_{Ls}^{\beta \top} C u_{Lt}^\gamma) $ \\ 
$\O_{eu^2dD^2}^{(5, SS)}$  &  $\epsilon_{\alpha\beta\gamma} (D^\mu e_{Lp}^\top C D_\mu u_{Lr}^\alpha)  (u_{Ls}^{\beta \top} C d_{Lt}^\gamma) $ \\ 
$\O_{eu^2dD^2}^{(6, SS)}$  &  $\epsilon_{\alpha\beta\gamma} (D^\mu e_{Rp}^\top C D_\mu u_{Rr}^\alpha)  (u_{Rs}^{\beta \top} C d_{Rt}^\gamma) $ \\ \hdashline
$\O_{\nu ud^2D^2}^{(4, SS)}$  &  $\epsilon_{\alpha\beta\gamma} (D^\mu \nu_{Lp}^\top C D_\mu u_{Lr}^\alpha)  (d_{Ls}^{\beta \top} C d_{Lt}^\gamma) $ \\
$\O_{\nu ud^2D^2}^{(3, ST)}$  &  $\epsilon_{\alpha\beta\gamma} (D_\mu \nu_{Lp}^\top C D_\rho d_{Lr}^\alpha)  (d_{Ls}^{\beta \top} C \sigma^{\mu\rho} u_{Lt}^\gamma) $ \\ 
$\O_{eu^2dD^2}^{(7, SS)}$  &  $\epsilon_{\alpha\beta\gamma} (D^\mu e_{Lp}^\top C D_\mu d_{Lr}^\alpha)  (u_{Ls}^{\beta \top} C u_{Lt}^\gamma) $ \\ 
$\O_{eu^2dD^2}^{(5, ST)}$  &  $\epsilon_{\alpha\beta\gamma} (D_\mu e_{Lp}^\top C D_\rho u_{Lr}^\alpha)  (u_{Ls}^{\beta \top} C \sigma^{\mu\rho} d_{Lt}^\gamma) $ \\ 
$\O_{eu^2dD^2}^{(8, SS)}$  &  $\epsilon_{\alpha\beta\gamma} (D^\mu e_{Rp}^\top C D_\mu d_{Rr}^\alpha)  (u_{Rs}^{\beta \top} C u_{Rt}^\gamma) $ \\
$\O_{eu^2dD^2}^{(6, ST)}$  &  $\epsilon_{\alpha\beta\gamma} (D_\mu e_{Rp}^\top C D_\rho u_{Rr}^\alpha)  (u_{Rs}^{\beta \top} C \sigma^{\mu\rho} d_{Rt}^\gamma) $ 
\end{tabular}
\end{minipage}
\end{adjustbox}
\end{center}
\caption{The dimension-8 LEFT operators of class $\psi^4 D^2$ with $\Delta B =  \Delta L = 1$. 
All of the operators in this table have distinct Hermitian conjugates.
The subscripts $p, r, s, t$ are weak-eigenstate indices.
Operators below the dashed lines vanish when there is only one generation of fermions.}
\label{tab:left8_psi4d2_dB1dL1}
\end{table}

\begin{table}[H]
\begin{center}
\begin{adjustbox}{width=0.97\textwidth,center}
\small
\begin{minipage}[t]{6.9cm}
\renewcommand{\arraystretch}{1.5}
\begin{tabular}[t]{c|c}
\multicolumn{2}{c}{\boldmath$\Delta B =  \Delta L = 1: \psi_L^2 \psi_R^2 X_L + \hc$} \\
\hline
$\O_{\nu ud^2G}^{(1, ST)}$  &  $(T^A)_\alpha^\delta \epsilon_{\delta\beta\gamma} (\nu_{Lp}^\top C d_{Lr}^\alpha)  (d_{Rs}^{\beta \top} C \sigma^{\mu\rho} u_{Rt}^\gamma) G_{\mu\rho}^A$ \\
$\O_{\nu ud^2G}^{(2, ST)}$  &  $(T^A)_\beta^\delta \epsilon_{\alpha\delta\gamma} (\nu_{Lp}^\top C d_{Lr}^\alpha)  (d_{Rs}^{\beta \top} C \sigma^{\mu\rho} u_{Rt}^\gamma) G_{\mu\rho}^A$ \\
$\O_{\nu ud^2G}^{(3, ST)}$  &  $(T^A)_\beta^\delta \epsilon_{\alpha\delta\gamma} (\nu_{Lp}^\top C u_{Lr}^\alpha)  (d_{Rs}^{\beta \top} C \sigma^{\mu\rho} d_{Rt}^\gamma) G_{\mu\rho}^A$ \\
$\O_{\nu ud^2G}^{(1, TS)}$  &  $(T^A)_\alpha^\delta \epsilon_{\delta\beta\gamma} (\nu_{Lp}^\top C \sigma^{\mu\rho} d_{Lr}^\alpha)  (d_{Rs}^{\beta \top} C u_{Rt}^\gamma) G_{\mu\rho}^A$ \\
$\O_{\nu ud^2G}^{(2, TS)}$  &  $(T^A)_\beta^\delta \epsilon_{\alpha\delta\gamma} (\nu_{Lp}^\top C \sigma^{\mu\rho} d_{Lr}^\alpha)  (d_{Rs}^{\beta \top} C u_{Rt}^\gamma) G_{\mu\rho}^A$ \\
$\O_{\nu ud^2G}^{(3, TS)}$  &  $(T^A)_\beta^\delta \epsilon_{\alpha\delta\gamma} (\nu_{Lp}^\top C  \sigma^{\mu\rho} u_{Lr}^\alpha)  (d_{Rs}^{\beta \top} C d_{Rt}^\gamma) G_{\mu\rho}^A$ \\
$\O_{\nu ud^2F}^{(1, ST)}$  &  $(\nu_{Lp}^\top C d_{Lr}^\alpha)  (d_{Rs}^{\beta \top} C \sigma^{\mu\rho} u_{Rt}^\gamma) F_{\mu\rho}$ \\
$\O_{\nu ud^2F}^{(2, ST)}$  &  $ (\nu_{Lp}^\top C u_{Lr}^\alpha)  (d_{Rs}^{\beta \top} C \sigma^{\mu\rho} d_{Rt}^\gamma) F_{\mu\rho}$ \\
$\O_{\nu ud^2F}^{(1, TS)}$  &  $(\nu_{Lp}^\top C \sigma^{\mu\rho} d_{Lr}^\alpha)  (d_{Rs}^{\beta \top} C u_{Rt}^\gamma) F_{\mu\rho}$ \\ 
$\O_{eu^2dG}^{(1, ST)}$  &  $(T^A)_\alpha^\delta \epsilon_{\delta\beta\gamma} (e_{Lp}^\top C u_{Lr}^\alpha)  (u_{Rs}^{\beta \top} C \sigma^{\mu\rho} d_{Rt}^\gamma) G_{\mu\rho}^A$ \\
$\O_{eu^2dG}^{(2, ST)}$  &  $(T^A)_\beta^\delta \epsilon_{\alpha\delta\gamma} (e_{Lp}^\top C u_{Lr}^\alpha)  (u_{Rs}^{\beta \top} C \sigma^{\mu\rho} d_{Rt}^\gamma) G_{\mu\rho}^A$ \\
$\O_{eu^2dG}^{(3, ST)}$  &  $(T^A)_\alpha^\delta \epsilon_{\delta\beta\gamma} (e_{Rp}^\top C u_{Rr}^\alpha)  (u_{Ls}^{\beta \top} C  \sigma^{\mu\rho} d_{Lt}^\gamma) G_{\mu\rho}^A$ \\
$\O_{eu^2dG}^{(4, ST)}$  &  $(T^A)_\beta^\delta \epsilon_{\alpha\delta\gamma} (e_{Rp}^\top C u_{Rr}^\alpha)  (u_{Ls}^{\beta \top} C  \sigma^{\mu\rho} d_{Lt}^\gamma) G_{\mu\rho}^A$ \\
$\O_{edu^2G}^{(1, ST)}$  &  $(T^A)_\beta^\delta \epsilon_{\alpha\delta\gamma}  (e_{Rp}^\top C d_{Rr}^\alpha)  (u_{Ls}^{\beta \top} C  \sigma^{\mu\rho} u_{Lt}^\gamma) G_{\mu\rho}^A$ \\
$\O_{edu^2G}^{(2, ST)}$  &  $(T^A)_\beta^\delta \epsilon_{\alpha\delta\gamma}  (e_{Lp}^\top C d_{Lr}^\alpha)  (u_{Rs}^{\beta \top} C \sigma^{\mu\rho} u_{Rt}^\gamma) G_{\mu\rho}^A$ \\
$\O_{eu^2dG}^{(1, TS)}$  &  $(T^A)_\beta^\delta \epsilon_{\alpha\delta\gamma} (e_{Rp}^\top C \sigma^{\mu\rho}u_{Rr}^\alpha)  (d_{Ls}^{\beta \top} C  u_{Lt}^\gamma) G_{\mu\rho}^A$ \\
$\O_{eu^2dG}^{(2, TS)}$  &  $(T^A)_\beta^\delta \epsilon_{\alpha\delta\gamma} (e_{Lp}^\top C \sigma^{\mu\rho} u_{Lr}^\alpha)  (d_{Rs}^{\beta \top} C u_{Rt}^\gamma) G_{\mu\rho}^A$ \\ 
$\O_{eu^2dG}^{(3, TS)}$  &  $(T^A)_\beta^\delta \epsilon_{\alpha\delta\gamma}  (e_{Rp}^\top C \sigma^{\mu\rho} d_{Rr}^\alpha)  (u_{Ls}^{\beta \top} C  u_{Lt}^\gamma) G_{\mu\rho}^A$ \\ 
$\O_{eu^2dG}^{(4, TS)}$  &  $(T^A)_\beta^\delta \epsilon_{\alpha\delta\gamma}  (e_{Lp}^\top C \sigma^{\mu\rho} d_{Lr}^\alpha)  (u_{Rs}^{\beta \top} C u_{Rt}^\gamma) G_{\mu\rho}^A$ \\
$\O_{eu^2dF}^{(1, ST)}$  &  $(e_{Lp}^\top C u_{Lr}^\alpha)  (d_{Rs}^{\beta \top} C \sigma^{\mu\rho} u_{Rt}^\gamma) F_{\mu\rho}$ \\
$\O_{eu^2dF}^{(2, ST)}$  &  $(e_{Rp}^\top C u_{Rr}^\alpha)  (d_{Ls}^{\beta \top} C  \sigma^{\mu\rho} u_{Lt}^\gamma) F_{\mu\rho}$ \\
$\O_{eu^2dF}^{(3, ST)}$  &  $ (e_{Rp}^\top C d_{Rr}^\alpha)  (u_{Ls}^{\beta \top} C  \sigma^{\mu\rho} u_{Lt}^\gamma) F_{\mu\rho}$ \\
$\O_{eu^2dF}^{(4, ST)}$  &  $ (e_{Lp}^\top C d_{Lr}^\alpha)  (u_{Rs}^{\beta \top} C \sigma^{\mu\rho} u_{Rt}^\gamma) F_{\mu\rho}$ \\
$\O_{eu^2dF}^{(1, TS)}$  &  $(e_{Rp}^\top C \sigma^{\mu\rho}u_{Rr}^\alpha)  (d_{Ls}^{\beta \top} C  u_{Lt}^\gamma) F_{\mu\rho}$ \\
$\O_{eu^2dF}^{(2, TS)}$  &  $(e_{Lp}^\top C \sigma^{\mu\rho} u_{Lr}^\alpha)  (d_{Rs}^{\beta \top} C u_{Rt}^\gamma) F_{\mu\rho}$ \\ \hdashline
$\O_{\nu ud^2F}^{(2, TS)}$  &  $ (\nu_{Lp}^\top C \sigma^{\mu\rho} u_{Lr}^\alpha)  (d_{Rs}^{\beta \top} C d_{Rt}^\gamma) F_{\mu\rho}$ \\
$\O_{eu^2dF}^{(3, TS)}$  &  $ (e_{Rp}^\top C \sigma^{\mu\rho} d_{Rr}^\alpha)  (u_{Ls}^{\beta \top} C  u_{Lt}^\gamma) F_{\mu\rho}$ \\ 
$\O_{eu^2dF}^{(4, TS)}$  &  $ (e_{Lp}^\top C \sigma^{\mu\rho} d_{Lr}^\alpha)  (u_{Rs}^{\beta \top} C u_{Rt}^\gamma) F_{\mu\rho}$ 
\end{tabular}
\end{minipage}
\hspace{1cm}
\begin{minipage}[t]{6.7cm}
\renewcommand{\arraystretch}{1.5}
\begin{tabular}[t]{c|c}
\multicolumn{2}{c}{\boldmath$\Delta B =  \Delta L = 1: \psi_L^4 X_L + \hc$} \\
\hline
$\O_{\nu ud^2G}^{(4, ST)}$  &  $(T^A)_\beta^\delta \epsilon_{\alpha\delta\gamma} (\nu_{Lp}^\top C u_{Lr}^\alpha)  (d_{Ls}^{\beta \top} C \sigma^{\mu\rho} u_{Lt}^\gamma) G_{\mu\rho}^A$ \\
$\O_{\nu ud^2G}^{(5, ST)}$  &  $(T^A)_\beta^\delta \epsilon_{\alpha\delta\gamma} (\nu_{Lp}^\top C u_{Lr}^\alpha)  (d_{Ls}^{\beta \top} C \sigma^{\mu\rho} d_{Lt}^\gamma) G_{\mu\rho}^A$ \\
$\O_{\nu ud^2G}^{(6, ST)}$  &  $(T^A)_\gamma^\delta \epsilon_{\alpha\beta\delta} (e_{Lp}^\top C u_{Lr}^\alpha)  (d_{Ls}^{\beta \top} C \sigma^{\mu\rho} u_{Lt}^\gamma) G_{\mu\rho}^A$\\
$\O_{\nu ud^2F}^{(3, ST)}$  &  $ (\nu_{Lp}^\top C u_{Lr}^\alpha)  (d_{Ls}^{\beta \top} C \sigma^{\mu\rho} u_{Lt}^\gamma) F_{\mu\rho}$ \\
$\O_{\nu ud^2F}^{(4, ST)}$  &  $ (\nu_{Lp}^\top C u_{Lr}^\alpha)  (d_{Ls}^{\beta \top} C \sigma^{\mu\rho} d_{Lt}^\gamma) F_{\mu\rho}$ \\
$\O_{eu^2dG}^{(5, ST)}$  &  $(T^A)_\beta^\delta \epsilon_{\alpha\delta\gamma} (e_{Lp}^\top C u_{Lr}^\alpha)  (u_{Ls}^{\beta \top} C \sigma^{\mu\rho} d_{Lt}^\gamma) G_{\mu\rho}^A$ \\
$\O_{eu^2dG}^{(6, ST)}$  &  $(T^A)_\beta^\delta \epsilon_{\alpha\delta\gamma} (e_{Rp}^\top C u_{Rr}^\alpha)  (u_{Rs}^{\beta \top} C \sigma^{\mu\rho} d_{Rt}^\gamma) G_{\mu\rho}^A$ \\
$\O_{edu^2G}^{(3, ST)}$  &  $(T^A)_\beta^\delta \epsilon_{\alpha\delta\gamma} (e_{Lp}^\top C d_{Lr}^\alpha)  (u_{Ls}^{\beta \top} C \sigma^{\mu\rho} u_{Lt}^\gamma) G_{\mu\rho}^A$ \\
$\O_{edu^2G}^{(4, ST)}$  &  $(T^A)_\gamma^\delta \epsilon_{\alpha\beta\delta} (e_{Lp}^\top C d_{Lr}^\alpha)  (u_{Ls}^{\beta \top} C \sigma^{\mu\rho} u_{Lt}^\gamma) G_{\mu\rho}^A$\\
$\O_{edu^2G}^{(5, ST)}$  &  $(T^A)_\beta^\delta \epsilon_{\alpha\delta\gamma} (e_{Rp}^\top C d_{Rr}^\alpha)  (u_{Rs}^{\beta \top} C \sigma^{\mu\rho} u_{Rt}^\gamma) G_{\mu\rho}^A$ \\
$\O_{edu^2G}^{(6, ST)}$  &  $(T^A)_\gamma^\delta \epsilon_{\alpha\beta\delta} (e_{Rp}^\top C d_{Rr}^\alpha)  (u_{Rs}^{\beta \top} C \sigma^{\mu\rho} u_{Rt}^\gamma) G_{\mu\rho}^A$\\
$\O_{eu^2dF}^{(5, ST)}$  &  $(e_{Lp}^\top C u_{Lr}^\alpha)  (u_{Ls}^{\beta \top} C \sigma^{\mu\rho} d_{Lt}^\gamma) F_{\mu\rho}$ \\
$\O_{eu^2dF}^{(6, ST)}$  &  $(e_{Rp}^\top C u_{Rr}^\alpha)  (u_{Rs}^{\beta \top} C \sigma^{\mu\rho} d_{Rt}^\gamma) F_{\mu\rho}$ \\
$\O_{eu^2dF}^{(7, ST)}$  &  $(e_{Lp}^\top C d_{Lr}^\alpha)  (u_{Ls}^{\beta \top} C \sigma^{\mu\rho} u_{Lt}^\gamma) F_{\mu\rho}$ \\
$\O_{eu^2dF}^{(8, ST)}$  &  $(e_{Rp}^\top C d_{Rr}^\alpha)  (u_{Rs}^{\beta \top} C \sigma^{\mu\rho} u_{Rt}^\gamma) F_{\mu\rho}$ 
\end{tabular}
\end{minipage}
\end{adjustbox}
\end{center}
\caption{The dimension-8 LEFT operators of class $\psi^4 X$ with $\Delta B =  \Delta L = 1$. 
All of the operators in this table have distinct Hermitian conjugates.
The subscripts $p, r, s, t$ are weak-eigenstate indices.
Operators below the dashed lines vanish when there is only one generation of fermions.}
\label{tab:left8_psi4x_dB1dL1}
\end{table}


%% file: sections/basis_dB1dLm1.tex

\begin{table}[H]
\begin{center}
\begin{adjustbox}{width=0.95\textwidth,center}
\small
\begin{minipage}[t]{6.6cm}
\renewcommand{\arraystretch}{1.5}
\begin{tabular}[t]{c|c}
\multicolumn{2}{c}{\boldmath$\Delta B = - \Delta L = 1: \psi_L^2 \psi_R^2 X_L + \hc$} \\
\hline
$\O_{\bar \nu ud^2G}^{(1)}$  &  $(T^A)_\alpha^\delta \epsilon_{\delta\beta\gamma} (\bar \nu_{Lp} d_{Rr}^\alpha)  (d_{Ls}^{\beta \top} C \sigma^{\mu\rho} u_{Lt}^\gamma) G_{\mu\rho}^A$ \\
$\O_{\bar\nu ud^2G}^{(2)}$  &  $(T^A)_\beta^\delta \epsilon_{\alpha\delta\gamma} (\bar \nu_{Lp} d_{Rr}^\alpha)  (d_{Ls}^{\beta \top} C \sigma^{\mu\rho} u_{Lt}^\gamma) G_{\mu\rho}^A$ \\
$\O_{\bar \nu ud^2G}^{(3)}$  &  $(T^A)_\beta^\delta \epsilon_{\alpha\delta\gamma} (\bar \nu_{Lp} u_{Rr}^\alpha)  (d_{Ls}^{\beta \top} C \sigma^{\mu\rho} d_{Lt}^\gamma) G_{\mu\rho}^A$ \\
$\O_{\bar \nu ud^2G}^{(4)}$  &  $(T^A)_\alpha^\delta \epsilon_{\delta\beta\gamma} (\bar \nu_{Lp} \sigma^{\mu\rho} d_{Rr}^\alpha)  (d_{Ls}^{\beta \top} C u_{Lt}^\gamma) G_{\mu\rho}^A$ \\
$\O_{\bar \nu ud^2G}^{(5)}$  &  $(T^A)_\beta^\delta \epsilon_{\alpha\delta\gamma} (\bar \nu_{Lp} \sigma^{\mu\rho} d_{Rr}^\alpha)  (d_{Ls}^{\beta \top} C u_{Lt}^\gamma) G_{\mu\rho}^A$ \\
$\O_{\bar \nu ud^2G}^{(6)}$  &  $(T^A)_\beta^\delta \epsilon_{\alpha\delta\gamma} (\bar \nu_{Lp} \sigma^{\mu\rho} u_{Rr}^\alpha)  (d_{Ls}^{\beta \top} C d_{Lt}^\gamma) G_{\mu\rho}^A$ \\
$\O_{\bar \nu ud^2F}^{(1)}$  &  $\epsilon_{\alpha\beta\gamma} (\bar \nu_{Lp} d_{Rr}^\alpha)  (d_{Ls}^{\beta \top} C \sigma^{\mu\rho} u_{Lt}^\gamma) F_{\mu\rho}$ \\
$\O_{\bar \nu ud^2F}^{(2)}$  &  $\epsilon_{\alpha\beta\gamma} (\bar \nu_{Lp} u_{Rr}^\alpha)  (d_{Ls}^{\beta \top} C \sigma^{\mu\rho} d_{Lt}^\gamma) F_{\mu\rho}$ \\
$\O_{\bar \nu ud^2F}^{(3)}$  &  $\epsilon_{\alpha\beta\gamma} (\bar \nu_{Lp} \sigma^{\mu\rho} d_{Rr}^\alpha)  (d_{Ls}^{\beta \top} C u_{Lt}^\gamma) F_{\mu\rho}$ \\
$\O_{\bar ed^3G}^{(1)}$  &  $(T^A)_\beta^\delta \epsilon_{\alpha\beta\gamma} (\bar e_{Lp} d_{Rr}^\alpha)  (d_{Ls}^{\beta \top} C \sigma^{\mu\rho} d_{Lt}^\gamma) G_{\mu\rho}^A$ \\
$\O_{\bar ed^3G}^{(2)}$  &  $(T^A)_\beta^\delta \epsilon_{\alpha\beta\gamma} (\bar e_{Rp} \sigma^{\mu\rho} d_{Lr}^\alpha)  (d_{Rs}^{\beta \top} C  d_{Rt}^\gamma) G_{\mu\rho}^A$ \\ 
$\O_{\bar ed^3G}^{(3)}$  &  $(T^A)_\beta^\delta \epsilon_{\alpha\beta\gamma} (\bar e_{Lp} \sigma^{\mu\rho} d_{Rr}^\alpha)  (d_{Ls}^{\beta \top} C  d_{Lt}^\gamma) G_{\mu\rho}^A$ \\
$\O_{\bar ed^3G}^{(4)}$  &  $(T^A)_\beta^\delta \epsilon_{\alpha\beta\gamma} (\bar e_{Rp} d_{Lr}^\alpha)  (d_{Rs}^{\beta \top} C \sigma^{\mu\rho}  d_{Rt}^\gamma) G_{\mu\rho}^A$ \\ 
$\O_{\bar ed^3F}^{(1)}$  &  $\epsilon_{\alpha\beta\gamma} (\bar e_{Rp} d_{Lr}^\alpha)  (d_{Rs}^{\beta \top} C  \sigma^{\mu\rho} d_{Rt}^\gamma) F_{\mu\rho}$ \\ 
$\O_{\bar ed^3F}^{(2)}$  &  $\epsilon_{\alpha\beta\gamma} (\bar e_{Lp} d_{Rr}^\alpha)  (d_{Ls}^{\beta \top} C \sigma^{\mu\rho} d_{Lt}^\gamma) F_{\mu\rho}$ \\ \hdashline
$\O_{\bar \nu ud^2F}^{(4)}$  &  $\epsilon_{\alpha\beta\gamma} (\bar \nu_{Lp} \sigma^{\mu\rho} u_{Rr}^\alpha)  (d_{Ls}^{\beta \top} C d_{Lt}^\gamma) F_{\mu\rho}$ \\
$\O_{\bar ed^3F}^{(3)}$  &  $\epsilon_{\alpha\beta\gamma} (\bar e_{Rp} \sigma^{\mu\rho} d_{Lr}^\alpha)  (d_{Rs}^{\beta \top} C  d_{Rt}^\gamma) F_{\mu\rho}$ \\ 
$\O_{\bar ed^3F}^{(4)}$  &  $\epsilon_{\alpha\beta\gamma} (\bar e_{Lp} \sigma^{\mu\rho} d_{Rr}^\alpha)  (d_{Ls}^{\beta \top} C d_{Lt}^\gamma) F_{\mu\rho}$ 
\end{tabular}
\end{minipage}
\hspace{1cm}
\begin{minipage}[t]{6.6cm}
\renewcommand{\arraystretch}{1.5}
\begin{tabular}[t]{c|c}
\multicolumn{2}{c}{\boldmath$\Delta B = - \Delta L = 1: \psi_L^4 X_L + \hc$} \\
\hline
$\O_{\bar \nu ud^2G}^{(7)}$  &  $(T^A)_\beta^\delta \epsilon_{\alpha\delta\gamma} (\bar \nu_{Lp} u_{Rr}^\alpha)  (d_{Rs}^{\beta \top} C \sigma^{\mu\rho} d_{Rt}^\gamma) G_{\mu\rho}^A$ \\
$\O_{\bar \nu ud^2G}^{(8)}$  &  $(T^A)_\beta^\delta \epsilon_{\alpha\delta\gamma} (\bar \nu_{Lp} d_{Rr}^\alpha)  (d_{Rs}^{\beta \top} C \sigma^{\mu\rho} u_{Rt}^\gamma) G_{\mu\rho}^A$ \\
$\O_{\bar \nu ud^2G}^{(9)}$  &  $(T^A)_\beta^\delta \epsilon_{\alpha\delta\gamma} (\bar \nu_{Lp} \sigma^{\mu\rho} u_{Rr}^\alpha)  (d_{Rs}^{\beta \top} C d_{Rt}^\gamma) G_{\mu\rho}^A$ \\
$\O_{\bar \nu ud^2F}^{(5)}$  &  $\epsilon_{\alpha\beta\gamma} (\bar \nu_{Lp} u_{Rr}^\alpha)  (d_{Rs}^{\beta \top} C \sigma^{\mu\rho} d_{Rt}^\gamma) F_{\mu\rho}$ \\
$\O_{\bar \nu ud^2F}^{(6)}$  &  $\epsilon_{\alpha\beta\gamma} (\bar \nu_{Lp} d_{Rr}^\alpha)  (d_{Rs}^{\beta \top} C \sigma^{\mu\rho} u_{Rt}^\gamma) F_{\mu\rho}$ \\
$\O_{\bar ed^3G}^{(5)}$  &  $(T^A)_\alpha^\delta \epsilon_{\delta\beta\gamma} (\bar e_{Lp} d_{Rr}^\alpha)  (d_{Rs}^{\beta \top} C \sigma^{\mu\rho} d_{Rt}^\gamma) G_{\mu\rho}^A$ \\
$\O_{\bar ed^3G}^{(6)}$  &  $(T^A)_\alpha^\delta \epsilon_{\delta\beta\gamma} (\bar e_{Rp} d_{Lr}^\alpha)  (d_{Ls}^{\beta \top} C \sigma^{\mu\rho} d_{Lt}^\gamma) G_{\mu\rho}^A$ \\
$\O_{\bar ed^3F}^{(5)}$  &  $\epsilon_{\alpha\beta\gamma} (\bar e_{Lp} d_{Rr}^\alpha)  (d_{Rs}^{\beta \top} C \sigma^{\mu\rho} d_{Rt}^\gamma) F_{\mu\rho}$ \\
$\O_{\bar ed^3F}^{(6)}$  &  $\epsilon_{\alpha\beta\gamma} (\bar e_{Rp} d_{Lr}^\alpha)  (d_{Ls}^{\beta \top} C \sigma^{\mu\rho} d_{Lt}^\gamma) F_{\mu\rho}$  \\ \hdashline
$\O_{\bar ed^3G}^{(7)}$  &  $(T^A)_\beta^\delta \epsilon_{\alpha\delta\gamma} (\bar e_{Lp} d_{Rr}^\alpha)  (d_{Rs}^{\beta \top} C \sigma^{\mu\rho} d_{Rt}^\gamma) G_{\mu\rho}^A$ \\
$\O_{\bar ed^3G}^{(8)}$  &  $(T^A)_\beta^\delta \epsilon_{\alpha\delta\gamma} (\bar e_{Rp} d_{Lr}^\alpha)  (d_{Ls}^{\beta \top} C \sigma^{\mu\rho} d_{Lt}^\gamma) G_{\mu\rho}^A$ 
\end{tabular}
\end{minipage}
\end{adjustbox}

\begin{adjustbox}{width=0.9\textwidth,center}
\small
\begin{minipage}[t]{6.4cm}
\renewcommand{\arraystretch}{1.5}
\begin{tabular}[t]{c|c}
\multicolumn{2}{c}{\boldmath$\Delta B = - \Delta L = 1: \psi_L^2 \psi_R^2 D^2 + \hc$} \\
\hline
$\O_{\bar \nu ud^2D^2}^{(1)}$  &  $\epsilon_{\alpha\beta\gamma} (D^\mu \bar \nu_{Lp} D_\mu d_{Rr}^\alpha)  (d_{Ls}^{\beta \top} C u_{Lt}^\gamma) $ \\
$\O_{\bar \nu ud^2D^2}^{(2)}$  &  $\epsilon_{\alpha\beta\gamma} (D_\mu \bar \nu_{Lp} D_\rho d_{Rr}^\alpha) (d_{Ls}^{\beta \top} C \sigma^{\mu\rho} u_{Lt}^\gamma) $ \\ 
$\O_{\bar \nu ud^2D^2}^{(3)}$  &  $\epsilon_{\alpha\beta\gamma} (D_\mu \bar \nu_{Lp} D_\rho u_{Rr}^\alpha) (d_{Ls}^{\beta \top} C \sigma^{\mu\rho} d_{Lt}^\gamma) $ \\ 
$\O_{\bar ed^3D^2}^{(1)}$  &  $\epsilon_{\alpha\beta\gamma} (D_\mu \bar e_{Lp} D_\rho d_{Rr}^\alpha) (d_{Ls}^{\beta \top} C \sigma^{\mu\rho} d_{Lt}^\gamma) $ \\ 
$\O_{\bar ed^3D^2}^{(2)}$  &  $\epsilon_{\alpha\beta\gamma} (D_\mu \bar e_{Rp} D_\rho d_{Lr}^\alpha) (d_{Rs}^{\beta \top} C \sigma^{\mu\rho} d_{Rt}^\gamma) $ \\ \hdashline
$\O_{\bar \nu ud^2D^2}^{(4)}$  &  $\epsilon_{\alpha\beta\gamma} (D^\mu \bar \nu_{Lp} D_\mu u_{Rr}^\alpha)  (d_{Ls}^{\beta \top} C d_{Lt}^\gamma) $ \\
$\O_{\bar ed^3D^2}^{(3)}$  &  $\epsilon_{\alpha\beta\gamma} (D^\mu \bar e_{Lp} D_\mu d_{Rr}^\alpha) (d_{Ls}^{\beta \top} C d_{Lt}^\gamma) $ \\ 
$\O_{\bar ed^3D^2}^{(4)}$  &  $\epsilon_{\alpha\beta\gamma} (D^\mu \bar e_{Rp} D_\mu d_{Lr}^\alpha) (d_{Rs}^{\beta \top} C  d_{Rt}^\gamma) $ 
\end{tabular}
\end{minipage}
\hspace{1cm}
\begin{minipage}[t]{6.1cm}
\renewcommand{\arraystretch}{1.5}
\begin{tabular}[t]{c|c}
\multicolumn{2}{c}{\boldmath$\Delta B =  - \Delta L = 1: \psi_L^4 D^2 + \hc$} \\
\hline
$\O_{\bar \nu ud^2D^2}^{(5)}$  &  $\epsilon_{\alpha\beta\gamma} (D^\mu \bar \nu_{Lp} D_\mu d_{Rr}^\alpha)  (d_{Rs}^{\beta \top} C u_{Rt}^\gamma) $  \\ \hdashline
$\O_{\bar \nu ud^2D^2}^{(6)}$  &  $\epsilon_{\alpha\beta\gamma} (D^\mu \bar \nu_{Lp} D_\mu u_{Rr}^\alpha)  (d_{Rs}^{\beta \top} C d_{Rt}^\gamma) $ \\
$\O_{\bar \nu ud^2D^2}^{(7)}$  &  $\epsilon_{\alpha\beta\gamma} (D_\mu \bar \nu_{Lp} D_\rho d_{Rr}^\alpha)  (d_{Rs}^{\beta \top} C \sigma^{\mu\rho} u_{Rt}^\gamma) $ \\ 
$\O_{\bar ed^3D^2}^{(5)}$  &  $\epsilon_{\alpha\beta\gamma} (D^\mu \bar e_{Lp} D_\mu d_{Rr}^\alpha)  (d_{Rs}^{\beta \top} C d_{Rt}^\gamma) $ \\
$\O_{\bar ed^3D^2}^{(6)}$  &  $\epsilon_{\alpha\beta\gamma} (D^\mu \bar e_{Rp} D_\mu d_{Lr}^\alpha)  (d_{Ls}^{\beta \top} C d_{Lt}^\gamma) $ 
\end{tabular}
\end{minipage}
\end{adjustbox}
\end{center}
\caption{The dimension-8 LEFT operators with $\Delta B = - \Delta L = 1$. 
All of the operators in this table have distinct Hermitian conjugates.
The subscripts $p, r, s, t$ are weak-eigenstate indices.
Operators below the dashed lines vanish when there is only one generation of fermions.}
\label{tab:left8_dB1dLm1}
\end{table}